# THE NEXUS OF QUANTUM TECHNOLOGY, INTELLECTUAL PROPERTY, AND NATIONAL SECURITY

DETERRENCE BY DENIAL FOR DEMOCRATIC RESILIENCE: AN LSI TEST FOR SECURING THE QUANTUM INDUSTRIAL COMMONS

Mauritz Kop[*]

## ABSTRACT

Our world of power and national security is increasingly probabilistic: like a quantum wavefunction, it encodes multiple plausible futures until policy choices—and shocks—collapse them into outcomes. In quantum mechanics, a "state" is a probability map of possible futures, encoded in complex amplitudes rather than certainties, and remaining in superposition until measurement collapses it into an observable result. Grand strategy in the quantum era shares this ontology. Policymakers cannot forecast a single trajectory for fault-tolerant qubits, quantum sensors at scale, or entanglement distribution across quantum networks; they must prepare for several plausible "eigenstates" of capability, diffusion, and systemic rivalry.

Great-power-competition scenario-planning methods, such as the Sullivan–Feldman "Eight Worlds" uncertainty framework, are useful precisely because they force governance to remain robust across divergent paths. Whatever the continuing debates about quantum foundations and their tension with classical intuitions, quantum systems are being engineered and deployed—so law and policy should focus on operational control points rather than abstractions.

Quantum technologies have moved from laboratory curiosities to strategic infrastructure, with an approaching "event horizon" reflected in recent U.S. strategic assessments—including the U.S.-China Economic and Security Review Commission's call for a "Quantum First" posture by 2030—and in parallel White House initiatives aimed at securing critical inputs and accelerating trusted innovation. A similar time-compression dynamic is now visible in frontier AI. Notably, Amodei warns that AI is entering a technological "adolescence" in which discontinuous capability gains can outpace post-hoc governance; in turn, quantum-AI convergence can shorten quantum engineering cycles (and thus threat horizons), while quantum-safe migration, verification, and measurement discipline can help mitigate AI-enabled national security risk. Recent U.S.-China Economic and Security Review Commission research further documents that China's quantum program is centrally mobilized and security-aligned, and that the most consequential advantages may arise not only from computing milestones but also from sensing and cryptanalytic applications, thereby sharpening the need for a deterrence-by-denial governance posture.

---

[*] [Mauritz Kop](#) is the Founder of the Stanford Center for Responsible Quantum Technology and a Senior Fellow at the Centre for International Governance Innovation (CIGI).

In the context of this Article, "responsible quantum technology" (RQT) is accordingly presented as values-based deterrence by denial, implemented through a legal, ethical, and institutional control plane that protects the industrial commons and enhances democratic resilience against authoritarian appropriation and exploitation. It also explains how quantum capabilities may reshape strategic stability—especially through sensing, navigation-and-timing, communications, and cryptanalysis—and argues that a values-driven deterrence-by-denial posture, illustrated through a Taiwan invasion-denial vignette, is the least escalatory way to preserve crisis stability while denying adversarial gains.

The Article's central claim is that the United States and its allies should pursue security-sufficient openness, operationalized through a least-trade-restrictive, security-sufficient, innovation-preserving (LSI) test that disciplines both state and private action. The LSI test integrates emerging instruments of economic statecraft, including Project Vault to strengthen the U.S. strategic critical-minerals reserve, a Quantum Criticality Index (QCI) for prioritizing quantum-decisive inputs, the White House Critical Minerals Presidential Order to secure the physical front; and the Genesis Mission–American Science and Security Platform (ASSP) architecture to create secure, closed-loop enclaves for high-sensitivity collaborative R&D.

LSI is applied across the quantum stack—spanning pillarized control points in computing, sensing, simulation, networking, communication, quantum–AI hybrids, and enabling inputs/materials. It is also applied across the quantum stack's upstream dependencies: intellectual property (IP) doctrine and strategy, where incentives must accelerate innovation without fueling the pendulum swing between stifling patent thickets and opaque trade secrecy; public funding and data-rights architecture; calibrated secrecy orders; dual-use export controls, investment screening, and services controls; and standards, certification, and cryptographic baselines (including PQC, crypto-agility, and quantum-network interoperability). Aligned with the 2025 U.S. National Security Strategy, the LSI framework treats supply chains and cloud platforms as entangled governance nodes—chokepoints in cryogenics, isotopes, lasers, and superconducting nanowire single-photon detector (SNSPD)-grade materials on the physical front, and identity, access control, logging, and auditability on the digital front.

The Article's contribution is an implementable coalition playbook, offering administrable, empirically anchored criteria, templates, and differentiated guardrails—including "red-zone" domains where denial is the default—to avoid both over-securitization (which chills publication, standards leadership, and venture formation) and under-securitization (which leaks crown-jewel capabilities that are slow to reacquire). Properly applied, LSI reduces the risk of a self-defeating "Silicon Curtain" while establishing standards-first interoperability as a stabilizing eigenstate of the international order and enabling RQT-by-design to shape trusted adoption pathways beyond the coalition, including in the majority world.

*Mauritz Kop - The Nexus of Quantum Technology, Intellectual Property & National Security*

*(Feb. 8, 2026)*

TABLE OF CONTENTS

Introduction

The nature of the universe is probabilistic, defined not by certainty but by distributions over position and momentum: quantum states, described by complex amplitudes rather than certainties, exist in superposition and unfold over time until measurement collapses them into an observable outcome. Our world of geopolitics is entering a quantum condition: power is defined less by certainty than by probabilities, and the futures that matter most are the ones policy can still force into being. States confront multiple plausible futures—rapid advances in error-corrected qubits, ubiquitous quantum sensors, and globally scaled quantum networks—and must choose legal instruments that remain robust across scenarios. As a matter of metaphor, the Sullivan–Feldman "Eight Worlds" geopolitical uncertainty framework usefully echoes the "many-worlds" image sometimes invoked in quantum foundations: until policy choices and shocks effectively "measure" the environment, strategy must be designed to perform across multiple plausible futures rather than a single forecast.[1] Today's policy environment is itself in superposition: decision-makers oscillate between open-science instincts and restrictive reflexes, between patent disclosure as signaling and secrecy as denial. When enforcement occurs—the "measurement" event—those choices collapse into a durable regulatory state that shapes diffusion, standards, and industrial structure. The governance analogue is that policy must be designed for discontinuities—capability jumps that collapse uncertainty into irreversible deployment—an intuition reinforced by AI's current trajectory and sharpened by AI–quantum convergence that can shorten R&D iteration cycles and pull forward national security threat horizons.[2] Because allied ecosystems are entangled through supply chains, cloud platforms, and standards bodies, a policy move on one node propagates system-wide effects. Whatever the philosophical debates over quantum foundations and the limits of classical intuition, the engineering is real. The task is to calibrate the legal Hamiltonian—the combined operator of intellectual property, export and services controls, standards, procurement, technology transfer, and research security—so the system converges on a stable, democratic, innovation-preserving equilibrium. A core claim, developed later in Section VIII.V, is that quantum's most destabilizing pathways are often indirect—through improved detection, targeting, and decision-speed—so governance must be designed to preserve strategic stability while denying adversarial operational advantages, including in a Taiwan contingency.

These dynamics are no longer conjectural. The U.S.-China Economic and Security Review Commission has recommended that Congress establish a "Quantum First" by 2030 national goal, explicitly tying quantum computational advantage to mission-critical domains such as cryptography, drug discovery, and materials science.[3] A 2025 staff research report by Joseph Federici for the U.S.-China Economic and Security Review Commission adds granular evidence on the competitive structure—emphasizing centralized mobilization, military-relevant pathways from research to capability, and the strategic salience of sensing and cryptanalysis.[4] In that setting, the relevant policy levers are operational control points rather than assumptions of market-driven diffusion. The 2025 U.S. National Security Strategy likewise calls for U.S. technology and standards—particularly in AI, biotech, and quantum computing—to drive the world forward.[5] In parallel, the White House's Genesis Mission order establishes the American Science and Security Platform (ASSP) as a trusted, integrated compute and data substrate spanning quantum information science and critical materials,[6] while a January 2026 proclamation under Section 232 treats processed critical minerals and their derivative products as a national-security concern.[7] Project Vault—introduced as a public–private critical-minerals stockpile to support American business—illustrates this same turn from monitoring exposure to pre-positioning supply-chain resilience.[8] A Quantum Criticality Index (QCI)[9] in this context could help prioritize which quantum-relevant inputs merit stockpiling through The Vault, and under what trigger thresholds. This Article's LSI test provides a method to translate security-sufficient openness from rhetoric into a replicable deterrence-by-denial posture across export controls, standards, research security, procurement, and allied technology access.

---

[1] *See* Sullivan, Jake & Feldman, Tal, *Geopolitics in the Age of Artificial Intelligence - Strategy and Power in an Uncertain AI Future,* FOREIGN AFF. (Jan. 27 2026), https://www.foreignaffairs.com/united-states/geopolitics-age-artificial-intelligence
[2] *See* Amodei, Dario, *The Adolescence of Technology: Confronting and Overcoming the Risks of Powerful AI* (Jan. 2026), https://www.darioamodei.com/essay/the-adolescence-of-technology
[3] U.S.-China Econ. & Sec. Rev. Comm'n, 2025 Report to Congress: Executive Summary and Recommendations 13–14, 48–49 (Nov. 2025) (recommending Congress establish a "Quantum First" by 2030 national goal), https://www.uscc.gov/sites/default/files/2025-11/2025_Executive_Summary.pdf

Three dynamics render such discipline urgent. First, quantum's path-dependent development is on track to deliver capabilities that will be difficult to constrain after they mature. In its post-quantum cryptography (PQC) effort, NIST has already finalized standards (FIPS 203, 204, 205)[10], NSM-10 frames quantum leadership as inseparable from mitigating risks to vulnerable cryptographic systems,[11] and the Office of Management and Budget (OMB) has directed federal agencies to inventory quantum-vulnerable cryptography and plan PQC migrations. Consistent with recent threat assessments, "harvest-now, decrypt-later" exposure should be treated as a present-tense risk that depends on adversary collection and retention decisions, not on speculative timelines for fault-tolerant quantum computing.[12] Yet PQC migration remains uneven, and quantum computing's future trajectory remains uncertain, implying that governance must operate under deep epistemic uncertainty and in a strategic environment characterized by systemic rivalry.[13] Second, IP and control tools are now being actively retooled to manage the competition for technological supremacy. This "control-plane turn" tracks the parallel AI-competition discourse and chip-industrial policy framing.[14] Meanwhile, more aggressive authoritarian regimes may exploit the diffusion process by integrating quantum-enhanced sensing and communication into domestic surveillance architectures—entrenching repression and exporting authoritarian governance models and spillovers through infrastructure and standards.[15] Third, state power is shifting from trade in finished goods to control of capital and capabilities: the Bureau of Industry and Security (BIS) created new quantum-related entries on the Commerce Control List (including Export Control Classification Number (ECCN) 4A906)[16] and introduced a plurilateral License Exception IEC (Implemented Export Controls) for destinations that have adopted aligned controls; the U.S. Treasury issued the first outbound-investment rule implementing E.O. 14105 with coverage for quantum information technologies;[17] and the European Commission recommended coordinated outbound-investment screening in semiconductors, AI, and quantum.[18]

---

[4] Joseph Federici, *Vying for Quantum Supremacy: U.S.-China Competition in Quantum Technologies* (U.S.-China Econ. & Sec. Rev. Comm'n, Staff Research Report, Nov. 18, 2025). https://www.uscc.gov/research/vying-quantum-supremacy-us-china-competition-quantum-technologies#_Toc214309247

[5] The White House, 2025 National Security Strategy 8 (Dec. 2025) (calling for U.S. technology and standards—particularly in AI, biotech, and quantum computing—to "drive the world forward"), https://www.whitehouse.gov/wp-content/uploads/2025/12/2025-National-Security-Strategy.pdf

[6] The White House, Launching the Genesis Mission (Nov. 24, 2025) (establishing the American Science and Security Platform (ASSP)), https://www.whitehouse.gov/presidential-actions/2025/11/launching-the-genesis-mission/

[7] The White House, Adjusting Imports of Processed Critical Minerals and Their Derivative Products into the United States (Proclamation) (Jan. 14, 2026), https://www.whitehouse.gov/presidential-actions/2026/01/adjusting-imports-of-processed-critical-minerals-and-their-derivative-products-into-the-united-states/

[8] Project Vault (White House video): The White House, Introducing Project Vault, a Critical Mineral Stockpile for American Businesses (Feb. 2, 2026), https://www.whitehouse.gov/videos/introducing-project-vault-a-critical-mineral-stockpile-for-american-businesses-%F0%9F%92%8E%F0%9F%87%BA%F0%9F%87%B8/

[9] Cho, Dongyoun, Mauritz Kop & Min-Ha Lee, *Strategic Governance of Quantum Supply Chains: A Criticality-Based Framework for Risk, Resilience, and Data-Driven Foresight* (2025), https://www.researchgate.net/publication/397511771_Strategic_Governance_of_Quantum_Supply_Chains_A_Criticality-Based_Framework_for_Risk_Resilience_and_Data-Driven_Foresight

[10] Nat'l Inst. of Standards & Tech., NIST Releases First 3 Finalized Post-Quantum Encryption Standards (Aug. 13, 2024), https://www.nist.gov/news-events/news/2024/08/nist-releases-first-3-finalized-post-quantum-encryption-standards. See also Federal Register notice: https://www.federalregister.gov/documents/2024/08/14/2024-17956/announcing-issuance-of-federal-information-processing-standards-fips-fips-203-module-lattice-based.

---

[11] The White House, National Security Memorandum on Promoting United States Leadership in Quantum Computing While Mitigating Risks to Vulnerable Cryptographic Systems (NSM-10) (May 4, 2022), https://bidenwhitehouse.archives.gov/briefing-room/presidential-actions/2022/05/04/national-security-memorandum-on-promoting-united-states-leadership-in-quantum-computing-while-mitigating-risks-to-vulnerable-cryptographic-systems/

[12] *See* e.g., Yaakov Weinstein and Brandon Rodenburg, Quantum computing: quantifying current state to assess cybersecurity threats, MITRE Corp., (Jan 29, 2025), https://www.mitre.org/news-insights/publication/intelligence-after-next-quantum-computing-quantifying-current-state-art; and Edward Parker, U.S.-Allied Militaries Must Prepare for the Quantum Threat to Cryptography, JUST SECURITY (May 28, 2025), https://www.justsecurity.org/113733/quantum-computing-crytopography.

[13] *See* Kop, Mauritz, *A Bletchley Park for the Quantum Age*, WAR ON THE ROCKS (Nov. 6, 2025), https://warontherocks.com/2025/11/a-bletchley-park-for-the-quantum-age/.

[14] *See* e.g., Daniel Castro, Michael McLaughlin & Eline Chivot, Who Is Winning the AI Race: China, the EU or the United States?, Ctr. for Data Innovation (Aug. 2019), https://datainnovation.org/2019/08/who-is-winning-the-ai-race-china-the-eu-or-the-united-states/; and Argyri Panezi, The Security Stakes in the Global Quantum Race, JUST SECURITY (July 15, 2025), https://www.justsecurity.org/116473/security-stakes-global-quantum-race /.

[15] *See* e.g., Albert Cevallos, How Autocrats Weaponize AI — And How to Fight Back, J. Democracy, March 2025, https://www.journalofdemocracy.org/online-exclusive/how-autocrats-weaponize-ai-and-how-to-fight-back/; and Valentin Weber, Data-Centric Authoritarianism: How China's Development of Frontier Technologies Could Globalize Repression, Nat'l Endowment for Democracy, Feb. 11 2025, https://www.ned.org/data-centric-authoritarianism-how-chinas-development-of-frontier-technologies-could-globalize-repression-2/.

[16] Cybersecurity & Infrastructure Sec. Agency, Nat'l Sec. Agency & Nat'l Inst. of Standards & Tech., Quantum-Readiness: Migration to Post-Quantum Cryptography (Aug. 21, 2023), https://www.cisa.gov/news-events/alerts/2023/08/21/cisa-nsa-and-nist-publish-factsheet-quantum-readiness. (PDF: https://media.defense.gov/2023/Aug/21/2003284212/-1/-1/0/CSI-QUANTUM-

Existing scholarship and agency practice too often treat IP and national security as parallel tracks. Patent-eligibility doctrine under 35 U.S.C. § 101—particularly after *Alice*,[19] and refined by *Enfish*,[20] *McRO*,[21] and *Berkheimer*[22]—is applied to quantum algorithms, compilers, and error-correction claims without sustained attention to research-security spillovers. Conversely, export and investment controls increasingly reach software, cloud access, and "as-a-service" models, yet rarely grapple with how Bayh–Dole,[23] FAR/DFARS data-rights clauses and secrecy orders[24] shape disclosure, licensing, and publication timelines that bear directly on security externalities.

The LSI framework gives institutions a common decision rule for tuning open and more closed innovation pathways to national security objectives in the quantum stack. This rubric disciplines state and private action by requiring any measure to be (i) the least trade-restrictive necessary to achieve its goal; (ii) demonstrably security-sufficient to mitigate a material threat; and (iii) carefully designed to be innovation-preserving for the underlying scientific commons. Under LSI, agencies, universities, sponsors, and firms select the least restrictive legal lever that still

---

READINESS.PDF). See also: OMB Memo M-23-02 (Nov. 18, 2022), https://www.whitehouse.gov/wp-content/uploads/2022/11/M-23-02-M-Memo-on-Migrating-to-Post-Quantum-Cryptography.pdf).

[17] Bureau of Indus. & Sec., Commerce Control List Additions and Revisions; Implementation of Controls on Advanced Technologies Consistent with Controls Implemented by International Partners, 89 Fed. Reg. 72,926 (Sept. 6, 2024), https://www.govinfo.gov/app/details/FR-2024-09-06/2024-19633 (BIS-hosted FR PDF: https://www.bis.doc.gov/index.php/documents/federal-register-notices-1/3521-89-fr-72926-quantum-c-1-ifr-0694-aj60-9-6-2024/file); and U.S. Dep't of the Treasury, Outbound Investment Security Program—Final Rule (Oct. 28, 2024) (implementing E.O. 14105), https://home.treasury.gov/system/files/206/TreasuryDepartmentOutboundInvestmentFinalRuleWEBSITEVERSION_0.pdf . (Treasury explainer/"Additional Information": https://home.treasury.gov/news/press-releases/jy2690).

[18] Comm'n Recommendation (EU) 2025/63, On Reviewing Outbound Investments (Jan. 15, 2025), 2025 O.J. (L 63), https://eur-lex.europa.eu/eli/reco/2025/63/oj/eng. (PDF: https://eur-lex.europa.eu/legal-content/EN/TXT/PDF/?uri=OJ%3AL_202500063).

[19] Alice Corp. v. CLS Bank Int'l, 573 U.S. 208 (2014), https://supreme.justia.com/cases/federal/us/573/208/

[20] Enfish, LLC v. Microsoft Corp., 822 F.3d 1327 (Fed. Cir. 2016), https://law.justia.com/cases/federal/appellate-courts/cafc/2015-1244/2015-1244-2016-05-12.html.

[21] McRO, Inc. v. Bandai Namco Games Am., Inc., 837 F.3d 1299 (Fed. Cir. 2016), https://law.justia.com/cases/federal/appellate-courts/cafc/2015-1080/2015-1080-2016-09-13.html.

[22] Berkheimer v. HP Inc., 881 F.3d 1360 (Fed. Cir. 2018), https://law.justia.com/cases/federal/appellate-courts/cafc/2017-1437/2017-1437-2018-02-08.html

[23] Bayh–Dole Act, 35 U.S.C. §§ 200–212, https://www.law.cornell.edu/uscode/text/35/part-II/chapter-18.

[24] Invention Secrecy Act, 35 U.S.C. §§ 181–188, https://www.law.cornell.edu/uscode/text/35/part-II/chapter-17. *See* also USPTO, 120 Secrecy Orders, https://www.uspto.gov/web/offices/pac/mpep/s120.html

(i) mitigates concrete national-security risks at the relevant layer of the quantum stack; (ii) preserves contestability and diffusion necessary for scientific progress and scale; and (iii) avoids offshoring research or driving it into trade secrecy in ways that degrade security. Crucially, this test is dynamic, not static. Following the Eight Worlds scenario framework, the application of LSI turns on the ease of adversaries catching up.[25] For hard catch-up technologies involving tacit knowledge and complex physical supply chains (e.g., lithography, cryogenics), strict export controls satisfy the Security-Sufficient prong. Conversely, for easy catch-up domains like algorithmic software—where diffusion is rapid and leakage inevitable—the strategy must shift from restriction to acceleration, prioritizing adoption speed over futile containment. Operationalizing LSI reduces Type I errors (over-securitization that chills open science) and Type II errors (under-securitization that hemorrhages non-recoupable capabilities) and provides an off-ramp from a trajectory toward a Silicon Curtain.

This article motivates LSI with three recurring fact patterns. (1) The Compiler Claim: A research group seeks protection for a compilation or error-mitigation method that measurably improves fidelity on identified hardware; the § 101 question ("abstract idea" vs. concrete technological improvement) intersects with export-review concerns about sharing control-pulse libraries, model weights, datasets, or device-calibration files across borders. (2) The Refrigerator Chokepoint: A startup scales dilution refrigerators and cryo-electronics indispensable to multi-qubit systems; ECCN thresholds and license exceptions must be harmonized with university collaboration needs and pre-publication disclosure obligations. (3) Cloud-as-a-Service: Foreign laboratories access U.S. quantum hardware via cloud; access controls, logging, and user screening intersect with IP licensing, data-rights (government-purpose vs. limited rights), and standards-based PQC-ready interfaces. Each scenario is currently governed piecemeal; LSI supplies a cross-cutting method to pick minimally distortive tools that still hit the risk. Yet effective application of LSI faces a "situational awareness gap": policymakers often lack real-time data on the true state of quantum progress.[26] You cannot regulate what you cannot measure. Therefore, a prerequisite for any control regime is a "National Quantum Readiness Index"—a standardized assessment framework across the allied nations that tracks industrial capacity, talent density, and supply chain resilience, replacing hype with empirical baselines.[27]

The analysis proceeds from doctrine to empirically anchored design. On the IP axis, it addresses patent eligibility for quantum software post-*Alice* through claims drafted to concrete, hardware-bounded improvements in compilers, control, and error mitigation. On the public-funding axis, it explains federal data-rights architecture under the Federal Acquisition Regulation (FAR) and the Defense Federal Acquisition Regulation Supplement (DFARS)[28]—the U.S. procurement rules that allocate what rights the government receives in technical data and computer software produced under grants and contracts (e.g., "unlimited," "government-purpose," "limited," and "restricted" rights), making early scoping and markings outcome-determinative for IP value capture. On the controls axis, it synthesizes a range of state instruments, including: treating the Invention Secrecy Act[12] as a narrow, time-bounded tool for genuinely sensitive disclosures; mapping export controls under the Export Administration Regulations (EAR)—the U.S. Commerce Department's rules governing dual-use items; and, where performance governs deployability, adopting engineering metrics such as Size, Weight, Power, and Cost (SWaP-C) to specify capability thresholds that matter in practice. On the standards axis, the PQC migration begins to anchor procurement and critical-infrastructure expectations. These instruments can be read coherently with TRIPS Article 73[29]—a guardrail that grounds necessity and proportionality—so that choices at one layer do not inadvertently break another.

---

[25] *See* Sullivan & Feldman, *supra* note …

[26] Forrest, Tracey, Paul Samson, S. Yash Kalash & Michael P.A. Murphy, *Quantum Technologies and Situational Awareness: The G7 and Beyond* (Ctr. for Int'l Governance Innovation, Special Report Draft, Dec. 2025),
https://www.cigionline.org/static/documents/Quantum_Technologies_and_Situational_Awareness_-_Draft.pdf

[27] Chakraborty, Ria, Kim de Laat & Raymond Laflamme, *Canada's Migration to Post-Quantum Cryptography: Public-Private Roles* (Ctr. for Int'l Governance Innovation, Special Report, 2025), https://www.cigionline.org/publications/canadas-migration-to-post-quantum-cryptography-public-private-roles/

Historically, great powers have risen through technological appropriation—from industrial espionage and catch-up manufacturing to standards setting and financial statecraft. In the networked economy, chokepoints and regulatory gravity—weaponized interdependence[30] and the Brussels Effect[31]—enable extraterritorial leverage without kinetic force. LSI reframes both: tuned chokepoint leverage to security sufficiency; favor open, contestable standards to preserve diffusion without gifting dominance. In this domain, values-based deterrence is not neutral: unlike authoritarian coercion premised on opaque leverage, democratic deterrence relies on resilient, open, rule-bound systems—standards, assurance, and procurement baselines—that are harder to corrupt and faster to adapt in allied and majority-world infrastructure.

The remainder of this Article proceeds as follows. Chapter I situates quantum governance inside the architecture of intellectual property in the knowledge economy, and frames the core dilemma as a pendulum swing between overprotection and underprotection. Chapter II maps the principal instruments of national power relevant to dual-use quantum systems—export controls, competition policy, standards governance, and supply-chain security—and introduces the governance paradox created by misaligned policy tools. Chapter III provides a historical exegesis of technological appropriation and its modern analogs in standards strategy and financial statecraft. Chapter IV differentiates the quantum pillars (computing, sensing, simulation, communication, networking, and quantum-AI hybrids) by maturity and governable

---

[28] FAR 52.227-14 (Rights in Data—General); DFARS 252.227-7013 (Rights in Technical Data—Other Than Commercial Products and Commercial Services); DFARS 252.227-7014 (Rights in Noncommercial Computer Software and Noncommercial Computer Software Documentation). FAR 52.227-14 (Rights in Data—General), see https://www.acquisition.gov/far/52.227-14, https://www.acquisition.gov/dfars/252.227-7013 and https://www.acquisition.gov/dfars/252.227-7014.
[29] Agreement on Trade-Related Aspects of Intellectual Property Rights (TRIPS) art. 73, Apr. 15, 1994, 1869 U.N.T.S. 299, https://www.wto.org/english/docs_e/legal_e/27-trips.pdf.
[30] Henry Farrell & Abraham L. Newman, Weaponized Interdependence: How Global Economic Networks Shape State Coercion, 44 Int'l Sec. 42 (2019), MIT Press page: https://direct.mit.edu/isec/article/44/1/42/12290.
[31] Anu Bradford, The Brussels Effect: How the European Union Rules the World (Oxford Univ. Press 2020).

"control points," enabling calibrated interventions rather than blunt decoupling. Chapter V compares U.S., EU, Chinese, Canadian, and key allied approaches to strategy, value-capture, and coalition leverage. Chapter VI then develops the core nexus analysis across IP doctrine, export controls, research security, and commercialization pathways. Chapter VII deepens this treatment through special issues—including government rights and security carve-outs—that routinely determine whether controls are effective or self-defeating. Chapter VIII connects these micro-design choices to grand strategy and the weaponization of interdependence— drawing out quantum's impact on strategic stability, the limits of classical deterrence analogies, and a Taiwan invasion-denial use case (Replicator-scale swarms plus quantum-resilient positioning, navigation, and timing)—and clarifies when chokepoint leverage preserves resilience and when it hardens into a Silicon Curtain. Chapter IX extracts governance blueprints from prior technological revolutions to propose implementable institutional designs for the quantum era. Chapter X translates the framework into a practical navigation guide for innovators, universities, and counsel. Chapter XI outlines a sequenced reform agenda—from incremental fixes to systemic options—and identifies open research questions. Chapter XII concludes by synthesizing deterrence by denial for democratic resilience as an implementable LSI governance program, and by specifying coalition architecture (including standards-first pathways and RQT-by-design) to secure the quantum industrial commons. Appendix A provides a one-page LSI Playbook that operationalizes the framework into a practitioner-facing checklist and governance matrix for agencies, universities, and firms.

## I. The Architecture of Intellectual Property in the Knowledge Economy

The knowledge economy rewards orchestration more than ownership.[32] In quantum—where cryogenics, control electronics, compilers, circuit libraries, and application-layer models are tightly coupled—law "bites" at different layers of the stack with different force. The mistake in much contemporary commentary is to search for a single master key in patent doctrine or, conversely, to abandon IP in favor of security instruments alone.[33] This Chapter treats intellectual property as one pillar in a broader governance architecture that also includes procurement data rights, secrecy orders, export classifications, standard-setting and FRAND, open-source licensing, and international constraints. Read holistically and disciplined by the LSI principle—least trade-restrictive, security-sufficient, and innovation-preserving— that architecture can avoid the familiar pendulum swing: the overprotection that produces thickets, stacking, and secrecy spillovers on the one hand, and the underprotection that drives capital flight and under-investment on the other.[34] This approach draws on the "innovation policy pluralism" paradigm, recognizing that no single instrument—patent, prize, or subsidy—is sufficient for a technology as protean as quantum.[35] Instead, governance must blend these mechanisms, relying on empirical verification rather than abstract models. This analysis therefore combines legal-doctrinal assessment with quantitative patent landscape data to distinguish genuine choke points from theoretical thickets. The operative aim, consistent with the Article's through-line, is security-sufficient openness: adjust legal tools narrowly to real risks while keeping the research commons and market contestability intact, thereby avoiding a drift toward a costly "Silicon Curtain."

---

[32] *See* Lemley, Mark A. and Melamed, Doug, Missing the Forest for the Trolls (May 23, 2013). 113 Columbia Law Review 2117 (2013), Stanford Law and Economics Olin Working Paper No. 443, http://dx.doi.org/10.2139/ssrn.2269087

[33] *See generally* Peter S. Menell, Tailoring Legal Protection for Computer Software, 39 Stan. L. Rev. 1329 (1987).

[34] *See* e.g., Aboy, M., Minssen, T. & Kop, M. Mapping the Patent Landscape of Quantum Technologies: Patenting Trends, Innovation and Policy Implications. *IIC* **53**, 853–882 (2022). https://doi.org/10.1007/s40319-022-01209-3

I. Patents

Patent law remains the most visible lever because it structures disclosure, contestability, and early-stage financing. This section focuses on subject-matter eligibility under § 101 and on drafting patents "to the stack"—claims tethered to concrete, hardware-bounded technical improvements rather than high-level mathematical goals. Yet subject-matter eligibility under § 101 has been a minefield for software-mediated inventions.[36] After *Alice*, claims that merely dress an abstract idea in generic computation fail; but subsequent Federal Circuit decisions have preserved a lane for software claims tethered to concrete, non-conventional technical improvements—e.g., *Enfish*, *McRO*—and cautioned, in *Berkheimer*, that "conventionality" often raises fact issues ill-suited to early disposition.[37] In quantum, the practical drafting heuristic is to tie claims to hardware-bounded improvements—compiler passes that measurably reduce two-qubit error on identified devices, control loops that raise fidelity under specified noise models, or mitigation pipelines benchmarked against reproducible datasets—rather than to claim mathematical goals at a high level of generality. An LSI-consistent drafting approach therefore ties claims to hardware-bounded improvements with reproducible, device-specific benchmarks, minimizing preemption risk while preserving patent's information-forcing function.[38]

II. Enablement

Eligibility, of course, is not the only gate. This segment addresses enablement and written description under § 112 and explains why a tiered disclosure posture—interfaces and validated performance claims publicly disclosed, with sensitive parameterizations and process know-how retained as trade secrets—can preserve both innovation incentives and research-security objectives. Enablement and written description under § 112 cabin the scope of permissible claims.[39] Quantum fields sharpen the perennial tension: disclosing the fragile "recipes" that actually deliver performance—calibration parameters, noise-floor characterizations, lab notebooks that capture tacit technique—can collapse competitive advantage and, in some cases, raise research-security risks. The governance solution is not blanket opacity but modularity:

---

[35] *See* Daniel J. Hemel & Lisa Larrimore Ouellette, *Innovation Policy Pluralism*, 128 Yale L.J. 544 (2019), https://yalelawjournal.org/article/innovation-policy-pluralism
[36] *See* Mojdeh Farsi, A Proposal to Revise the Alice Test for Software Patents 5 (2020) (Ph.D. dissertation, University of California, Berkeley).
[37] *See* 35 U.S.C. § 101.
[38] Mauritz Kop, Quantum Computing and Intellectual Property Law, Berkeley Tech. L.J. (Feb. 21, 2022), https://btlj.org/2022/02/quantum-computing-and-intellectual-property-law/
[39] 35 U.S.C. § 112.

disclose interfaces and validated performance claims sufficient to enable the invention at the level claimed, while reserving highly specific parameterizations and process know-how as trade secrets guarded by rigorous reasonable-measures programs. That tiered posture can be sequenced with filing and clearance calendars, minimizing premature enabling publications while still putting rivals on notice and enabling cumulative innovation.

### III. Trade Secrets and Data

Trade-secret protection under the Defend Trade Secrets Act (DTSA) and state law complements patents where disclosure would destroy advantage or eligibility is doubtful.[40] The core design choice is modular secrecy rather than blanket opacity: patent interfaces and reproducible gains, while protecting calibration datasets, parameterizations, and tacit procedures through audited reasonable-measures programs and sequenced publication after filing and classification. Under U.S. law, the Defend Trade Secrets Act creates a federal civil cause of action for misappropriation.[41] Empirically, early-stage firms often prefer trade secrecy over patenting for speed, cost, and financing strategy—especially when disclosure would invite fast-following by better-capitalized rivals.[42] In a rivalry setting, this raises the salience of trade-secret theft as a strategic pathway for capability acquisition.[43] In practice, many firms in pre-commercial, fault-tolerance pathways curate "modular secrecy": they patent interfaces and validated performance gains while holding parameterizations, calibration datasets, and tacit procedures under tightly audited secrecy, then sequence publication after filing and export classification.[44] This secrecy, while commercially rational, complicates national and international efforts to assess technological progress and manage security risks. For scholars and policymakers, the implication is methodological: patent counts alone understate capability; triangulation with standards participation, procurement-grade conformance tests, and audited reproducibility metrics becomes essential. LSI also counsels "security-by-contract"—confidential annexes, escrow of critical artifacts, and time-bounded publication windows—to preserve diffusion without leaking crown-jewel know-how.

---

[40] Defend Trade Secrets Act of 2016, 18 U.S.C. § 1836; see also Kewanee Oil Co. v. Bicron Corp., 416 U.S. 470 (1974); Mark A. Lemley, The Surprising Virtues of Treating Trade Secrets as IP Rights, 61 Stan. L. Rev. 311 (2008); and Tait Graves, Trade Secrets as Property: Theory and Consequences, 15 J. Intell. Prop. L. & Tech. 39 (2007), https://digitalcommons.law.uga.edu/jipl/vol15/iss1/2

[41] Defend Trade Secrets Act of 2016, *supra* note …

[42] *See* David S. Levine & Ted Sichelman, Why Do Startups Use Trade Secrets?, 94 NOTRE DAME L. REV. 751 (2019), https://scholarship.law.nd.edu/ndlr/vol94/iss2/6/

[43] Comm'n on the Theft of Am. Intell. Prop., The IP Commission Report: The Report of the Commission on the Theft of American Intellectual Property (May 2013), https://www.k-state.edu/research/faculty/other-resources/facility-security-office/documents/ip-commission-report.pdf. [Update (Feb. 2017): Nat'l Bureau of Asian Research, IP Commission Report Update, https://www.nbr.org/wp-content/uploads/pdfs/publications/IP_Commission_Report_Update.pdf]

[44] *See* e.g. Mauritz Kop, AI & Intellectual Property: Towards an Articulated Public Domain, 28 Tex. Intell. Prop. L.J. 297, 332 (2020), https://papers.ssrn.com/sol3/papers.cfm?abstract_id=3409715.

### IV. Copyright and Interfaces

Adjacent bodies of law refine the boundaries. This section explains why interfaces—APIs, intermediate representations, and invocation layers—are the natural "control points" for interoperability and standardization, and why copyright's idea–expression limits (and fair-use space for interface re-use) make it a complement to patent and trade secret rather than a substitute. Copyright protects expressive code and documentation but not ideas, methods, or systems; *Baker v. Selden* and *Feist* mark the idea–expression line and the originality floor.[45] *Google v. Oracle* confirms a significant fair-use space for interface re-use in software, especially where interoperability promotes progress.[46] In a quantum stack, these distinctions matter because interfaces—the intermediate representations compilers target, the APIs through which error-mitigation or control routines are invoked—are the natural surfaces for standardization.[47] A sophisticated portfolio therefore treats copyright as a complement to patent and trade secret: document and license interfaces in ways that encourage diffusion and standard-setting participation, while keeping the crown-jewel implementation details under tighter control.

### V. Government-Funded IP and Data Rights

Government-funded research adds a determinative layer. The following addresses how Bayh–Dole and FAR/DFARS data-rights architecture can determine downstream licensing, publication timing, and IP value capture. Who may use, modify, and disclose code and data often turns not on Title 35 but on data-rights allocations under Bayh–Dole and procurement clauses. Bayh–Dole permits contractors to elect title to subject inventions subject to disclosure, government license, and U.S. manufacturing preferences; it also preserves "march-in" as a policy lever.[48] FAR 52.227-14 and DFARS 252.227-7013/-7014 then allocate "unlimited," "government-purpose," "limited," and "restricted" rights in technical data and computer software according to funding source, deliverable status, and proper markings.[49] In practice, early scoping of deliverables and legends is outcome-determinative: carelessly commingled source code or datasets can default to broad government rights that complicate later licensing, while over-asserted restrictions can slow mission work and trigger disputes. The LSI frame helps here: specify the minimum government rights necessary to accomplish the programmatic objective; segregate background from foreground IP; and align publication review windows with patent filing and security screens so that openness is sequenced rather than suppressed.

---

[45] Baker v. Selden, 101 U.S. 99 (1880); Feist Publ'ns, Inc. v. Rural Tel. Serv. Co., 499 U.S. 340 (1991); Paul Goldstein & P. Bernt Hugenholtz, International Copyright: Principles, Law, and Practice 4–5 (4th ed. 2019).

[46] Google LLC v. Oracle Am., Inc., 141 S. Ct. 1183 (2021).

[47] Peter S. Menell, Rise of the API Copyright Dead?: An Updated Epitaph for Copyright Protection of Network and Functional Features of Computer Software, 31 Harv. J.L. & Tech. 305 (2018), https://jolt.law.harvard.edu/assets/articlePDFs/v31/31HarvJLTech305.pdf

[48] Lisa Larrimore Ouellette & Bhaven N. Sampat, Using Bayh-Dole Act March-In Rights to Lower US Drug Prices, 5 JAMA Health Forum e243775 (2024), https://jamanetwork.com/journals/jama-health-forum/fullarticle/2825385

[49] *See* FAR 52.227-14 (Rights in Data—General); DFARS 252.227-7013 (Rights in Technical Data—Other Than Commercial Products and Commercial Services); DFARS 252.227-7014 (Rights in Noncommercial Computer Software and Noncommercial Computer Software Documentation).

## VI. Secrecy Orders

Secrecy orders under the Invention Secrecy Act are the sharpest tool in the disclosure toolkit, but they should remain rare, time-bounded, and calibrated- narrowly scoped to specific, demonstrable risks.[50] Empirical analysis indicates that secrecy orders have reached a recent high, underscoring how innovation policy can silently become control policy.[51] A secrecy order can pause publication and issuance where disclosure might be detrimental to national security, and it can be paired with compensation if government use ensues. Consistent with LSI, secrecy orders should be rare, narrow in scope, and time-bounded, with scheduled review and compensation where government use occurs; overuse pushes research into indefinite opacity and starves the standards process of high-value contributions.

## VII. Standards, Interoperability, and Competition

Standard-setting brings both opportunity and hazard. By crystallizing interfaces, it multiplies interoperability, investment, and learning; by creating standard-essential patents (SEPs), it can also create hold-up risk and stacking. The canonical concern is patent "hold-up": once a protocol becomes locked in, a patent holder can demand supracompetitive royalties.[52] And where many overlapping rights touch a single standard, the resulting "patent thicket" can be mitigated through cross-licensing, patent pools, and carefully designed SSOs that reduce transaction costs and strategic blockage.[53] Antitrust and competition policy remain the backstop for disciplining collusion, exclusion, and opportunistic licensing strategies.[54] U.S. courts have treated Fair, Reasonable, and Non-Discriminatory (FRAND) commitments as enforceable contracts (*Microsoft v. Motorola*), instructed juries to apportion value to the patented contribution rather than to the standard as a whole (*Ericsson v. D-Link*), and declined expansive antitrust theories that would re-engineer core licensing models (*FTC v. Qualcomm*).[55] For quantum software and communication—think PQC profiles, quantum-safe network protocols, or compiler intermediate representations (IRs)—those principles counsel disciplined essentiality determinations and licensing offers that travel across jurisdictions. A portfolio aligned to LSI contributes early and openly to standards while modeling FRAND stacks that are neither exclusionary nor performative, thus reducing the temptation for regulators to intervene *ex post*.

---

[50] 35 U.S.C. §§ 181–188.

[51] *See* Steven Aftergood, Invention Secrecy Hits Recent High, Fed'n of Am. Scientists (Oct. 31, 2018), https://fas.org/publication/invention-secrecy-2018 /.

[52] Joseph Farrell, John Hayes, Carl Shapiro & Theresa Sullivan, Standard Setting, Patents, and Hold-Up, 74 Antitrust L.J. 603 (2007), https://www.jstor.org/stable/i27897557

[53] Carl Shapiro, Navigating the Patent Thicket: Cross Licenses, Patent Pools, and Standard Setting, in Innovation Policy and the Economy, Vol. 1, at 119 (Adam B. Jaffe, Josh Lerner & Scott Stern eds., 2001), https://www.nber.org/books-and-chapters/innovation-policy-and-economy-volume-1/navigating-patent-thicket-cross-licenses-patent-pools-and-standard-setting

[54] *See* Robert Pitofsky, Antitrust and Intellectual Property: Unresolved Issues at the Heart of the New Economy, Remarks Before the Antitrust, Tech. & Intell. Prop. Conf. (Univ. of Cal., Berkeley) (Mar. 2, 2001), https://www.ftc.gov/news-events/news/speeches/antitrust-intellectual-property-unresolved-issues-heart-new-economy; Herbert J. Hovenkamp, Intellectual Property and Competition, Univ. of Pa. Carey L. Sch., Faculty Scholarship (last visited Feb. 3, 2026), https://scholarship.law.upenn.edu/faculty_scholarship/1807/; and Mauritz Kop, Mateo Aboy, Timo Minssen, Intellectual property in quantum computing and market power: a theoretical discussion and empirical analysis, *Journal of Intellectual Property Law & Practice*, Volume 17, Issue 8, August 2022, Pages 613–628, https://doi.org/10.1093/jiplp/jpac060

[55] Microsoft Corp. v. Motorola, Inc., 795 F.3d 1024 (9th Cir. 2015); Ericsson, Inc. v. D-Link Sys., Inc., 773 F.3d 1201 (Fed. Cir. 2014); FTC v. Qualcomm Inc., 969 F.3d 974 (9th Cir. 2020).

## VIII. Open-Source Posture and Export Controls

Open-source posture is another accelerator that must be tuned rather than romanticized. *Jacobsen v. Katzer* confirms that violating license conditions can sound in copyright, making compliance more than etiquette.[56] Open-sourcing non-sensitive libraries can widen the talent funnel and speed testing; open-sourcing everything can amount to unintended enabling disclosures under both patent and export law. The prudent course is to sequence publications and releases behind filing and classification decisions; to adopt contributor license agreements that harmonize with the patent strategy; and to apply export-control screening where code, models, or datasets meet EAR (Export Administration Regulations) definitions of controlled "software" or "technology." That approach pairs openness with responsibility and keeps diffusion from bleeding into leakage.

## IX. TRIPS Security Exceptions

Finally, the international layer sets the envelope conditions. Because TRIPS sits within the WTO legal order established by the Marrakesh Agreement, security exceptions operate as constrained carve-outs rather than blank checks: tailoring and evidentiary justification are central to legitimacy.[57] TRIPS Article 73 recognizes essential-security exceptions while constraining opportunism through good-faith limits developed in WTO practice.[58] When read through LSI, Article 73 supports time-limited restrictions—on specific capability thresholds, for example—while disfavoring broad, indefinite bans that fracture standards or smother early-stage research. For quantum, the better path is not a field-wide carve-out but LSI-calibrated use of Article 73: narrowly tailored, time-limited measures tied to demonstrable capability

---

[56] Jacobsen v. Katzer, 535 F.3d 1373 (Fed. Cir. 2008).
[57] *See* Marrakesh Agreement Establishing the World Trade Organization, Apr. 15, 1994, 1867 U.N.T.S. 154, https://www.wto.org/english/docs_e/legal_e/04-wto_e.htm; Abbott, Frederick M. (2020): The TRIPS Agreement Article 73 Security Exceptions and the COVID-19 pandemic, Research Paper, No. 116, South Centre, Geneva, https://www.econstor.eu/bitstream/10419/232239/1/south-centre-rp-116.pdf; and Jorge L. Contreras, Is the National Security Exception in the TRIPS Agreement a Realistic Option in Confronting COVID-19?, EJIL: Talk! (Apr. 24, 2020), https://www.ejiltalk.org/is-the-national-security-exception-in-the-trips-agreement-a-realistic-option-in-confronting-covid-19/
[58] TRIPS Agreement art. 73: Agreement on Trade-Related Aspects of Intellectual Property Rights art. 73, Apr. 15, 1994, 1869 U.N.T.S. 299.

thresholds—guardrails that protect essential security interests without precipitating standards fracture or wholesale retreat from openness.[59]

Chapter II turns from this IP-centered architecture to the broader instruments of national power—export and services controls, investment screening, procurement baselines, and supply-chain security—and shows why misalignment among these tools produces the governance paradox that LSI is designed to discipline.

II. The Instruments of National Power and Economic Competition

The development and control of quantum technology is not merely another sectoral wager; it is a systems contest in which states and firms compete to shape the conditions under which knowledge is created, capitalized, and diffused. The legal frameworks of IP and trade are embedded within a larger geopolitical context where nations wield a variety of instruments— from export controls to industrial policy—to secure their interests. Joint doctrine identifies these instruments as the DIME quartet—diplomatic, informational, military, and economic—while contemporary practice expands the set to include finance, intelligence, and law enforcement (DIME-FIL).[60]

Framed through this Article's core commitment to security-sufficient openness, the question is not whether to restrict or to liberalize, but how to calibrate each instrument to be the least trade-restrictive, yet nonetheless security-sufficient and innovation-preserving. Here, the LSI test offers a principled off-ramp from a drift toward a costly "Silicon Curtain"—a bifurcation of global technology into incompatible, value-laden stacks, represented by a potential "quantum splinternet". Treated piecemeal, these instruments produce the pathologies of over- and under-securitization. Treated as a coherent toolbox disciplined by the LSI test, they yield a portable decision rule for statecraft.

I. Industrial Policy and Subsidies

States have coupled export restrictions with affirmative industrial policy, using subsidies not only to build capacity but also to impose security obligations through "security-by-contract." The CHIPS and Science Act of 2022, for instance, appropriates historic incentives for domestic semiconductor R&D and authorizes regionally distributed "Tech Hubs" that explicitly include quantum.[61] Crucially, Commerce's 2023 "guardrails" rule conditions these subsidies on recipients' restraint from expanding certain advanced capacity in "countries of concern," functioning as quasi-regulatory instruments that bind firms through funding agreements.[62]

---

[59] This LSI-calibrated approach refines our earlier proposal for a potential field-wide security exception for quantum technologies under TRIPS Article 73. *See* Mauritz Kop & Mark Brongersma, *Integrating Bespoke IP Regimes for Quantum Technology into National Security Policy* (Aug. 8, 2021) Stanford Law School Working Paper https://purl.stanford.edu/fx255fg2947. The present analysis, reflecting an evolution in our thinking, argues for more granular, capability-specific measures over a broad categorical exclusion.

[60] Joint Pub. 1, Doctrine for the Armed Forces of the United States I-3 (2013) (DIME); U.S. Air Force, Purple Book: The AF Joint Team Doctrine 10 (2024) (DIME-FIL).

[61] CHIPS and Science Act of 2022, Pub. L. No. 117-167, 136 Stat. 1366.

The European Union has launched symmetric tools. The EU's Economic Security Strategy prioritizes risk assessment in quantum and other critical technologies; a Critical Raw Materials Act, in force since May 2024, targets supply-chain chokepoints; and a 2025 Commission Recommendation urges Member States to review outbound investments.[63] The enabling stack for quantum hardware depends on critical minerals and materials whose extraction, processing, and refining are geographically concentrated, creating chokepoints that translate commercial dependence into strategic leverage.[64] In Europe, the Critical Raw Materials framework codifies a security-of-supply approach that is directly relevant to quantum inputs and processing chokepoints.[65] Recent analyses frame quantum supply chains as a test case for a broader shift toward economic security and a less permissive global trade posture.[66] To make these measures operational and align them with the White House Critical Minerals Presidential Order, I propose integrating a Quantum Criticality Index (QCI)[67]—a tri-axial metric assessing supply risk, substitutability, and strategic significance—to objectively determine when a specific material (e.g., Helium-3 or isotopically enriched silicon) crosses the threshold from commodity trade to a national security "Red Zone."[68] Project Vault supplies a concrete implementation layer for this data-driven approach: critical mineral stockpile design (inclusion, drawdown, replenishment) can be tied to QCI-based thresholds, once adopted as a federal index.[69] Together, these measures amount to a transatlantic industrial-security architecture that applies the LSI principle of security sufficiency directly into the fabric of public funding.

## II. Trade Controls and Technology Denial

Export controls remain the most visible lever of geoeconomic statecraft. Under the Export Administration Regulations (EAR), the Bureau of Industry and Security (BIS) created new Export Control Classification Numbers (ECCNs)—such as 4A906—for quantum computing items and their associated toolchains, imposing worldwide license requirements.[70] Export controls under EAR restrict the transfer of certain commodities, software, and technology to foreign countries or entities for national security reasons.[71] These are refined through end-user restrictions, such as the addition of PRC entities to the Entity List, signaling a durable licensing posture of "presumption of denial."[72] The scope includes intangible technology transfers (e.g., technical data, cloud access, and services) and is increasingly salient for quantum computing components and enabling stacks.[73]

---

[62] Preventing the Improper Use of CHIPS Act Funding, 88 Fed. Reg. 65,599 (Sept. 25, 2023).
[63] European Comm'n, Economic Security Strategy & Risk-Assessment Recommendation (Oct. 3, 2023); Regulation (EU) 2024/1252, 2024 O.J. (L) 1 (Critical Raw Materials Act).
[64] *See* Jane Nakano, The Geopolitics of Critical Minerals Supply Chains, Ctr. for Strategic & Int'l Stud. (Mar. 11, 2021), https://www.csis.org/analysis/geopolitics-critical-minerals-supply-chains.
[65] Regulation (EU) 2024/1252 of the European Parliament and of the Council of 11 Apr. 2024 Establishing a Framework for Ensuring a Secure and Sustainable Supply of Critical Raw Materials, https://eur-lex.europa.eu/eli/reg/2024/1252/oj
[66] *See* Ulrich Mans, Quantum Supply Chains: A Test Case for a New Economic World Order, JUST SECURITY (Aug. 27, 2025), https://www.justsecurity.org/115813/quantum-new-economic-world-order/.
[67] *See* Cho, Kop & Lee, *supra* note …
[68] *See* The White House, Adjusting Imports of Processed Critical Minerals and Their Derivative Products into the United States, *supra* note …
[69] *See* The White house, Project Vault, *supra* note …. *See also* Project Vault (New York Times report): Alan Rappeport & Tony Romm, Trump Unveils $12 Billion Critical Minerals Stockpile, N.Y. Times (Feb. 2, 2026), https://www.nytimes.com/2026/02/02/business/trump-critical-minerals-stockpile.html

An LSI-grounded approach deploys these controls with engineering specificity. Capability thresholds should be expressed in operational terms like SWaP-C (size, weight, power, and cost) rather than abstract metrics. This precision is critical because, as Michael McFaul argues, broad decoupling is strategically self-defeating; the objective must be "selective economic decoupling" that restricts only the narrowest set of dual-use technologies necessary for national security while preserving the broader commercial ecosystem.[74] Export controls must therefore function as a scalpel, not a hammer, to avoid severing the very innovation networks that sustain democratic resilience. Furthermore, restrictions should be paired with reciprocity-based license exceptions for trusted partners that implement aligned controls, such as the plurilateral "Implemented Export Controls (IEC)" pathway. This approach preserves the scientific commons while hardening defenses against strategic rivals.

### III. Capital Controls

Investment screening complements trade controls by modulating the financial channels through which capabilities are scaled. Inbound, the Committee on Foreign Investment in the United States (CFIUS) reviews foreign investments in U.S. businesses holding "critical technologies," a category that automatically expands with the addition of new quantum-related ECCNs.[75] Outbound investment restrictions now operate as a complementary control layer for certain national-security technologies and products in countries of concern.[76] Outbound, the U.S. Treasury's rule implementing E.O. 14105, effective January 2, 2025, prohibits or requires notification for certain U.S. investments in quantum information technologies when transacting with "countries of concern."[77] From an LSI perspective, these regimes should steer capital away from the most sensitive applications while leaving room for the broader research economy to function—a clockwise torque on the pendulum, not a freezing of cross-border flows.

---

[70] Commerce Control List: Implementation of Controls on Quantum Computing Items, 89 Fed. Reg. ___ (Sept. 6, 2024) (to be codified at 15 C.F.R. pt. 774).

[71] *See* Christopher Casey and Paul Kerr, The U.S. Export Control System and the Export Control Reform Act of 2018, Cong. Rsch. Serv., R46814 (updated Dec. 6, 2022), https://crsreports.congress.gov/product/pdf/R/R46814; U.S. Export Controls, Int'l Trade Admin. (last visited Feb. 3, 2026), https://www.trade.gov/us-export-controls; and Export Control Basics, Bureau of Indus. & Sec. (last visited Feb. 3, 2026), https://www.bis.gov/articles/export-control-basics

[72] Bureau of Indus. & Sec., Commerce Adds 37 PRC Entities to the Entity List for Enabling PRC Quantum and Aerospace Programs (May 9, 2024) (press release).

[73] See e.g., Peter Alexander Earls Davis, Mateo Aboy & Timo Minssen, Regulatory Challenges and Opportunities of Export Controls on Quantum Computing, in QUANTUM TECHNOLOGY GOVERNANCE: LAW, POLICY AND ETHICS IN THE QUANTUM ERA (Mateo Aboy, Marcello Corrales Compagnucci & Timo Minssen eds., 2026), https://papers.ssrn.com/sol3/papers.cfm?abstract_id=5404548; and Hannah Kelley, Dual-Use Technology and U.S. Export Controls, Ctr. for a New Am. Sec. (CNAS), June 15, 2023, https://www.cnas.org/publications/reports/dual-use-technology-and-u-s-export-controls

[74] Michael McFaul, *The Case for Selective Economic Decoupling with Autocracies*, (Dec. 2025), McFaul's World, https://michaelmcfaul.substack.com/p/the-case-for-selective-economic-decoupling

### IV. Sanctions and Law Enforcement

Sanctions and law enforcement serve as powerful accountability backstops. The Treasury's Office of Foreign Assets Control (OFAC) administers sanctions that function as a capital-markets complement to export controls,[78] while the joint DOJ-Commerce Disruptive Technology Strike Force coordinates prosecutions against illicit procurement networks.[79]

LSI disciplines these tools with proportionality and accountability. This includes requiring adjudicable LSI statements for restrictive measures—public, reason-giving justifications demonstrating least-restrictiveness, security sufficiency, and innovation preservation. Such a requirement creates a pathway to review and lift measures as risks recede, preventing emergency actions from becoming permanent fixtures.

### V. Standards and Certification

Standard-setting is a core instrument of economic and ideological competition, and embedding values in code is the mechanism by which technical baselines become governance. The U.S.-led migration to NIST-finalized post-quantum cryptography (PQC) standards (FIPS 203, 204, and 205) leverages federal procurement to counter "harvest-now, decrypt-later" risks.[80] In parallel, the EU's leadership in bodies like ETSI projects[81] regulatory gravity—the "Brussels Effect"—to globalize its preferred technical norms for things like Quantum Key Distribution (QKD).[82] LSI insists on a principled separation: keep standards open and focused on interoperability, while housing explicit trade restrictions in trade law, subject to transparent justification. However, this does not imply neutrality. By embedding democratic requirements—auditability, privacy-preserving telemetry, and transparent error-correction—into the technical baseline, standards function as *values-based deterrence*. They create a market reality where authoritarian systems that rely on opaque data exfiltration simply cannot interoperate, effectively deterring their adoption in allied infrastructure.

---

[75] 31 C.F.R. pt. 800 (CFIUS regulations); 31 C.F.R. § 800.215 (defining "critical technologies" by reference to the Commerce Control List).).

[76] Exec. Order No. 14105, 88 Fed. Reg. 54,867 (Aug. 9, 2023) (Addressing United States Investments in Certain National Security Technologies and Products in Countries of Concern), https://www.federalregister.gov/documents/2023/08/14/2023-17449/addressing-united-states-investments-in-certain-national-security-technologies-and-products

[77] U.S. Dep't of the Treasury, Outbound Investment Security Program—Final Rule 2024, *supra* note …

[78] 50 U.S.C. §§ 1701–1706 (International Emergency Economic Powers Act); 31 C.F.R. ch. V (OFAC sanctions programs).

[79] U.S. Dep't of Just. & U.S. Dep't of Com., Disruptive Technology Strike Force—Creation Announcement (Feb. 16, 2023).

[80] Nat'l Inst. of Standards & Tech., Announcing Issuance of FIPS 203/204/205 (Aug. 14, 2024); Cybersecurity & Infrastructure Sec. Agency, Nat'l Security Agency & NIST, Migration to Post-Quantum Cryptography: Planning for the Future (Aug. 21, 2023).

[81] ETSI GS QKD 014 (REST-based key-delivery API); ETSI GS QKD 016 (Common Criteria Protection Profile).

## VI. A Comparative Interlude: The Canada Middle-Power Paradox

The LSI framework is not merely a tool for great powers. It offers a strategic playbook for middle powers navigating a contested landscape, as illustrated by the Canada middle-power paradox: a nation with world-class research and a vibrant startup ecosystem that nonetheless struggles with domestic IP value capture and commercial scale-up. An LSI-grounded strategy for Canada would leverage its strengths through targeted, coalition-centric plays: using public procurement to create predictable domestic demand for PQC and quantum sensing; conditioning innovation funding on transparent IP and data-rights plans; and relying on reciprocity-based license-exception compacts like the IEC to ensure that international collaboration is a two-way street. Linking this to alliance capability planning, dual-use quantum prioritization should be treated as an explicit component of allied force-development and target-setting—rather than an implicit spillover from civilian excellence.[83] However, Canada faces a specific structural risk: governance fragmentation, where federal guidance (like NIST standards) fails to penetrate provincial jurisdiction over health and education data. An LSI-consistent playbook for middle powers must therefore include centralized "Quantum Readiness" offices empowered to bridge the technical-political gap across federal systems.[84] This converts research leadership into durable sovereign capability without resorting to protectionism and—critically—avoid treating 'quantum sovereignty' as a slogan detached from concrete supply-chain, benchmarking, and procurement capability.[85]

## VII. A Practitioner's Playbook: Applying the LSI Framework

For counsel and decision-makers, the LSI test provides a coherent workflow. It requires embedding pre-publication export screens for research; classifying components against the

---

[82] Anu Bradford, The Brussels Effect: How the European Union Rules the World (Oxford Univ. Press 2020); Henry Farrell & Abraham L. Newman, Weaponized Interdependence: How Global Economic Networks Shape State Coercion, 44 Int'l Sec. 42 (2019).

[83] *See* Michael P.A. Murphy, Tracey Forrest & Paul Samson, From Peace Dividend to Defence Dividend: Dual-Use, Quantum and NATO Targets, CIGI Paper No. 333 (Ctr. for Int'l Governance Innovation Sept. 17, 2025), https://www.cigionline.org/publications/from-peace-dividend-to-defence-dividend-dual-use-quantum-and-nato-targets/; and Michael P. A. Murphy, Paul Samson, Tracey Forrest, Innovation as Deterrence: How Canada's New NATO Target Can Thrive in a Tech-Driven World, CIGI, 23 Oct 2025, https://www.cigionline.org/articles/innovation-as-deterrence-how-canadas-new-nato-target-can-thrive-in-a-tech-driven-world/

[84] Chakraborty, de Laat & Laflamme, *supra* note …

[85] *See* Michael Krelina, Opinion: Quantum Sovereignty—Reality Check, LinkedIn, Oct 30, 2025, https://www.linkedin.com/pulse/opinion-quantum-sovereigntyreality-check-michal-krelina-fmg3e/

Commerce Control List (**CCL**)[86] using SWaP-C thresholds; diligencing CFIUS exposure and structuring deals to avoid outbound triggers; treating standards participation as a competitive front; and hard-wiring security-by-contract in all grants and procurements through disciplined Bayh–Dole and FAR/DFARS data-rights planning. When deployed systematically, the DIME-FIL toolkit can support both national security and the vibrancy required for innovation. The goal is not to litigate the morality of openness or the inevitability of decoupling, but to ask at every control point: What is the narrowest legal instrument that arrests the risk without smothering diffusion? In quantum, those who answer with precision will set the tempo.

### A. LSI as a Decision Operator Across Three Fronts

In practice, LSI functions as a decision operator applied to a multi-dimensional system whose variables include qubit performance (fidelity, connectivity, and error-correction overhead), sensitivity and deployment constraints, and the feasibility of entanglement distribution at scale. Three "fronts" repeatedly matter for governance: (1) the physical front (materials, components, and manufacturing capabilities, from isotopically enriched inputs and cryogenic supply chains to SNSPD-grade thin films); (2) the digital front (software, compilers, error-mitigation routines, and access to QCaaS platforms that translate scarce quantum hardware into widely usable services); and (3) the collaborative front (the research networks, standards bodies, and industrial alliances through which knowledge and talent diffuse). Treating these fronts as coupled—often entangled—helps avoid policies that "solve" one dimension while destabilizing another, such as controls that protect hardware but leave services, standards, or upstream chokepoints as ungoverned attack surfaces.

### B. Deterrence by Denial and Red-Zone Defaults

The tone and posture of the LSI test should be explicitly realist. For narrowly defined "red-zone" capabilities—where diffusion would predictably and materially strengthen adversarial military or intelligence capacity—the default should flip from "permit with guardrails" to "deny unless a bounded exception is justified." This aligns with the U.S.-China Economic and Security Review Commission's 2025 recommendation to shift from a "presumption of denial" to a strict "policy of denial" for mission-critical quantum technologies, specifically in cryptography and materials science.[87] This is deterrence by denial in legal form: raising the cost of adversarial learning and shortening the window for appropriation. Recent policy directions that treat quantum as critical infrastructure—including supply-chain initiatives such as the Genesis Mission and the Advanced Semiconductor Supply Chain Program (ASSP)—underscore the emerging shift from narrow export-control thinking to resilience-first governance.[88] Verification and compliance become part of the design, akin to a "quantum Zeno" analogy: constant, credible observation (identity assurance, audit logs, export-control end-use checks, and research-security protocols) can freeze risky pathways and stabilize

---

[86] *See* Bureau of Industry and Security, Commerce Control List, https://www.bis.gov/regulations/ear/interactive-commerce-control-list/
[87] U.S.-China Econ. & Sec. Rev. Comm'n, 2025 Report to Congress, *supra* note …
[88] The White House, Launching the Genesis Mission (Nov. 24, 2025), *supra* note …

permitted ones. The LSI test then forces specificity—what capability is being denied (e.g., error-corrected compute, quantum navigation-grade sensing, or entanglement distribution at scale), what evidence triggers tightening or relaxation, and how the measure avoids unnecessary fragmentation of allied innovation ecosystems.

*C. Innovation-Preserving Enclaves and Standards-First Stability*

Innovation preservation does not require indiscriminate openness. A coalition can preserve scientific dynamism through secure enclaves: tightly governed research environments, purpose-limited access to QCaaS, and reciprocity-based collaboration channels that enable legitimate experimentation while denying adversaries frictionless scale-up. Standards participation is the complementary outward-facing strategy. Treating interoperability, testability, and crypto-agility as "standards-first" requirements turns standards into a stability mechanism—anchoring a shared technical baseline that supports both market formation and security review. In this posture, standards operate as administrable "eigenstates": once a secure, interoperable baseline is adopted across a coalition, it becomes harder for authoritarian preferences—opaque interfaces, unverifiable security assumptions, or coercive data access—to travel through the technical rulebook.

*D. The "Bletchley Park" Architecture for Allied Governance*

To prevent a "Quantum Splinternet," allies require an institutional architecture that bridges science, operations, and diplomacy. I propose a "Bletchley Park" model for the quantum age: a federated structure that combines shared threat assessment, declassification of quantum risks, and harmonized standards.[89] This architecture supports "One Test, Many Markets" mutual recognition for PQC and quantum hardware certification, reducing compliance friction while raising the security baseline across the alliance. By treating allied interoperability as a force multiplier, this model satisfies the LSI test's least-trade-restrictive prong while maintaining a unified front against authoritarian standards projection.

III. How Great Powers Rose: An Exegesis of Technological Appropriation

I. Appropriation as the Unwritten Doctrine of Power

Every ascendant power has mastered a repertoire of appropriation. The term is not a pejorative so much as a description of how states convert the world's stock of know-how—codified and tacit, legal and illicit—into domestic capacity. The contemporary framing of the quantum race as a morality play between "innovators" and "appropriators" is historically thin; technological appropriation has been a routine instrument of ascent for nearly every great power. In the early British takeoff, Parliament's bans on artisan emigration and machine export tried to dam the flow; well before closure, Britain itself cultivated *brain gain*—most famously Flemish weavers

---

[89] Kop, Mauritz, *A Bletchley Park for the Quantum Age*, WAR ON THE ROCKS (Nov. 6, 2025), https://warontherocks.com/2025/11/a-bletchley-park-for-the-quantum-age/, *supra* note …

and other skilled artisans—whose relocation seeded early textile capabilities.[90] still, knowledge seeped outward through itinerant mechanics, pirated designs, and the migration of capital seeking lower costs and laxer rules.[91] Continental imitators—and the nascent United States— paired reverse engineering with state support for technical schools, importing foreign experts even as they adapted machines to local factor endowments and, at times, practiced industrial espionage.[92] French agents and continental manufacturers organized systematic acquisition of British designs, including bribery of skilled workers—a normalized practice in the era's competitive arsenal.[93] This grammar of catch-up, repeated in later centuries, is the prologue to our present moment: appropriation persists, but its modalities have shifted from the theft of drawings and lathes to the capture of standards, the manipulation of supply-chain chokepoints, and the strategic use of finance, data, and law. If history has a moral, it is not that appropriation can be stopped. It is that the means by which states channel, tax, and condition appropriation— what this Article calls security-sufficient openness, disciplined by LSI—determine whether the innovation flywheel accelerates at home or slips beyond reach. Appropriation sits on a continuum—from open scientific exchange and reverse engineering, through compulsory licensing and wartime vesting under emergency powers, to illegal industrial espionage—with law evolving to channel rather than abolish it.[94]

In the nineteenth-century United States, appropriation was both a national project and a legal design choice. Early American leaders welcomed British artisans and their tacit know-how even as Westminster criminalized the export of twist frames and spinning mules. Samuel Slater's transplantation of Arkwright techniques is the classroom example; Francis Cabot Lowell's replication and improvement of the power loom after touring English mills is the other.[95] The deeper story is institutional. Congress framed a patent system that reduced fees, simplified procedures, and circulated specifications widely, democratizing access to the "useful arts" for small inventors and frontier mechanics.[96] Beyond patents, the United States delayed adherence to the Berne Convention and conditioned foreign authors' rights on a domestic "manufacturing clause," an openly nationalist posture later scrutinized by the Court when works were restored to protection; only in 1989 did the United States finally accede to Berne, formalizing its shift from "pirate" to IP hegemon.[97] As latecomers mature, they predictably seek to *kick away the ladder*: having leveraged permissive regimes while catching up, they advocate stronger international IP to entrench their lead.[98] The Paris Convention of 1883 and its progeny can thus be read—at least in part—as codifying incumbents' preferences once industrial maturity was achieved.[99] Appropriation here worked through law's publicity function: disclosure and contestability were not the price of monopoly but the engine of learning. The state did not eliminate imitation; it structured it.

---

[90] Eric J. Hobsbawm, *Industry and Empire: The Birth of the Industrial Revolution* (Weidenfeld & Nicolson 1968).
[91] David S. Landes, *The Unbound Prometheus: Technological Change and Industrial Development in Western Europe from 1750 to the Present* (1969).
[92] Friedrich List, *The National System of Political Economy by Friedrich List,* trans. Sampson S. Lloyd, with an Introduction by J. Shield Nicholson (London: Longmans, Green and Co., 1909).
[93] David J. Jeremy, *Transatlantic Industrial Revolution: The Diffusion of Textile Technologies Between Britain and America, 1790–1830s* (1981), https://mitpress.mit.edu/9780262100229/transatlantic-industrial-revolution/
[94] Thorstein Veblen, *Imperial Germany and the Industrial Revolution* (1915).
[95] Jeremy, *supra* note …
[96] Zorina B. Khan, *The Democratization of Invention: Patents and Copyrights in American Economic Development, 1790–1920* (Cambridge Univ. Press 2005).
[97] See WIPO Lex, *Berne Convention—U.S. Accession (1989)*, https://www.wipo.int/wipolex/en/text/582639; and *Golan v. Holder*, 565 U.S. 302 (2012).

Germany's industrial ascent recast the toolset. Germany's catch-up exemplified Gerschenkron's latecomer thesis and Abramovitz's convergence logic, where disciplined borrowing and complementary investments drive rapid technological ascent.[100] Friedrich List's "national system" married infant-industry protection to railways, technical education, and targeted importation of foreign machinery and experts.[101] German firms were not simply pirates; they learned, refined, and, within a generation, set their own quality benchmarks in chemicals, optics, and electrical equipment. Japan's Meiji transformation repeated the pattern with bureaucratic precision. The state hired *oyatoi gaikokujin* to bootstrap factories and schools, then used a developmental bureaucracy—what Chalmers Johnson later called "plan rationality"—to steer private capital into priority sectors while modulating foreign technology inflows through licensing, procurement, and tariff policy.[102] Absorptive capacity, not access alone, proved decisive; firms and states that built organizational routines for learning converted imitation into mastery.[103]

## II. Lawful Imitation, Emergency Powers, and State-Engineered Diffusion

Anglo-American law preserved channels for such lawful imitation. Reverse engineering remains a "proper means" under trade-secret law, and state-law anti-copying regimes that collide with the federal patent scheme are preempted; courts have likewise protected intermediate copying and decompilation for interoperability in *Sega* and *Sony v. Connectix*.[104] At times, antitrust decrees functioned as appropriation catalysts; the 1956 AT&T consent decree compelled broad licensing of the Bell Labs portfolio and, by many accounts, spurred wave after wave of downstream innovation.[105]

---

[98] Ha-Joon Chang, *Kicking Away the Ladder: Development Strategy in Historical Perspective* (Anthem Press 2002).

[99] Paris Convention for the Protection of Industrial Property, Mar. 20, 1883, 828 U.N.T.S. 305.

[100] *See* Alexander Gerschenkron, *Economic Backwardness in Historical Perspective* (1962); and Moses Abramovitz, Catching Up, Forging Ahead, and Falling Behind, 46 J. Econ. Hist. 385 (1986).

[101] List, *supra* note …

[102] Chalmers Johnson, *MITI and the Japanese Miracle: The Growth of Industrial Policy, 1925–1975* (Stanford Univ. Press 1982).

[103] Wesley M. Cohen & Daniel A. Levinthal, Absorptive Capacity: A New Perspective on Learning and Innovation, 35 Admin. Sci. Q. 128 (1990).

[104] *See Sega Enters. Ltd. v. Accolade, Inc.*, 977 F.2d 1510 (9th Cir. 1992); and *Sony Computer Entm't, Inc. v. Connectix Corp.*, 203 F.3d 596 (9th Cir. 2000).

The coercive edge of appropriation—seizure, clandestine collection, and outright expropriation—has always coexisted with contractual and institutional channels. During the First World War, the United States invoked the Trading with the Enemy Act to vest and auction thousands of enemy-owned patents, a program later upheld in *United States v. Chemical Foundation*.[106] The Supreme Court reiterated that framework in *Farbwerke v. Chemical Foundation*,[107] and economic historians trace capability effects—including in materials and metallurgy—from those vestings; auctions and compulsory licensing associated with World War I also shifted inventive activity in measurable ways.[108] Not all appropriation was adversarial. The 1940 "Tizard Mission" transferred British radar, cavity magnetrons, and know-how to American laboratories, while "Operation Paperclip" relocated German rocket expertise after the war.[109] In parallel, Soviet atomic espionage compressed learning cycles for its nuclear program.[110] In the networked era, cyber intrusions have scaled illicit appropriation, with state-linked campaigns exfiltrating terabytes of data from commercial and defense firms—APT1 and "Cloud Hopper" as canonical dossiers;[111] the 2015 U.S.–China "common understanding" to refrain from cyber-enabled IP theft for commercial gain produced transient reductions with contested compliance.[112] These episodes reveal appropriation as an instrument of both alliance and rivalry.

## III. Networked Leverage and the Rules-Based Exception Space

What is new about the twenty-first-century repertoire is its networked leverage. Henry Farrell and Abraham Newman call it weaponized interdependence: states that sit at central nodes in finance, logistics, standards, and data flows can project coercion outward without firing a

---

[105] Martin Watzinger, Thomas A. Fackler, Markus Nagler & Monika Schnitzer, How Antitrust Enforcement Can Spur Innovation: Bell Labs and the 1956 Consent Decree, 12 Am. Econ. J.: Pol'y 256 (2020).
[106] *United States v. Chemical Foundation, Inc.*, 272 U.S. 1 (1926).
[107] *Farbwerke Vormals Meister Lucius & Brüning v. Chemical Found., Inc.*, 283 U.S. 152 (1931).
[108] *See* W.F. Mueller, *The Origins of the Basic Inventions Underlying Du Pont* (NBER 1962); and Petra Moser, *Compulsory Licensing—Did Patent Violations During the Great War Discourage Invention?* (2014).
[109] *See* U.K. Nat'l Archives, *The Tizard Mission and the Cavity Magnetron*; and *Records Relating to Operation Paperclip* (archival collection).
[110] Atomic Heritage Found., *Espionage and the Manhattan Project*, https://ahf.nuclearmuseum.org/ahf/history/espionage/.
[111] *See* Mandiant, *APT1: Exposing One of China's Cyber Espionage Units* (2013); and PwC & BAE Systems, *Operation Cloud Hopper: Exposing a Systematic, Multi-Year Espionage Campaign* (2017).
[112] *See* White House, *Fact Sheet: President Xi's State Visit to the United States* (Sept. 25, 2015); and Cong. Rsch. Serv., *Cybersecurity: U.S.–China Cyber Agreement and Issues for Congress* (2015).

shot.¹¹³ Anu Bradford's Brussels Effect describes the complementary dynamic: a jurisdiction with regulatory gravity can export its rules by making access contingent on compliance, setting default architectures for global markets.¹¹⁴ Appropriation in this ecosystem is as much about bending the curve of standards and certification as it is about copying designs. A firm that captures the interface—an API, a protocol, a test suite—can force rivals to design to its rhythm, tax their participation through standard-essential patents, and embed its preferences in the reference implementation.¹¹⁵ States, in turn, can do geopolitics through procurement profiles and conformity-assessment regimes, pulling allied supply chains into alignment and raising the fixed costs of switching to an alternative stack.

China's rise has been interpreted, sometimes caricatured, through the lens of forced technology transfer, joint-venture requirements, cyber theft, and the civil-military fusion that moves advances across organizational boundaries. That is part of the story. The rest is more prosaic and more enduring: long-horizon investment in engineering education; a dense manufacturing ecosystem capable of world-class yield learning; ambitious participation in standards bodies; and a statecraft that bundles finance, infrastructure, and digital services in initiatives like the Digital Silk Road to seed external dependence on Chinese equipment and protocols. As Michael Horowitz emphasizes, catching up in military-relevant technologies depends not only on blueprints but on organizational emulation and complementary industries.¹¹⁶ Appropriation policy, in other words, must be paired with the slow work of building firms and feedback loops. Scale and sophistication today may be unprecedented, but the strategic logic is not: China's mix of talent programs, standards participation, and illicit acquisition echoes nineteenth-century American and continental tactics,¹¹⁷ while the U.S. response—export controls on semiconductors and quantum items, investment screening, and listings—rhymes with Britain's earlier efforts to police artisan flows and machinery exports.¹¹⁸ Recent U.S. prosecutions in China-related cases illustrate how research-security and counterintelligence enforcement has been operationalized at the university–lab interface.¹¹⁹ This enforcement posture is consistent with long-standing assessments of IP theft as a strategic acquisition channel, including state-directed efforts to close the AI-chip gap.[120] The early British experience shows both the reach and limits of restricting skilled labor and machinery diffusion during industrial catch-up.[121]

---

¹¹³ Farrell & Newman, *supra* note …
¹¹⁴ Bradford, *supra* note …
¹¹⁵ Tim Büthe & Walter Mattli, *The New Global Rulers: The Privatization of Regulation in the World Economy* (Princeton Univ. Press 2011).
¹¹⁶ Michael C. Horowitz, *The Diffusion of Military Power: Causes and Consequences for International Politics* (Princeton Univ. Press 2010).
¹¹⁷ *See* The Commission on the Theft of American Intellectual Property, *IP Commission Report* (2017).
¹¹⁸ *See* e.g., Bureau of Indus. & Sec., *Export Controls on Advanced Computing and Semiconductor Manufacturing Items*, 87 Fed. Reg. 62,186 (Oct. 13, 2022) (and subsequent amendments).
¹¹⁹ *See* e.g., U.S. Dep't of Justice, Harvard University Professor and Two Chinese Nationals Charged in Three Separate China Related Cases (Jan. 28, 2020), https://www.justice.gov/archives/opa/pr/harvard-university-professor-and-two-chinese-nationals-charged-three-separate-china-related

The post-1995 trade regime reframed appropriation's legal boundaries by hard-wiring minimum IP standards into the WTO while preserving a security valve in TRIPS Article 73.[122] Beyond security exceptions, TRIPS Articles 31 and 31bis codify compulsory licensing (including export to address public-health needs), supplying a lawful, procedurally bounded diffusion channel alongside Article 73.[123] Recent WTO panels interpreting the parallel GATT provision confirm that the security exception is bounded by good-faith limits, necessity, and proportionality, even as it recognizes sovereign space to act.[124] Read through LSI, the implication is straightforward. Modern appropriation policy—export and outbound screening calibrated to capability thresholds; research-security and procurement clauses that sequence disclosure and protect genuinely sensitive parameters; standards participation that ties FRAND licensing to interoperable, contestable interfaces—can be reconciled with the international trade order when it is disciplined, justified, and revisable.

History also cautions against the seductions of financial statecraft unmoored from capability analysis. Great powers have often wielded capital controls, sanctions, and access to payment rails to shape industrial trajectories abroad; sometimes they buy time for domestic learning, sometimes they ossify incumbents and starve ecosystems of rivalry. The test, as Robert Blackwill and Jennifer Harris put it, is whether geoeconomic tools advance strategy rather than substitute for it.[125] LSI forces specificity. What exactly is the capability you are trying to arrest or slow? How will you know if a restriction is working, and at what cost to your own diffusion and scale? When will you lift it? A history of appropriation teaches that nonspecific, indefinite measures tend to metastasize into doctrine and, in the process, dull the very edge they claimed to sharpen.

Canada's experience, as a middle power with world-class quantum research and a smaller domestic market, illustrates appropriation's two faces. On one face, open collaboration and immigration channels create dense scientific networks that feed discovery; on the other, weak domestic scale-up and foreign-controlled IP vehicles strain value capture.[126] A historically informed LSI posture would leverage procurement to generate predictable demand for quantum-safe migration and sensing in public infrastructure, couple grants to transparent IP and data-rights plans, and negotiate reciprocity-based license-exception compacts with allies so that collaboration is partnered rather than one-way. The lesson is not that middle powers must choose between openness and restriction; it is that their sovereignty in standards and supply chains will be decided by whether they can translate appropriation into home-grown capability faster than rivals can translate their openness into extraction.

---

[120] *See* e.g., Fanny Potkin, Exclusive: How China Built Its 'Manhattan Project' to Rival the West in AI Chips, REUTERS (Dec. 18, 2025), https://www.reuters.com/world/china/exclusive-how-china-built-its-manhattan-project-rival-west-ai-chips-2025-12-17/

[121] *See* David J. Jeremy, Damming the Flood: British Government Efforts to Check the Outflow of Technicians and Machinery, 1780–1843, 51 Bus. Hist. Rev. 1 (1977), https://www.cambridge.org/core/journals/business-history-review/article/damming-the-flood-british-government-efforts-to-check-the-outflow-of-technicians-and-machinery-17801843/C2351C757F8433D0B3701BDBF4D1BFB6; and Jeremy, *supra* note …

[122] Agreement on Trade-Related Aspects of Intellectual Property Rights art. 73, Apr. 15, 1994, 1869 U.N.T.S. 299.

[123] WTO, *Analytical Index—TRIPS: Art. 31*; *Protocol Amending the TRIPS Agreement*, WT/L/641 (2005) (art. 31bis, in force Jan. 23, 2017).

[124] Panel Report, Russia—Measures Concerning Traffic in Transit, WTO Doc. WT/DS512/R (adopted Apr. 26, 2019) (interpreting GATT art. XXI).

[125] Robert D. Blackwill & Jennifer M. Harris, *War by Other Means: Geoeconomics and Statecraft* (Harvard Univ. Press 2016).

Historically, great powers rose through technological appropriation, industrial espionage and catch-up manufacturing, yes, but also standards-setting and financial statecraft. The networked economy magnifies those levers. The risk is to mistake leverage for strategy. Security-sufficient openness, implemented through LSI, does not deny the need to slow adversarial learning at choke points; it insists that the discipline of least-restrictiveness, sufficiency, and innovation preservation be made explicit, auditable, and time-bounded. That is how appropriation is bent toward domestic capability without decoupling into brittle parallel stacks. The history is not simply prologue; it is a method.

IV. Differentiating Quantum Technology Pillars, Maturity, and Governance Challenges

History's repertoire of appropriation is not an antiquarian prologue; it is a method for allocating legal friction where it changes real capabilities. As Chapter III established, states have always mixed lawful imitation, contractual transfer, emergency takings, and, at times, illicit acquisition to bend the arc of learning. In the quantum era, the governance problem is not unitary. To treat "quantum" as a monolith is a category error. The domain is a stack of distinct pillars—computing, sensing, simulation, communication, networking, quantum-AI hybrids, and enabling materials/devices—each with different maturity curves, strategic risks, and control surfaces. The discipline this Article offers—least trade-restrictive, security-sufficient, and innovation-preserving (LSI)—must therefore be applied with pillar sensitivity. Anchored in the realities of weaponized interdependence and regulatory externalities, the task is to calibrate, not to proclaim.[127]

Part I of this Chapter maps quantum pillars and maturity. Part II applies LSI by pillar/domain. Part III consolidates conformity assessment and claims verification. Part IV distills cross-cutting compliance patterns. Part V synthesizes the nexus across pillars. Part VI reads the geopolitical nexus—United States, European Union, China, and middle powers—through the same lens. Part VII closes with a practitioner's maturity–governance matrix and a concise "LSI audit" to make measures reviewable and, crucially, liftable.

---

[126] *See* e.g., Jeremy de Beer, Building Canada's Capacity to Use Intellectual Property for Innovation, Dec. 5, 2025, CIGI, https://www.cigionline.org/articles/building-canadas-capacity-to-use-intellectual-property-for-innovation/

[127] *See* Farrell & Newman, *supra* note ….; and Bradford, *supra* note …

## I. Pillarized Taxonomy and Maturity Map

A coherent governance strategy begins with a differentiated understanding of the pillars and their relative maturity. We assess maturity as a composite of publicly anchored indicators: (1) Technology Readiness Level (TRL)–like guideposts in official roadmaps, (2) procurement-readiness signaled by finalized standards and certifications (e.g., PQC FIPS), (3) the existence of auditable, application-oriented benchmarks, and (4) overall ecosystem readiness, including workforce and testbeds.[128] On this basis, a comparative profile reads as follows. Sensing and timing score highest (mid-to-high TRLs for defined use cases); communication is bifurcated between software-ready PQC and certification-emergent QKD; and computing remains in the pre-fault-tolerant NISQ era (low-to-mid TRLs) with major verification campaigns underway.[129] Public anchors in U.S. and U.K. quantum roadmaps, the U.S. Department of Energy's 2024 applications roadmap, and the UK POST briefing corroborate this ordering and the need for auditable verification before policy treats "advantage" claims as procurement-relevant.[130]

To govern these diverse maturities, we must move beyond the traditional binary "dual-use" classification, which is increasingly obsolete for omnipresent quantum technologies where barriers to entry are lowering and civilian utility is inextricably linked to military potential.[131] Instead, we adopt the more granular Vaynman-Volpe civilian integration framework that assesses technologies along two distinct dimensions: the "distinguishability" between military and civilian applications, and the "degree of integration" within the civilian economy.[132] This heuristic allows for precise calibration: technologies with low distinguishability and high integration (like PQC and error-mitigation software) require standards-based governance to avoid a "dead zone" of cooperation, whereas technologies with high distinguishability and low integration (like nuclear-grade quantum sensing or specialized cryogenics) remain viable candidates for traditional export controls.

## II. Governance by Pillar

---

[128] Off. of Sci. & Tech. Pol'y, Exec. Off. of the President, *National Quantum Initiative Annual Report* (FY 2024), https://www.quantum.gov/wp-content/uploads/2023/12/NQI-Annual-Report-FY2024.pdf
[129] Cong. Rsch. Serv., IF11836, *Defense Primer: Quantum Technology* (2024); Def. Innovation Unit, *Quantum Sensing Enters the DoD Landscape* (Sept. 27, 2023).
[130] U.S. Dep't of Energy, *Quantum Information Science Applications Roadmap* (Dec. 2024); UK Parliament, POST, *Quantum computing, sensing and communications* (POSTnote 742, Apr. 22, 2025).
[131] Brendan Walker-Munro, *Moving Beyond "Dual Use": Quantum Technologies and the Need for New Research Security Paradigms*, 12 EPJ Quantum Tech. 136 (2025), https://doi.org/10.1140/epjqt/s40507-025-00448-w
[132] Jane Vaynman & Tristan A. Volpe, *Dual Use Deception: How Technology Shapes Cooperation in International Relations*, 77 Int'l Org. 599 (2023), https://ideas.repec.org/a/cup/intorg/v77y2023i3p599-632_4.html

This section offers an LSI decision framework per quantum pillar —a pillar-sensitive method for mapping distinct technology maturity curves and exposure surfaces to administrable control points, so governance can be calibrated rather than treated as blunt "quantum" decoupling. For each pillar, the decision rule is the same: (i) state the specific capability risk you are targeting; (ii) select the least trade-restrictive lever (IP, contracting, export/investment control, standards/conformity) that still arrests that risk; (iii) bind the measure to verification and sunset so it remains proportionate and revisable.

*A. Quantum Computing*

The computing pillar concentrates the imagination, sitting closest to the cryptanalytic threat of "Q-Day." This distant but profound risk triggers anticipatory state responses and shapes a bifurcated IP strategy: foundational algorithms (e.g., Shor's, Grover's)[133] are treated as public goods, while hardware architectures and control stacks are aggressively patented, creating dense thickets. Simultaneously, firms often hold calibration assets, fabrication recipes, and device-characterization corpora as trade secrets, producing information asymmetries and complicating publication timelines.

To make the risk surface explicit: the principal exposures are (i) unmonitored cloud access to sensitive hardware, where remote QCaaS sessions can de facto bypass traditional export geographies unless identity, purpose, and rate limits are enforced; (ii) leakage of calibration files, device models, and error-mitigation assets, which can collapse an adversary's learning curve by substituting costly, tacit characterization with turnkey recipes; and (iii) premature claims of "advantage" untethered to auditable benchmarks, which distort policy, procurement, and capital allocation by rewarding marketing over measured capability.

Each of these exposures maps to a different lever—governed cloud access and telemetry for (i), data-rights and pre-publication controls for (ii), and QED-C/DARPA-QBI verification for (iii).

Defensive posture: Governments have decoupled cryptographic hygiene from compute timelines by finalizing PQC baselines (FIPS 203/204/205) so that migration proceeds now under procurement and certification mandates, driven by immediate "harvest-now, decrypt-later" risk.[134] This urgency is mathematically grounded in "Mosca's Theorem": if the time to migrate infrastructure plus the shelf-life of the data exceeds the time until a quantum computer arrives, the system is already broken.[135] NIST's PQC standards are a governance anchor, but migration timelines should be evaluated against adversary collection incentives and allied operational requirements, particularly for military and critical infrastructure.[136] Therefore, PQC migration must be framed not merely as a security upgrade, but as a business continuity requirement—partners who are not quantum-safe will be effectively excluded from the allied supply chain. Decoupling moves the locus of compliance from speculation about when fault-tolerance arrives to concrete program management: inventories of cryptographic dependencies, crypto-agility in design controls, Cryptographic Module Validation Program (CMVP) validations, and staged cutovers with rollback plans—an administrative state of the art that rewards documentation and punishes bravado.[137] Ideally, regulation should mandate "crypto-agility"—the systemic capacity to switch algorithms rapidly—rather than hard-coding specific static algorithms. As Rand and Rand demonstrate, static mandates create brittleness, whereas agility mandates are LSI-compliant because they are less trade-restrictive yet more resilient against future cryptanalytic breakthroughs.[138]

---

[133] *See* Peter Shor, *Polynomial-Time Algorithms for Prime Factorization and Discrete Logarithms on a Quantum Computer*, SIAM J.Sci.Statist.Comput. 26 (1997) 1484, https://doi.org/10.1137/S0097539795293172; and Lov K. Grover, A fast quantum mechanical algorithm for database search, May 29, 1996, arXiv:quant-ph/9605043v3

[134] *See* e.g., Cybersecurity & Infrastructure Sec. Agency, Nat'l Security Agency & Nat'l Inst. of Standards & Tech., *Quantum-Readiness: Migration to Post-Quantum Cryptography* (Aug. 21, 2023).

Offensive posture: Export and capital controls are targeting future capabilities. Commerce's September 2024 rule added ECCN 4A906 for quantum computing items (with cognate software/technology entries),[139] while Treasury's outbound-investment rule implementing E.O. 14105 modulates capital and governance rights in quantum information technologies.[140] Because CFIUS defines "critical technologies" by reference to the Commerce Control List, these ECCN additions also widen inbound review for certain foreign investments in U.S. quantum businesses.[120] The resulting perimeter is not merely jurisdictional but architectural: remote access to QCaaS, export of calibration corpora, and the provision of integration services become licensable touchpoints, which—if drafted with capability thresholds and coalition exceptions—can slow adversarial learning without freezing legitimate collaboration.

LSI implications: IP claims should be drafted to hardware-bounded, reproducible improvements—compiler passes that reduce two-qubit errors on specified devices; control strategies validated against declared noise models—so that eligibility becomes engineering rather than metaphysics. Controls should key to capability thresholds (logical-qubit counts, gate

---

[135] Bindel, Nina, Michele Mosca & Bill Munson, *The Quantum Threat to Cybersecurity and Privacy*, in THE BOUNDARIES OF DATA 46 (Bart van der Sloot & Rolf van Schendel eds., 2024), https://www.taylorfrancis.com/chapters/oa-edit/10.5117/9789463729192_CH03/quantum-threat-cybersecurity-privacy-nina-bindel-michele-mosca-bill-munson
[136] Weinstein and Rodenburg, *supra* note …
[137] See NIST, *Cryptographic Module Validation Program (CMVP)* (FIPS 140-3 validations); NIST, *Post-Quantum Cryptography: Selection and Standardization of Public-Key Algorithms*; FIPS PUBS 203–205 (2024).
[138] Lindsay Rand & Theodore Rand, *The "Prime Factors" of Quantum Cryptography Regulation*, 3 Notre Dame J. on Emerging Tech. 37 (2022), https://ndlsjet.com/rand-the-prime-factors-of-quantum-crytopgraphy-regulation/
[139] Bureau of Indus. & Sec., *Commerce Control List Additions and Revisions; Implementation of Controls on Advanced Technologies Consistent with Controls Implemented by International Partners*, 89 Fed. Reg. 72,926 (Sept. 6, 2024) (adding, inter alia, ECCN 4A906).
[140] U.S. Dep't of the Treasury, *Outbound Investment Security Program—Final Rule* 2024, *supra* note …

fidelities, verified algorithmic performance) with clear license exceptions for allied partners and gates on remote access and model-weight export above defined performance envelopes. In publicly funded work, FAR/DFARS scoping and markings should prevent unintended "government-purpose" leakage of proprietary models and toolchains. Finally, correctness must be publicly legible: DARPA's QBI and QED-C's application-oriented benchmarks provide verification grammar that procurement officers can cite, litigation can withstand, and markets can internalize.[141]

*B. Quantum Sensing & Timing*

The nexus for sensing and metrology is defined by immediate, tangible dual-use applications. The trigger is not a future threat but the current military utility of hypersensitive devices for navigation, surveillance, and detection. Consequently, state responses are more mature and reactive; multilateral lists already control certain high-performance gravimeters and SQUID (superconducting quantum interference device) magnetometers, with thresholds requiring regular revision as miniaturization proceeds.[142]

Two distinct leak paths matter. One is uncontrolled diffusion of militarily significant, yet compact, dual-use instruments (e.g., backpackable gravimeters or robust, chip-scale atomic clocks) whose SWaP-C profile makes field deployment realistic outside controlled programs. The other is component diversion: high-precision sub-assemblies (e.g., low-phase-noise timing references, narrow-line lasers, cryo-electronics) re-routed through gray channels or "service" exports that effectively deliver end-use capability piecemeal. An LSI posture therefore keys controls to deployability (SWaP-C and mission profile), not raw sensitivity alone, and pairs thresholds with decontrol notes for academic-scale instruments to keep metrology and collaboration alive.

LSI implications: Because deployability, not raw sensitivity, drives strategic impact, controls should be keyed to SWaP-C and mission profiles, with transparent decontrol notes for academic-scale instruments to preserve collaboration and metrology.[143] Where defense classification or secrecy orders are invoked, scheduled review under an LSI cadence can protect sensitive parameters without starving the standards process. Repeated assessments by defense, intelligence, and parliamentary bodies converge on sensing/timing as earliest for mission use; OECD[144] and CNAS/DIU syntheses reinforce the policy salience of neutral metrology and test-and-evaluation capacity as demand-pull accelerants.[145] Procurement can hasten maturity without distortion by funding open measurement protocols, publishing calibration references, and insisting that claims travel across labs—techniques that keep markets honest without closing doors.

---

[141] *See* Def. Advanced Research Projects Agency (DARPA), *Quantum Benchmarking (QBI) Program* (program page, updated 2025); and Quantum Econ. Dev. Consortium (QED-C), *Quantum Algorithm Exploration Using Application-Oriented Performance Benchmarks* (Feb. 16, 2024).
[142] The Wassenaar Arrangement, *List of Dual-Use Goods and Technologies* (controls on gravimeters and SQUID magnetometers; periodic threshold updates).
[143] U.S. Dep't of Def., *Defense Acquisition Guidebook* (defining SWaP-C considerations in deployability and design trades).
[144] OECD, *A Quantum Technologies Policy Primer* (2025).

*C. Quantum Simulation*

Quantum simulation is the higher-TRL, nearer-term slice of the computing pillar, and governance that focuses only on "Q-Day" risks missing the earliest consequential capability jumps.[146] Unlike universal fault-tolerant machines, simulators are typically special-purpose (analog or digital) devices built to replicate specific quantum systems—precisely the domain where quantum mechanics defeats classical approximation. A large global base already exists: hundreds of quantum simulators operate across platforms (cold atoms, trapped ions, superconducting circuits, photonics), with rapid progress toward more programmable, application-oriented prototypes.[147] Their near-term value proposition is practical: modeling molecular interactions, materials, and complex quantum many-body dynamics relevant to catalysts, batteries, superconductors, and drug discovery—use cases explicitly highlighted in defense-facing assessments.[148] Here, the national-security risk is not "code-breaking" but acceleration: simulation can compress discovery cycles for strategically sensitive materials and processes (including energetics, stealth, and nuclear-adjacent chemistry), while also speeding the enabling stack that makes later universal quantum computing more feasible.[149] As with QCaaS, simulation often arrives as a service, with the sensitive "export" being controlled access to high-performance runs, calibration assets, and workflow know-how rather than a shippable box.[150] An LSI-consistent baseline therefore preserves open science for general methods, benchmarks, and non-sensitive workloads, while layering targeted, verifiable constraints ((contractual use restrictions, Know Your Customer/Know Your World (KYC/KYW), and service licensing)) for red-zone workloads where the marginal acceleration is predictably militarizable.[151] Verification can be made operational through auditable workload declarations, logging/telemetry for high-risk runs, and exception processes that maintain allied and academic collaboration under defined guardrails. Finally, procurement and standards should treat simulation claims as application-oriented performance statements—tied to reproducible problem definitions and cross-lab validation—so that investment and regulation track measured capability rather than hype.

---

[145] Def. Innovation Unit, *Quantum Sensing Enters the DoD Landscape* (Sept. 27, 2023); and Cong. Rsch. Serv., IF11836, *Defense Primer: Quantum Technology* (2024).
[146] Simson L. Garfinkel & Chris J. Hoofnagle, Quantum Computing and Simulation: Policy Implications, Ass'n for Computing Mach., TechBriefs, Issue 4, at 1–3 (July 2022), https://www.acm.org/public-policy/techbriefs/quantum-computing-and-simulation
[147] Ehud Altman et al., Quantum Simulators: Architectures and Opportunities, arXiv:1912.06938v2, at 3–6 (Dec. 20, 2019), https://arxiv.org/abs/1912.06938
[148] *See* Federici, supra note …
[149] Chris Jay Hoofnagle & Simson L. Garfinkel, Law and Policy for the Quantum Age (Cambridge Univ. Press 2021).
[150] Michal Krelina, Military and Security Dimensions of Quantum Technologies: A Primer, at 16, 62 (Stockholm Int'l Peace Rsch. Inst. (SIPRI), July 2025), https://doi.org/10.55163/ZVTL1529
[151] Ibid.

LSI therefore treats quantum simulation as an "early-arrival" governance problem: preserve open publication of general methods, error analysis, and non-sensitive benchmarks, while gating the scarce, militarizable margin—high-fidelity runs, calibration assets, and workflow know-how—through narrowly tailored access controls and auditable service terms. Because simulation advantage is use-case specific, restrictions should be triggered by *workload risk* (e.g., red-zone materials, energetics, or weapon-adjacent chemistry) rather than by a generalized "quantum" label, with coalition exceptions and reviewable sunsets tied to verified capability and demonstrated misuse. In practice, LSI's least trade-restrictive posture is contractual and telemetry-based (KYC/KYW, logging, and exception processes) first, with export or licensing escalation only where the incremental acceleration is predictably militarizable and cannot be managed through verifiable service governance.

*D. Quantum Communication*

On one axis lie PQC standards—the algorithmic hardening of classical networks against future quantum adversaries—now consolidated in FIPS 203/204/205 and moving into procurement and certification. On another axis lies quantum key distribution (QKD) and, further out, repeater-mediated architectures. The governance error would be to treat these as fungible substitutes. PQC migration is the near-term floor for national cyber hygiene; QKD is, at best, a specialized overlay that must clear questions of cost, deployment, and systems integration.

The near-term risk profile is threefold. First, interim cryptographic fragility: slow or uneven PQC migration leaves high-value systems exposed to harvest-now-decrypt-later campaigns even as standards exist. Second, non-audited QKD deployments can create *opaque zones* of asserted security that are operationally brittle—absent certified integration with key-management, incident response, and lifecycle maintenance, assurance fails when it is needed most. Third, vendor lock-in arises when proprietary interfaces and standard-essential patents are not tempered by FRAND-honest licensing, making later interoperability or migration prohibitively costly.

LSI implications: ETSI's QKD interface and protection profiles[152] provide early reference stacks, but their downstream effects will turn on conformity assessment in procurement, FRAND-honest SEP licensing, and avoiding standards that smuggle export policy by other means.[153] Western assurance bodies have questioned the operational benefits of broad QKD overlays; an LSI-consistent response pairs PQC procurement mandates with auditable QKD certification where adopted (ETSI GS QKD 014/016) and insists on interoperable interfaces to prevent vendor lock-in.[154] The guiding discipline is simple to state and hard to fake: if a QKD overlay cannot document its integration into key-management systems, incident response, and lifecycle maintenance—on terms a red team can interrogate—it has no claim on public money.

---

[152] ETSI GS QKD 014, *Quantum Key Distribution (QKD); REST-based Key Delivery API*; ETSI GS QKD 016, *Common Criteria Protection Profile for a QKD Module*.
[153] *See generally Microsoft Corp. v. Motorola, Inc.*, 795 F.3d 1024 (9th Cir. 2015) (contractual enforceability of FRAND); *Ericsson, Inc. v. D-Link Sys., Inc.*, 773 F.3d 1201 (Fed. Cir. 2014) (apportionment and jury instructions in SEP damages).

*E. Quantum Networking*

Quantum communication (point-to-point state transfer, e.g., QKD, teleportation) is distinct from quantum networking, the distributed architecture that coordinates many such links (entanglement routing, repeaters, swapping, and control planes) to enable wide-area networking, networked sensing, and distributed computation. Teleportation's 1993 proof fixed the conceptual stakes—states can be transmitted only by destroying the original and using classical side-channels—complicating legal notions of "transfer" and "possession" across borders.[155] Chinese deployments—e.g., the 2,000-km Beijing–Shanghai backbone and Micius satellite experiments[156]—demonstrate scale and sharpen "protocol politics" in ITU fora.[157]

Two 2026 milestones from Jian-Wei Pan's group sharpen what it means to be "ahead" in long distance quantum networking beyond deployed quantum key distribution (QKD) trunks: they report memory–memory entanglement between trapped-ion nodes linked by 10 km of optical fiber that survives longer than the average time required to establish it, directly relieving a core quantum-repeater bottleneck in which decoherence has historically outpaced entanglement generation and purification.[158] They further use the same platform to demonstrate device-independent quantum key distribution (device-independent QKD) at experimental distances (including proof-of-principle device-independent QKD with finite-size analysis over 10 km and a positive asymptotic key rate over 101 km), while a related Science paper reports device-independent QKD over 100 km with single atoms, pushing device-independent QKD toward city-scale regimes and reducing exposure to implementation side channels.[159] Although both results remain laboratory demonstrations and scaling to deployed fiber, field maintenance, and repeater-chained architectures will take sustained engineering, taken together they underscore China's momentum not only in "network scale" (backbones and satellites) but also in the repeater-grade entanglement primitives required for a quantum internet.

LSI implications: Because networking maturity is best assessed through replicable integration environments, governance should track testbeds using the NQIAC definition (multi-user,

---

[154] *See* ETSI GS QKD 016, *Common Criteria Protection Profile for a QKD Module* (2023); European Commission, *EuroQCI—European Quantum Communication Infrastructure.*
[155] *see* Charles H. Bennett et al., Teleporting an Unknown Quantum State via Dual Classical and Einstein–Podolsky–Rosen Channels, 70 *Phys. Rev. Lett.* 1895 (1993).
[156] Jian-Wei Pan et al., Satellite-Based Entanglement Distribution Over 1200 Kilometers, 356 *Science* 1140 (2017).
[157] ITU-T Y.3800-series Recommendations (foundational docs for quantum information networking and QKD frameworks).
[158] Wen-Zhao Liu et al., Long-Lived Remote Ion-Ion Entanglement for Scalable Quantum Repeaters, *Nature* (2026), https://doi.org/10.1038/s41586-026-10177-4
[159] Bo-Wei Lu et al., Device-Independent Quantum Key Distribution over 100 km with Single Atoms, 391 *Science* 592 (2026), https://doi.org/10.1126/science.aec6243

Page 36 of 111

rigorous, replicable systems-integration platforms) and treat testbed access as both a security and innovation lever; current public mapping identifies 13 testbeds in the United States and 15 in Europe (including the UK).[160] Therefore, treat networking as critical infrastructure: emphasize interconnect, resilience, and jurisdictional clarity; use ITU-T Y.3800-series frameworks for taxonomy; and resist premature global overlays absent audited deployments, robust key-management integration, and open interfaces.[161] Cross-border entanglement creates "jurisdictional entanglement" problems that existing territorial doctrines do not neatly resolve; interim fixes—choice-of-law clauses in network peering, auditable provenance of entanglement generation, and sovereign "break-glass" procedures—should be designed before, not after, the first incident.

*F. Quantum-AI Hybrids*

Quantum–AI hybrids are best treated as a cross-pillar accelerator rather than a separate pillar: near-term effects are likely to come from AI improving quantum control, calibration, error mitigation, and workflow optimization, while longer-horizon claims concern quantum-enhanced optimization and quantum machine learning for pattern recognition and multi-source intelligence fusion.[162] Security risk follows the *acceleration pathway*: hybrids can compress design–test cycles for sensitive materials and sensing, reduce operational frictions in contested environments, and lower the expertise barrier for effective use of QCaaS/simulation services. At the same time, the hybrid can be defensive: AI can strengthen anomaly detection, verification workflows, and compliance automation for controlled quantum services, improving auditability without freezing collaboration.[163]

LSI implications: the governing unit should be the *task accelerated* (what the hybrid measurably speeds up), not the marketing label "quantum-AI." Accordingly, controls should (i) key to defined performance envelopes and red-zone workloads, (ii) treat remote service access, model weights, and calibration corpora as the sensitive transfer surfaces, and (iii) require verification grammar—benchmarks, telemetry, and reproducible evaluation protocols—before procurement or regulation credits "hybrid advantage." Where acceleration is credible and militarizable, LSI prefers tiered, auditable service governance (KYC/KYW, workload declarations, logging, and exception processes with coalition carve-outs) over blunt bans; where acceleration is speculative, LSI prefers standards-first evaluation and open testbeds to separate signal from theatre.[164]

*G. The Enabling Stack (Materials, Cryogenics, Lasers, Control Electronics, Fabrication)*

---

[160] *See* Ruane et al., *supra* note …
[161] *Ibid.*
[162] Michal Krelina, Military and Security Dimensions of Quantum Technologies: A Primer, at 16, 62 (Stockholm Int'l Peace Rsch. Inst. (SIPRI), July 2025), https://doi.org/10.55163/ZVTL1529, *supra* note …
[163] Hoofnagle & Garfinkel (2021), *supra* note …
[164] *See* Aboy et al (2025), *supra* note …

The enabling-materials and components pillar is the least glamorous and the most leveraged. Dilution refrigerators, cryo-CMOS and mixed-signal control electronics, single-photon detectors, vacuum systems, narrow-line lasers, photonics and RF components, isotopically enriched feedstocks, and the helium value chain are where supply-chain fragility translates into capability delay. Industrial policy has recognized this: the CHIPS and Science Act mobilizes incentives for upstream tooling and R&D consortia;[165] the European Union's Critical Raw Materials Act targets supply-side resilience.[166] The Wassenaar Arrangement remains a venue for coordination, but its consensus model and adaptation lag make plurilateral alignment (and Commerce's practice of implementing controls consistent with partners) the practical vector.[167]

LSI implications: Because every pillar draws on shared enablers, surgical controls on cryogenics, lasers, timing sources, photonics, and control electronics frequently outperform blunt end-item bans—provided thresholds are explicit, license-exception pathways for trusted coalitions keep allied research, service, and spares ecosystems healthy, and SWaP-C inflects deployability analysis. Stockpiles, service access, and spare-parts logistics are policy variables, not exogenous facts; governments should treat maintenance contracts and field-service rights as national-capability levers and draft them accordingly.

III. Conformity, Verification, and Claims Management

To prevent advantage theatre, governance must privilege independently verifiable performance claims. Non-expert opacity is structural—multiple modalities, a lack of independently verified universal metrics, and context-dependent links between devices and algorithms—so "verification" must be treated as an administrative-capacity problem as much as a technical one.[168] Three mechanisms should be institutionally central: (1) auditable, application-oriented benchmarks (QED-C) that measure quality-of-result, runtime, and resource usage; (2) programmatic verification (DARPA's Quantum Benchmarking Initiative) that tests whether approaches can reach utility within a declared horizon; and (3) formal certification where applicable (e.g., FIPS validations via CMVP for PQC implementations; Common Criteria evaluations for QKD modules).[169] These instruments discipline marketing narratives, guide procurement, and reduce policy error. When agencies cite these instruments in solicitations and debriefings, they convert technical culture into administrative law—the subtle but decisive move that turns "best practice" into enforceable practice.

---

[165] CHIPS and Science Act of 2022, Pub. L. No. 117-167, 136 Stat. 1366.
[166] Regulation (EU) 2024/1252, *Critical Raw Materials Act*, 2024 O.J. (L) 1.
[167] The Wassenaar Arrangement on Export Controls for Conventional Arms and Dual-Use Goods and Technologies, Public Documents (participation and list-update process).
[168] *See* Ruane et al., *supra* note …
[169] *See* Quantum Econ. Dev. Consortium (QED-C), *Quantum Algorithm Exploration Using Application-Oriented Performance Benchmarks* (Feb. 16, 2024); and Def. Advanced Research Projects Agency (DARPA), *Quantum Benchmarking (QBI) Program* (program page, updated 2025).

IV. Cross-Cutting Compliance Patterns

By "compliance" here I mean the full stack of regulatory and contractual obligations that attach to quantum R&D and commercialization—not just export control formalities, but also sanctions screening (OFAC), inbound and outbound investment reviews (CFIUS/Treasury), government procurement requirements (e.g., FIPS/CMVP validations for PQC), federal data-rights architecture (FAR/DFARS), research-security and disclosure rules (e.g., NSPM-33–style policies at universities and agencies), cybersecurity and privacy duties where sensitive data is handled, and the intellectual-property policies of standards bodies (FRAND/SEP disclosures).[170] The practitioners who carry this load are concrete: university research compliance officers and export control officers; principal investigators and lab managers; in-house counsel and compliance teams at startups and primes; outside trade, IP, and government-contracts counsel; standards delegates and patent professionals; program managers and contracting officers in government; and, increasingly, cloud "trust & safety" or compliance engineers who gate QCaaS access. For example, imagine a U.S. spinout preparing a preprint on trapped-ion control while onboarding an international postdoc and negotiating a federal prototype contract: the research compliance officer drafts a Technology Control Plan (TCP) and runs an EAR classification to confirm whether control electronics and calibration files implicate ECCN 4A906 "technology"; outside trade counsel evaluates a deemed-export license for the postdoc and whether any license exceptions apply; in-house counsel maps deliverables and markings to FAR/DFARS to avoid inadvertent "government-purpose" rights in proprietary toolchains; the security team stands up a controlled data environment and role-based access for device models; the product lead schedules CMVP validation steps to align with a customer's FIPS 203/204/205 procurement gate; standards counsel files timely IPR disclosures in an ETSI working group; and corporate counsel screens a strategic investment for CFIUS sensitivity while confirming no outbound-investment notification is triggered.[171] That integrated workflow is what "compliance" operationalizes—sequencing science, contracts, and controls so innovation can proceed without tripping legal tripwires.

For practitioners, pillar sensitivity translates into a coherent workflow. Pre-publication and pre-release reviews should extend to manuscripts, repositories, calibration files, device-characterization corpora, and cloud notebooks; export screening should account for "deemed exports" and user provenance; contracts should hard-wire crypto-agility, log retention, secure enclaves, and FAR/DFARS-consistent data-rights maps; standards strategy should position participants early in PQC profiles and QKD interoperability groups with aligned patent disclosures and FRAND commitments; and capital screens should diligence inbound (CFIUS) exposure via ECCN cross-walks and outbound prohibitions/notifications under Treasury's rule.[172] Open-source posture accelerates diffusion if sequenced behind filing and classification decisions and if contributor-license agreements harmonize with patent strategy; it becomes leakage when enabling know-how is turned into a global public good by inadvertence.[173] A simple operational rule carries far: draft publication plans and export plans on the same day; if you cannot write the latter, you are not ready for the former.

---

[170] *See* e.g. U.S. Dep't of the Treasury, *Outbound Investment Security Program—Final Rule* 2024, *supra* note …

[171] *See* e.g. Bureau of Indus. & Sec. 2024, *supra* note …

[172] *See* 15 C.F.R. pts. 730–774 (Export Administration Regulations) (defining "technology," "software," deemed exports, license exceptions); and 31 C.F.R. § 800.215 (defining "critical technologies" by reference to the Commerce Control List).

## V. The Nexus Across Quantum Pillars

The governance nexus at the intersection of quantum technology, intellectual property, and national security manifests differently in each pillar because maturity, deployability, and threat profiles vary across the stack. It therefore demands simultaneous navigation of dual-use risks, export and investment controls, and globally distributed supply chains—and, crucially, the feedback loop that couples them.

By "feedback loop," I mean a recurring causal cycle in which visible dual-use capability prompts policymakers to impose controls; those controls reprice information and incentives inside firms, pushing them toward different IP postures (e.g., more patenting at interfaces or deeper trade secrecy) and toward standards strategies that seek defensible moats; that shift, in turn, increases the strategic value of chokepoint inputs (cryogenics, lasers, timing sources), inviting both regulatory tightening and adversary leverage through supply-chain coercion or stockpiling; the resulting scarcity, operational work-arounds, and compliance data then flow back into the next round of rulemaking, exceptions, and benchmark definitions. Put simply, this is an iterative cycle: capability assessments inform controls; controls shape IP and standards strategy; those strategies reallocate supply-chain pressure; and the resulting shifts in capability and chokepoints require recalibrated controls over time. The loop can be virtuous when LSI calibrates measures to evidence, builds in sunsets, and uses benchmarks to relax gates as risks abate; it turns vicious when blunt rules drive activity into opacity, fragment standards, or provoke retaliatory mineral and capital restrictions. For computing, the feedback loop is driven by a long-term, high-impact dual-use threat of breaking legacy encryption and enabling faster optimization and simulation workflows—and the eventual arrival of Q-Day. Public demonstrations of programmable superconducting processors are often treated as proof-of-principle 'capability signals'—useful for narrative mobilization, but insufficient as governance benchmarks absent application-relevant metrics.[174] Export controls are nascent, targeting future capabilities (e.g., logical-qubit counts) rather than current systems; governments have decoupled cryptographic hygiene from compute timelines by finalizing PQC baselines so migration proceeds now under procurement and certification mandates.[175] Cyber agencies have underscored harvest-now-decrypt-later risk, which shifts immediate pressure to inventory, crypto-agility, and conformance rather than speculative timelines. In practice, that means independent benchmarks and incident telemetry operate as the "signals" that tighten or relax access gates over time, so the loop is driven by measured performance and observed misuse rather than by hype or fear. The compliance center of gravity thus moves into the cloud: governed access to QCaaS by user provenance and use case; telemetry and logging sufficient to reconstruct misuse; and graduated restrictions keyed to verified device performance, not vendor claims. Recent proposals emphasize cross-sector benchmarking regimes—linking governments, industry, and academia—to reduce hype-driven procurement and enable comparable, testable claims across platforms.[176] Benchmarks and QBI-style challenges provide the evidentiary backbone for those gates, while FAR/DFARS scoping prevents inadvertent transfer of crown-jewel know-how under broad "government purpose" rights.[177]

---

[173] *See Jacobsen v. Katzer*, 535 F.3d 1373 (Fed. Cir. 2008).

[174] *See* e.g., Frank Arute et al., Quantum Supremacy Using a Programmable Superconducting Processor, 574 Nature 505 (2019).

[175] *See* e.g., Bureau of Indus. & Sec. 2024, *supra* note …

Simultaneously, some firms opt for deep trade secrecy, and supply chains hinge on upstream chokepoints in cryogenics, lasers, and control electronics. These chokepoints now sit in formal control architectures: Commerce added quantum entries to the CCL (e.g., 4A906) and aligned posture with partners; CFIUS inbound screens and Treasury's outbound rule modulate capital and governance rights tied to sensitive compute.[178] Because learning curves in compute collapse fastest when data and calibration methods leak, pre-publication review must treat repositories, weights, and device models as export-relevant artifacts; university policies that screen only manuscripts while ignoring code and notebooks will fail in practice and in principle.[111]

For sensing, the trigger is present-tense military utility; controls are anchored in multilateral regimes with performance metrics; classification and secrecy orders can short-circuit diffusion for genuinely sensitive devices; and LSI argues for SWaP-C-keyed thresholds and decontrol notes for academic-scale instruments to preserve collaboration and metrology. Procurement can double as policy: mission-profile test and evaluation in defense and civil pipelines, with publishable calibration standards, compresses time-to-truth and accelerates safe diffusion, while targeted secrecy orders with scheduled review keep the narrowest secrets narrow.

For quantum simulation, the feedback loop is driven by near-term capability and service delivery: as simulators move from laboratory demonstrations to practical use, policy attention will concentrate less on qubit counts and more on who can run which high-value workloads, under what contractual and auditing conditions.[179] Controls and compliance then reshape IP posture: firms and labs will increasingly protect calibration assets, domain-specific workflows, and high-value problem instances as trade secrets, while publishing general methods to preserve scientific priority and standards influence. That shift increases the strategic value of enabling inputs (control electronics, lasers, cryogenics, detectors) and of secure service operations, inviting both tighter governance and adversary efforts to obtain "recipe-level" know-how through access, services, or talent channels.[180] LSI keeps the loop virtuous by tying any escalation (from contractual controls to licensing or export restrictions) to workload-specific, verifiable risk—especially where simulation predictably accelerates militarizable materials or weapon-adjacent processes—while maintaining coalition exceptions, audit trails, and sunsets so measures remain proportionate and liftable.

---

[176] *See* Benchmarking Quantum Technology Performance: Governments, Industry, Academia and their Role in Shaping our Technological Future, Eur. Ctr. for Int'l Pol. Econ. (ECIPE) (last visited Feb. 3, 2026), https://ecipe.org/publications/benchmarking-quantum-technology-performance/
[177] *See* FAR 52.227-14 (Rights in Data—General); DFARS 252.227-7013 (Rights in Technical Data—Other Than Commercial Products and Commercial Services); DFARS 252.227-7014 (Rights in Noncommercial Computer Software and Noncommercial Computer Software Documentation). FAR 52.227-14 (Rights in Data—General), *supra* note …
[178] Bureau of Indus. & Sec., *Commerce Control List Additions and Revisions; Implementation of Controls on Advanced Technologies Consistent with Controls Implemented by International Partners*, 89 Fed. Reg. 72,926 (Sept. 6, 2024), *supra* note …
[179] *See* Altman et al, *supra* note …
[180] *See also* Hoofnagle & Garfinkel (2021), *supra* note …

For communication and networking, the competition centers on secure infrastructure and protocol politics.[181] PQC is the near-term floor; QKD is a specialized overlay that raises assurance and cost questions. NSA skepticism about broad QKD overlays runs in parallel with China's demonstration effects (Beijing–Shanghai backbone; *Micius* satellite) and standards pushes in ITU and ETSI.[182] An LSI-consistent response pairs PQC procurement mandates with auditable QKD certification (ETSI GS QKD 014/016) where adopted and insists on FRAND-honest interfaces to avoid vendor lock-in and geopolitical bifurcation.[183] Crucially, "communication" differs from "networking": the former secures point-to-point key exchange; the latter builds an entanglement fabric whose control plane raises first-order questions of interconnect, resilience, and jurisdiction. Policy that conflates them will either over-invest in overlays that fail operational audits or under-invest in the hard engineering of repeaters and routing that the future network will actually require.

For quantum-AI hybrids, the nexus is emergent and acts as a threat multiplier. The prudential course is capability-based triggers (what is actually accelerated), layered compliance (AI model governance plus quantum access controls), and sandboxed testbeds with audited benchmarks to separate genuine acceleration from theatre.[184] Where the hybrid demonstrably accelerates model training or search, export and procurement should hinge on the specific task accelerated and the downstream use case, not on a generic label.

## VI. The Geopolitical Nexus

United States. The U.S. posture is best understood as market speed coupled to an expanding security perimeter, in which upper-stack dominance—software, QCaaS, and venture capital—functions as a chokepoint strategy that must be disciplined by LSI to avoid overreach and collateral damage. CHIPS guardrails condition subsidies on limits to expansion in countries of concern; Commerce's 2024 quantum controls and Treasury's 2024 outbound rule extend the perimeter to tooling, cloud access, and capital—tools that require LSI discipline to avoid overreach and collateral damage.[185] In practice, this looks like a three-layer architecture. At the code layer, NIST and CMVP make PQC a procurement-enforced default, turning cryptography into a compliance exercise rather than a forecast of Q-Day. At the platform layer, governed access to QCaaS—with provenance checks, performance-threshold gates, and incident logging—becomes the operational hinge of export control. At the capital layer, CFIUS and Treasury's outbound rule redraw who can finance or govern quantum businesses that cross capability thresholds. The risk is familiar: without LSI discipline, Washington can produce the "worst of both worlds"—over-securitized research that leaks anyway and under-invested supply chains that break under stress. The remedy is equally familiar: capability-based thresholds, coalition exceptions with verification, and sunset logic tied to auditable milestones.[186]

---

[181] *See* DeNardis, Laura. 2022. "Quantum Internet Protocols." Presentation at the Transatlantic Quantum Forum (hybrid event) September 16, https://www.youtube.com/watch?v=2IT9hefrNIE
[182] Nat'l Sec. Agency (NSA), *Quantum Key Distribution (QKD) and Quantum Cryptography* (Cybersecurity Advisory, Oct. 2020) (raising assurance, cost, and deployment concerns).
[183] ETSI GS QKD 014 V1.1.1, *Protocol and Data Format of REST-Based Key-Delivery API* (2019); ETSI GS QKD 016 V2.1.1, *Common Criteria Protection Profile—Pair of Prepare-and-Measure QKD Modules* (Nov. 27, 2023).
[184] *See* also Kop, M.*, Regulating Transformative Technology in The Quantum Age: Intellectual Property, Standardization & Sustainable Innovation* (October 7, 2020). Stanford - Vienna Transatlantic Technology Law Forum, Transatlantic Antitrust and IPR Developments, Stanford University, Issue No. 2/2020, https://purl.stanford.edu/ng658sy7588

European Union. The EU approach is best understood as regulatory orchestration for strategic autonomy, combining the Brussels Effect with collaborative R&D (Quantum Flagship), EuroQCI, certification-forward assurance (ETSI/Common Criteria), and an emerging pivot from pure standard-setting toward supply-side resilience and capital governance. The Critical Raw Materials Act and the Commission's 2025 outbound-investment recommendation signal a pivot from pure standard-setting to supply-side resilience and capital governance—an emerging industrial Brussels Effect that should remain interoperable with allied schemes.[187] In operational terms, Brussels' comparative advantage is rule orchestration: PQC migration proceeds through procurement, conformity assessment, and guidance; QKD deployments, where chosen, ride on ETSI profiles and Common Criteria evaluations; and EuroQCI stitches national pilots into a continental backbone. The open question is commercialization at speed. A European Quantum Act, properly framed, could link funding to certification roadmaps, FRAND-honest SEP policies, and reciprocity with allied regimes—trading regulatory coherence for time-to-market without sacrificing values.[188]

China. China's strategy is best understood as a closed, state-directed military-civil fusion loop: it bundles investment, domestic patenting, standards participation, and illicit acquisition; demonstration effects in quantum communication amplify standards leverage in ITU fora, while episodic mineral export restrictions (gallium, germanium) underscore supply-chain coercion capacity. These gains carry reciprocity costs if partners harden export, outbound-investment, and procurement screens in response.[189] Beijing's playbook is multi-modal: scale up domestic fabrication and component supply, publish standard-friendly reference implementations where it advantages Chinese IP, and press for protocol primacy in ITU while building showcase infrastructure—the Beijing–Shanghai backbone and *Micius*—that converts technical lead into diplomatic leverage.[190] The counter-strategy is not rhetorical condemnation but calibrated reciprocity: coalition procurement that rewards audited interoperability, export and outbound controls that focus on deployable capabilities rather than broad categories, and standards diplomacy that keeps forums open while refusing to ratify unaudited claims.

---

[185] *See* Bureau of Indus. & Sec., *Commerce Control List Additions and Revisions; Implementation of Controls on Advanced Technologies Consistent with Controls Implemented by International Partners*, 89 Fed. Reg. 72,926 (Sept. 6, 2024) *supra* note … ; and U.S. Dep't of the Treasury, *Outbound Investment Security Program—Final Rule* (Oct. 28, 2024) (implementing E.O. 14105; effective Jan. 2, 2025), *supra* note …
[186] *See* e.g., 31 C.F.R. § 800.215, *supra* note …
[187] Regulation (EU) 2024/1252, *Critical Raw Materials Act*, 2024 O.J. (L) 1, *supra* note …
[188] *See* Kop, Mauritz, *Towards a European Quantum Act: A Two-Pillar Framework for Regulation and Innovation,* Columbia Journal of European Law 31 (1), Sept 9, 2025, https://cjel.law.columbia.edu/
[189] *See* e.g., *Preventing the Improper Use of CHIPS Act Funding*, 88 Fed. Reg. 65,599 (Sept. 25, 2023) (CHIPS "guardrails" rule).

Middle powers. For middle powers, the governing logic is niche mastery translated into coalition scale. Canada can convert world-class research into value capture via procurement-led PQC migration and transparent IP plans; the UK can serve as translation hub across software/middleware layers and Five Eyes counterintelligence; Japan can become the indispensable foundry, securing allied supply chains in cryogenics, lasers, and photonics. In each case, LSI provides the common yardstick so coalition instruments (standards, procurement, controls, capital screens) are interoperable, proportionate, and revisable. For Canada, sovereign patent funds, procurement-anchored scale-ups, and reciprocity-based license-exception compacts with allies can turn openness into absorption rather than extraction.[191] For the UK, NQTP's hub structure, POST's cross-pillar briefings, and its AUKUS[192] and Five Eyes roles create leverage at the middleware and research-security layers.[193] For Japan, deep industrial competence in precision manufacturing makes friend-shored supply chains tangible—cryogenics service networks, laser and photonics capacity, and long-term parts agreements that translate resilience from slogan into schedule.[194]

Synthesizing these dynamics, it becomes clear that winning the global race for quantum-AI dominance is not merely a question of achieving a single scientific breakthrough. It is a contest to master this complex governance nexus. Long-term leadership will belong to the actor that can most effectively manage the inherent contradictions: innovate at speed while ensuring innovation is secure; protect IP while permitting knowledge flows fast enough to fuel the next round of discovery; and secure physical and digital supply chains against coercion. The success of the Western bloc will depend not only on the U.S. setting the pace, but on its ability to integrate the specialized, indispensable contributions of its key allies into a resilient democratic coalition. Put differently: the winner sets not only the standard, but the cadence—and does so with controls, contracts, and certification that pass an LSI audit in public.

VII. Practitioner's Maturity–Governance Matrix & LSI Audit Close

The matrix reads as follows. Quantum computing remains in the low-to-mid TRL band of the NISQ era; governance should therefore emphasize capability-keyed export controls on scalable hardware and associated "technology" (ECCN 4A906), governed cloud access whose gates are tied to independently verified device performance, and verification through QED-C benchmarks and DARPA's QBI so that claims become procurement-legible. Patent strategy should favor hardware-bounded, reproducible improvements, while trade-secret protection is reserved for calibration assets and device models; FAR/DFARS scoping and markings must be drafted to avoid unintended government-purpose leakage.

---

[190] *See* e.g., Jian-Wei Pan et al., *supra* note …
[191] *See* e.g., De Beer 2025, *supra* note …
[192] *See* e.g., David E. Sanger and Sarah Cahalan, Quantum Tech Will Transform National Security. It's Testing U.S. Alliances Now, NYT, 28 July 2023, https://www.nytimes.com/2023/07/28/world/australia/quantum-technology-aukus.html
[193] UK Parliament, POST 2025, *supra* note …
[194] *See* also Chalmers Johnson, *supra* note …

In quantum communications, governance bifurcates between PQC as the baseline and QKD as a specialized overlay. For PQC—high maturity on the software side—the levers are procurement mandates tied to FIPS 203/204/205, crypto-inventory and migration roadmaps, CMVP validations, and interoperability profiles that prevent lock-in and ease incident response. For QKD—medium maturity, certification-emergent—policy should require ETSI GS QKD 014/016 conformance, Common Criteria evaluations, and EuroQCI-style interoperability pilots, with FRAND-honest SEP licensing and audited deployments to satisfy assurance bodies before scale-out.[195]

In the quantum sensing domain, sensing and timing sit at mid-to-high TRL in defined use cases; export controls should be keyed to SWaP-C thresholds—deployability, not raw sensitivity—and accompanied by academic decontrols; mission-profile test-and-evaluation and neutral metrology testbeds should validate claims and compress certification timelines; secrecy orders, where truly necessary, should be targeted and time-bounded with scheduled review.[196]

Finally, in quantum networks, networking is early but foundational; governance should adopt ITU-T Y.3800 frameworks, regulate the network as critical infrastructure with resilience and interconnect as first principles, and run repeater and entanglement-routing trials under open interfaces and clear jurisdictional rules—keeping global overlays contingent on audited assurance and operational value.[197] To operationalize this compliance burden for small and medium enterprises (SMEs) lacking deep legal teams, we recommend the adoption of "Security Threat Discovery Cards" (STDCs).[198] These allow firms to self-assess socio-technical risks through a structured value-sensitive design process.

Furthermore, procurement officers should align requirements with the emerging standardization roadmap, specifically referencing active standards such as IEEE P7130 (Quantum Technologies Definitions) and ISO/IEC JTC 3 workstreams. This ensures that "security sufficiency" is measured against consensus engineering metrics rather than vague statutory definitions.

---

[195] *See* European Commission, *EuroQCI—European Quantum Communication Infrastructure, supra* note ...
[196] U.S. Dep't of Def., *Defense Acquisition Guidebook* (defining SWaP-C considerations in deployability and design trades).
[197] ITU-T Y.3800-series Recommendations, *supra* note ...
[198] *See* Umbrello, Steven, Pieter E. Vermaas, Indika Kumara, Joost Alleblas, Stefan Driessen & Willem-Jan van den Heuvel, *Quantum Security Threat Discovery: A Value Sensitive Design Approach to Discovering Security Risks of Quantum Sensing at the Port of Moerdijk*, 19 NANOETHICS 8 (2025), https://link.springer.com/article/10.1007/s11569-025-00475-y

LSI audit close: For any contemplated measure, an agency or general counsel should be able to write—briefly and publicly—(1) the capability at issue, (2) why the chosen lever is the least trade-restrictive that is nonetheless security-sufficient, (3) how the measure preserves the scientific commons, and (4) what verification or milestone will trigger narrowing or sunset. That documentation is not bureaucratic ornament; it is the operating receipt for a regime that must be security-serious and growth-compatible at once.

## V. A Comparative Analysis of Global Quantum Strategies

The contest over quantum technologies is, at bottom, a contest over institutional design: who can choreograph legal, industrial, and scientific instruments so that discovery becomes capability, capability becomes standard, and standard becomes leverage. Across jurisdictions, the same ingredients recur—post-quantum cryptography (PQC) migration, export and investment controls, standards positioning, and building upstream component capacity—but they are combined in different proportions. Comparative work on national innovation systems underscores that sustained capability arises from institutional complementarities—education, manufacturing learning curves, finance, and standard-setting participation—not from single-policy 'race' metaphors.[199]

A fourth cross-cutting ingredient, which this Chapter makes explicit, is the governance of cloud access for QCaaS—through identity, purpose, and performance-threshold gates—along with robust certification and FRAND discipline. In practice, these levers translate paper controls into operational constraint. This Chapter compares these strategies using the analytic lens developed in Chapters III–IV: security-sufficient openness disciplined by the LSI test (least trade-restrictive, security-sufficient, innovation-preserving) and situated within the realities of weaponized interdependence and the Brussels Effect. The result is not a league table, but a policy grammar for understanding the strategic logic of each major tech bloc and the coalition designs that can convert unilateral frictions into shared guardrails. The comparative stakes also include a stubborn governance paradox: controls must be tight enough to matter without becoming revenue instruments or academic choke points—a balance that turns on auditable thresholds, governed cloud access, and certification/FRAND discipline rather than blunt prohibitions.

### I. Analytical Frame: LSI Meets National Innovation Systems

An LSI-consistent strategy identifies a capability risk, chooses the least trade-restrictive lever that still arrests that risk, and binds it to verification and sunset so it remains proportionate and revisable. This is always filtered through national innovation systems. As established, network hubs can project coercion through chokepoints in finance and logistics (weaponized interdependence), while a jurisdiction with regulatory gravity can export its rules by making market access contingent on compliance (the Brussels Effect).[200] These two logics are the background radiation of the comparative landscape that follows. They also explain why similarly worded controls bite differently across blocs: in a hub-and-spoke dollar system or a single market with certification gravity, the same text produces far more practical leverage.

---

[199] *See* e.g., Robert Atkinson et al., Understanding and Comparing National Innovation Systems: The U.S., Korea, China, Japan, and Taiwan, Info. Tech. & Innovation Found. (Feb. 20, 2025), https://itif.org/publications/2025/02/20/understanding-comparing-national-innovation-systems-us-korea-china-japan-taiwan/

## II. United States

The U.S. approach—defined by market speed, an expanding security perimeter, and chokepoints in the quantum toolchain—couples a fast, venture-driven innovation engine to an increasingly articulated security perimeter. Guided by the 2025 National Security Strategy, which identifies quantum computing as a domain that will "decide the future of military power," the U.S. posture is shifting toward "overmatch." This is explicitly codified in the U.S.-China Economic and Security Review Commission's 2025 recommendation for a statutory "Quantum First" by 2030 national goal, specifically prioritizing cryptography, drug discovery, and materials science.[201] However, this pursuit of primacy must be tempered by strategic realism. As Colin Kahl observes, the notion of a decisive "race" to total dominance is a myth; neither the United States nor China can achieve hermetic control over diffusion.[202] Consequently, the U.S. strategy should focus less on futile attempts at total exclusion and more on maintaining a resilient lead in critical nodes—a posture of "managed competition", or "safeguarding through advancing" rather than absolute decoupling.

The national legislative centerpiece is the National Quantum Initiative (NQI)[203] Act of 2018[204] and its proposed reauthorization[205], which aims to coordinate and fund quantum R&D across government agencies like the National Science Foundation (NSF), the National Institute of Standards and Technology (NIST), and the Department of Energy (DOE) to accelerate breakthroughs, build a skilled workforce, and foster a robust public-private ecosystem. This effort was significantly bolstered by the CHIPS and Science Act of 2022, which authorized substantial new funding for NQI programs and broader high-tech manufacturing.[206] A key feature of the U.S. ecosystem is the Quantum Economic Development Consortium (QED-C), an industry-driven consortium managed by SRI International, established to identify and address the needs of the emerging quantum industry.[207] The U.S. (permissionless and *laissez-faire*-style) innovation ecosystem is further characterized by a vibrant venture capital market that funds a rich ecosystem of startups (e.g., Sygaldry, IonQ) and better-known scale-ups like PsiQuantum, Quantinuum and Rigetti, with SandboxAQ anchoring the software/security layer, and a culture of university spin-offs facilitated by the Bayh-Dole Act.[208] Complementary legislation, like the proposed DOE Quantum Leadership Act, seeks to further bolster DOE's role in developing quantum infrastructure, materials science, and supply chain resilience.[209]

---

[200] *See* Farrell & Newman, *supra* note ….; and Bradford, *supra* note …
[201] U.S.-China Econ. & Sec. Rev. Comm'n, 2025 Report to Congress, *supra* note …
[202] Colin H. Kahl, *The Myth of the AI Race: Neither America Nor China Can Achieve True Tech Dominance*, Foreign Aff. (Jan. 12, 2026), https://www.foreignaffairs.com/united-states/myth-ai-race
[203] National Quantum Initiative, https://www.quantum.gov
[204] National Quantum Initiative Act, Pub. L. No. 115-368, 132 Stat. 5094 (2018).
[205] *Cantwell, Young, Colleagues Introduce Bipartisan National Quantum Initiative Reauthorization Act*, U.S. Senate Comm. on Com., Sci. & Transp. (Jan. 8, 2026), https://www.commerce.senate.gov/2026/1/cantwell-young-colleagues-introduce-bipartisan-national-quantum-initiative-reauthorization-act.
[206] CHIPS and Science Act of 2022, Pub. L. No. 117-167, 136 Stat. 1366.

This promotional, accelerationist aspect is counterbalanced by an aggressive security posture. The sequence runs from code to cloud to capital. At the code layer, PQC is being normalized as an operational baseline, with NIST's FIPS 203, 204, and 205 fixing algorithmic anchors and CMVP validations turning conformance into a procurement-auditable obligation. OMB's cryptographic inventory mandate[210] and NSA's CNSA 2.0 guidance[211] push this normalization into day-to-day program management: agencies map dependencies, adopt crypto-agility, and schedule cutovers keyed to validated modules rather than speculative "Q-Day" timelines. This decouples cryptographic hygiene from the uncertain timelines of fault tolerance.[212] At the cloud (or platform) layer, export control migrates from physical crates to digital gates: who may access which devices, at what performance envelope, and under what telemetry is now the live perimeter. Practically, that means identity-verified QCaaS access, purpose limitations in terms-of-use, retention of audit-quality logs, and withholding of calibration corpora/model weights above declared performance thresholds—controls that can be tuned and, crucially, relaxed when benchmarks verify lower risk. BIS's September 2024 rule adding ECCN 4A906 for quantum-computing items defines this line in formal doctrine.[213] Platform governance is not optional, as remote access can function as a de facto export channel, so identity gating, purpose restrictions, and audit-quality logs must be treated as control surfaces on par with hardware shipments. And because academic labs are part of the U.S. engine, the regime needs guardrails that protect collaboration—over-licensing of routine lab access or "deemed export" overreach for foreign graduate students risks chilling precisely the diffusion that makes the ecosystem competitive. At the capital layer, inbound (CFIUS) and outbound (Treasury's E.O. 14105 rule) screening completes the triangle, modulating who can finance or govern consequential firms.[214]

---

[207] *See* Quantum Economic Development Consortium, https://quantumconsortium.org/
[208] *See* 35 U.S.C. § 203 (march-in rights). *See* also Ouellette & Sampat 2024, *supra* note …
[209] Department of Energy (DOE) Quantum Leadership Act of 2025, https://www.durbin.senate.gov/imo/media/doc/doe_quantum_leadership_act_one_pager_2025.pdf
[210] Off. of Mgmt. & Budget, Exec. Off. of the President, M-24-21, *Transition to Post-Quantum Cryptography* (Oct. 24, 2024).
[211] Nat'l Sec. Agency, *Commercial National Security Algorithm Suite 2.0* (Cybersecurity Advisory, Sept. 2022).
[212] *See* e.g., Nat'l Inst. of Standards & Tech. (NIST), *Announcing Approval of Three Federal Information Processing Standards (FIPS) for Post-Quantum Cryptography* (Aug. 13, 2024) (FIPS 203/204/205), *supra* note …
[213] Bureau of Indus. & Sec., 89 Fed. Reg. 72,926 (Sept. 6, 2024), *supra* note …
[214] *See* U.S. Dep't of the Treasury, *Outbound Investment Security Program—Final Rule* (Oct. 28, 2024), *supra* note …; 31 C.F.R. § 800.215, *supra* note …; and Exec. Order No. 14105, 88 Fed. Reg. 54867 (Aug. 9, 2023).

This architecture is knitted together by procurement and industrial policy. Pending U.S. legislative proposals would operationalize this logic via a quantum sandbox for near-term applications, targeted quantum manufacturing support, and supply-chain measures—each of which can be framed as 'governance-through-capability' rather than subsidy alone.[215] This aligns with emerging scholarship that treats quantum production choices—materials, fabs, metrology, and qualification pathways—as legally constitutive of the sector's governance architecture.[216] To secure the physical foundation of this ecosystem, a January 2026 proclamation under Section 232 designates processed critical minerals and their derivative products as a national security concern.[217] In parallel, the White House's "Genesis Mission" executive order establishes the American Science and Security Platform (ASSP), a trusted, integrated compute and data substrate designed to create secure, closed-loop enclaves for high-sensitivity collaborative R&D across quantum information science and critical materials.[218] The CHIPS guardrails condition subsidies on limits to expansion in countries of concern, while initiatives like the Defense Advanced Research Projects Agency's (DARPA) Quantum Benchmarking Initiative (QBI) exemplify this approach. QBI uses a defense-led program to drive the development of commercially relevant benchmarks, effectively setting the agenda for the industry.[219] However, participation comes at a cost for companies, often requiring them to grant the government "Government Purpose Rights" (GPR) to their IP, a significant trade-off between gaining credibility and ceding exclusivity.[220] What's more, the Bayh-Dole Act of 1980, which allows universities and small businesses to retain ownership of inventions made with federal funding, also contains "march-in rights." Though never exercised, this provision theoretically allows the government to force an IP holder to license their patent to other applicants if the invention is not being made available to the public on reasonable terms or if action is needed to address health or safety needs.

This combined strategic posture can be understood through the lens of what some in Washington have begun to style a nascent "American digital silk road"—a coordinated, multi-technology initiative aimed at establishing a U.S-led techno-economic sphere of influence.[221] In that frame, governed cloud access and certification become the connective tissue that lets the United States export not only technology but compliance architectures to allies, and with that, its norms and values. As articulated in strategic documents like the Winning the AI Race: America's AI Action Plan, this approach views the entire innovation ecosystem—from energy and data centers to talent and cybersecurity—as a single, integrated strategic asset, or a "full-stack" that can be leveraged for geopolitical advantage.[222] This involves not just fostering domestic innovation, but also an ambitious international strategy of exporting "full-stack AI export packages" to allies, building a coalition based on American technology and standards.[223]

---

[215] Quantum Sandbox for Near-Term Applications Act of 2025, H.R. 2480, 119th Cong. (2025), https://www.congress.gov/bill/119th-congress/house-bill/2480; Advancing Quantum Manufacturing Act of 2025, H.R. 2481, 119th Cong. (2025), https://www.congress.gov/bill/119th-congress/house-bill/2481; and Support for Quantum Supply Chains Act of 2025, H.R. 2482, 119th Cong. (2025), https://www.congress.gov/bill/119th-congress/house-bill/2482

[216] *See* e.g., Anushka Mittal, Probing the Production of Quantum Technologies to Imagine Its Legal Framework, 3 RSCH. DIRECTIONS: QUANTUM TECHS. e1 (2025), https://doi.org/10.1017/qut.2024.6

[217] The White House, Adjusting Imports of Processed Critical Minerals and Their Derivative Products into the United States (Proclamation) (Jan. 14, 2026), *supra* note …

[218] The White House, Launching the Genesis Mission (Nov. 24, 2025), *supra* note …

[219] DARPA QBI, *supra* note …

[220] *See, e.g.*, Defense Federal Acquisition Regulation Supplement (DFARS) 252.227-7013, Rights in Technical Data—Noncommercial Items, *supra* note …

[221] Kop, Towards a European Quantum Act (2025), *supra* note …

Internationally, the U.S. seeks to build a coalition of like-minded nations. The U.S.-EU Trade & Technology Council created a Quantum Task Force in 2024 to coordinate on research and standards, and similar collaborations exist with the UK and other Five Eyes partners to ensure alignment on export controls and quantum-resistant cryptography. The LSI cure for inevitable over-securitization risks is threefold: capability-based thresholds that can be independently verified, reciprocity-based license exceptions for trusted partners, and published sunsets tied to benchmarks so measures are visibly liftable when risk abates. However, experts have identified potential execution challenges and internal policy contradictions within the U.S. plan. Pro-innovation goals that advocate for deregulation may clash with restrictive trade and immigration policies, and a push to accelerate permitting for infrastructure like data centers could create friction with other regulatory frameworks, highlighting a core challenge for its market-driven, fragmented model.[224]

III. European Union

The EU's posture is best understood as standards power in service of strategic autonomy, now paired with a supply-side rebuild. Its approach relies less on unilateral coercion and more on regulatory orchestration and coalition choreography. The EU's foundational initiative, the Quantum Flagship, is a €1 billion, ten-year R&D program designed to coordinate research across member states and translate scientific excellence into technological innovation. This is being scaled up through the recently announced Quantum Europe Strategy, which aims to make Europe a global leader by 2030 by focusing on research, infrastructure, and ecosystem support.[225] The strategy envisions a future EU Quantum Act, expected by 2026, which will provide a comprehensive regulatory framework combining industrial policy modeled on the EU Chips Act with a risk-based approach to governance inspired by the EU AI Act.[226] A parallel Europe-origin "Responsible Quantum Technologies" (ResQT) community agenda—initiated largely by European researchers but including contributors from other regions—advances a broader responsibility vision, emphasizing sustainability (including supply-chain transparency

---

[222] *See America's AI Action Plan*, AI.GOV, https://www.ai.gov/action-plan.
[223] Exec. Order on Promoting the Export of the American AI Technology Stack (July 2025).
[224] *See* Navin Girishankar et al., *Experts React: Unpacking the Trump Administration's Plan to Win the AI Race*, Ctr. for Strategic & Int'l Stud. (July 2025).
[225] European Commission, *EU's plan to become a global leader in quantum by 2030* (Press Release, July 2, 2025), https://commission.europa.eu/news-and-media/news/eus-plan-become-global-leader-quantum-2030-2025-07-02_en
[226] *See* Proposal for a Regulation... (Artificial Intelligence Act), COM(2021) 206 final (Apr. 21, 2021). *See* also Kop, Towards a European Quantum Act (2025), *supra* note …

and energy realism), inclusive workforce development, and stakeholder-facing roadmaps for adoption, while often leaving under-specified how these commitments hold under acute strategic rivalry and technology-denial dynamics.[227]

Its comparative advantage is the Brussels Effect consistent with 'the European way' blueprint for reclaiming the digital future:[228] the ability to export rules by tying them to its massive market. This is visible in its certification-forward posture: ETSI specifications for QKD interfaces (GS QKD 014) and Common Criteria protection profiles (GS QKD 016) supply the conformity apparatus that allows buyers to demand audited deployments.[229] The EuroQCI program layers a continental networking vision over that assurance regime.[230] Procurement and emerging EU cybersecurity regulation that anticipates a 'quantum age' threat model then turns that paper assurance into market muscle, as ministries and utilities specify certification as a condition of award, pulling suppliers—European and foreign—into audited conformance.[231] This is complemented by a growing ecosystem of startups, such as IQM in Finland and Pasqal in France, and a network of research consortia funded through programs like Horizon Europe. This approach is increasingly framed as a bid to close the sovereignty gap vis-à-vis U.S. platform dominance.[232]

However, the EU faces a significant "valley of death" in commercialization, lagging far behind the U.S. in venture capital investment. To narrow that gap, Brussels has begun pairing standard-setting with supply-side ballast—the Critical Raw Materials Act for key inputs and a 2025 recommendation on outbound investment to align capital with security policy—while insisting that FRAND-honest SEP policies in quantum communication and middleware keep markets contestable.[233] The EU's comparative advantage in certification travels farthest when paired with realism about legacy regimes: Wassenaar's export control consensus model is slow, key competitors are outside the tent, and intangible capabilities travel at the speed of the cloud—

---

[227] *See* Adrian Schmidt et al., *Current and Future Directions for Responsible Quantum Technologies: A ResQT Community Perspective* (2025), https://arxiv.org/abs/2509.19815
[228] Kai Zenner et al., The 'European way': a blueprint for reclaiming our digital future (2025), https://papers.ssrn.com/sol3/papers.cfm?abstract_id=5251254
[229] ETSI GS QKD 014 (API); ETSI GS QKD 016 (Common Criteria Protection Profile), *supra* note …
[230] European Commission, *EuroQCI—European Quantum Communication Infrastructure, supra* note …
[231] *See* e.g., Peter Alexander Earls Davis, Mateo Aboy, Lee A. Bygrave, Marcello Corrales Compagnucci & Timo Minssen, EU Cybersecurity Regulation in the Quantum Age, in QUANTUM TECHNOLOGY GOVERNANCE: LAW, POLICY AND ETHICS IN THE QUANTUM ERA (Mateo Aboy, Marcello Corrales Compagnucci & Timo Minssen eds., 2026), https://papers.ssrn.com/sol3/papers.cfm?abstract_id=5383838
[232] *See* Padraig Nolan, Europe's Quantum Leap Challenges U.S. Dominance, Ctr. for Eur. Pol'y Analysis (CEPA) (last visited Feb. 3, 2026), https://cepa.org/article/europes-quantum-leap-challenges-us-dominance/
[233] Regulation (EU) 2024/1252, 2024 O.J. (L) 1 (Critical Raw Materials Act), *supra* note …; and Comm'n Recommendation (EU) 2025/63, 2025 O.J. (L 63) 1 (Outbound Investments), *supra* note …

facts that elevate procurement-tied certification and reciprocity corridors over purely list-based controls.

## IV. China

Recent hard-power assessments describe China's quantum push as state-orchestrated acceleration coupled with standards projection. Unlike the distributed, venture-backed U.S. model, China's program is framed as a centrally coordinated, security-aligned mobilization in which the Party-state concentrates talent, funding, and infrastructure in a small number of prioritized pathways. Federici's U.S.-China Economic and Security Review Commission staff research report emphasizes the close integration among state research laboratories, defense-affiliated firms, and the People's Liberation Army acquisition system, creating direct pathways for scientific progress to inform military procurement and for defense requirements to steer R&D.[234] The Commission further assumes that China is pursuing cryptographically relevant quantum computing while obscuring where its most sophisticated programs are located and how far they have progressed—an asymmetry that sharpens the Article's "Policy of Denial" logic under conditions of strategic uncertainty and the familiar harvest-now/decrypt-later risk.[235]

China's strategy is a vertically integrated, state-directed loop under military-civil fusion. As Stone & Wood emphasize from Chinese strategist writings, military-civil fusion is treated not as a narrow "program" but as a Party-led governance approach intended to weave civilian science & technology and industrial capacity into defense modernization and long-term strategic competition.[236] It is characterized by its long-term vision, massive scale, and tight integration of state and commercial actors. Quantum technology has been a national priority in successive Five-Year Plans, with public investment estimates reaching $15 billion, dwarfing those of the U.S. and EU.[237] In 2017, China established the Hefei National Laboratory for Quantum Information Sciences with a reported budget exceeding $10 billion, allowing it to establish world-leading facilities.[238]

The strategy of military-civil fusion ensures that the People's Liberation Army is a primary beneficiary of these advancements. This state-centric model relies on "patient capital" from government-led funds, particularly at the local level as seen in the "Hefei model," which provides holistic support to incubate a local quantum industrial chain. This ecosystem is seeded by leading research institutions like the University of Science and Technology of China (USTC), which actively transfers IP to spin-off companies like QuantumCTEK in exchange for equity.[239] China directs subsidized capital to national labs and favored firms, blending indigenous invention with a systematic campaign to acquire foreign technology, which U.S. and allied officials have stated includes forced technology transfer, talent recruitment programs, and widespread industrial espionage.[240] Complementing these acquisition vectors, Joske's work on the CCP's united front system and external influence operations clarifies how political-influence mechanisms can be used to shape foreign research environments, diaspora engagements and narratives around technology cooperation—complicating open-science assumptions in high-sensitivity quantum domains.[241] China's approach to IP involves an aggressive state-driven patenting strategy, leveraging its vast pool of STEM talent and government subsidies for patent filings, aimed at building a massive domestic portfolio that foreign firms must navigate. This strategy serves multiple purposes: it creates a "patent thicket" to deter foreign competitors, it projects an image of indigenous innovation leadership, and it can be weaponized to create legal friction and raise costs for Western firms.

---

[234] Joseph Federici, *Vying for Quantum Supremacy: U.S.-China Competition in Quantum Technologies* at 3–4 (U.S.-China Econ. & Sec. Rev. Comm'n, Staff Research Report, Nov. 18, 2025). https://www.uscc.gov/research/vying-quantum-supremacy-us-china-competition-quantum-technologies#_Toc214309247

[235] Federici, *supra* note …

[236] Alex Stone & Peter Wood, *China's Military-Civil Fusion Strategy: A View from Chinese Strategists* (China Aerospace Studies Institute, June 15, 2020), https://www.airuniversity.af.edu/Portals/10/CASI/documents/Research/Other-Topics/2020-06-15%20CASI_China_Military_Civil_Fusion_Strategy.pdf

[237] *See*, e.g., Elsa B. Kania and John Costello, *Quantum Hegemony? China's Ambitions and the Challenge to U.S. Innovation Leadership*, Ctr. for a New Am. Sec. (Sept. 2018), https://www.cnas.org/publications/reports/quantum-hegemony

[238] *See* e.g., Chen, Na, China's Top Quantum Tech Center Founded in Hefei, (Jan. 2014), https://english.cas.cn/newsroom/archive/news_archive/nu2014/201502/t20150217_140623.shtml

The demonstration effect is central: the 2,000-km Beijing–Shanghai QKD backbone and Micius satellite experiments render abstract standards politics concrete by demonstrating scale, which Beijing leverages in ITU-T fora to push its architectural preferences.[242] That demonstration narrative is anchored in peer-reviewed results on satellite-based entanglement distribution at continental scale[243], and it is increasingly interpreted (including by analysts interviewed in China-focused fora) as a high-stakes "quantum gamble" in which technical uncertainty is tolerated to secure standards leverage and strategic narrative advantage.[244] In parallel with the Beijing–Shanghai backbone and *Micius* demonstration effects, Jian-Wei Pan's team has now reported repeater-relevant trapped-ion entanglement over 10 km fiber with coherence exceeding entanglement-generation time and device-independent quantum key distribution at

---

[239] *See* e.g., Joseph Federici, *Vying for Quantum Supremacy: U.S.-China Competition in Quantum Technologies* (U.S.-China Econ. & Sec. Rev. Comm'n, Staff Research Report, Nov. 18, 2025), https://www.uscc.gov/research/vying-quantum-supremacy-us-china-competition-quantum-technologies#_Toc214309247

[240] *See* e.g., U.S. Dep't of Justice, *U.S. Charges Five Chinese Military Hackers for Cyber Espionage* (May 19, 2014).

[241] *See* Alex Joske, *The Party Speaks for You: Foreign Interference and the Chinese Communist Party's United Front System* (Austl. Strategic Pol'y Inst. 2020), https://www.aspi.org.au/report/party-speaks-you/; and Alex Joske, *Seeing the CCP's External Influence Work Through Beijing's Eyes*, The Strategist (Austl. Strategic Pol'y Inst.) (Aug. 26, 2022), https://www.aspistrategist.org.au/seeing-the-ccps-external-influence-work-through-beijings-eyes/

[242] *See* ITU-T Y.3800-series Recommendations (QKD frameworks).

[243] Juan Yin et al., Satellite-Based Entanglement Distribution Over 1200 Kilometers, 356 Science 1140 (2017), https://www.science.org/doi/10.1126/science.aan3211.

[244] *See* e.g., Lily Ottinger, *China's Quantum Gamble*, China Talk (Nov. 19, 2024) (interview with Elias X. Huber), https://www.chinatalk.media/p/chinas-quantum-gamble

100 km scale, signaling progress not only in network deployment but also in the entanglement primitives needed for a scalable quantum internet.[245]

As major internet firms like Alibaba and Baidu retrench from capital-intensive hardware and donate equipment to state institutes, Beijing further centralizes capability under public control, tightening the civil–military fuse. A notable recent trend is the strategic retreat of major private tech companies like Alibaba and Baidu from capital-intensive in-house quantum hardware research. These firms have shut down their quantum labs and donated valuable equipment to state-run institutions, pivoting their strategies toward the less capital-intensive software and cloud platform layers. This move can be interpreted as a rational business decision, a response to escalating U.S. sanctions, or a reflection of state pressure to consolidate vital research under more direct government control.[246]

Where Western controls bite, Beijing wields its most potent leverage: control over the processing of upstream inputs, such as the 2023 export restrictions on gallium and germanium.[247] Furthermore, China is actively working to shape the global technological landscape in its favor, using its Digital Silk Road initiative to export its technology and technical standards, potentially creating a "Beijing Effect" where its state-centric, surveillance-friendly norms become entrenched. In its global AI governance initiatives, Beijing has mirrored Western language on security while embedding its core interests of protecting national sovereignty and ensuring equal access to technology, criticizing Western "technological monopolies" and positioning itself as a champion for the "Global South."[248] The impact of being placed on the U.S. Entity List has spurred China to double down on its quest for indigenous innovation and technological self-sufficiency, accelerating efforts to create domestic alternatives for every part of the quantum supply chain.[249] The reciprocity cost is real: visible supply-chain coercion strengthens allied coalitions, hardens outbound screens and procurement exclusions, and drives standards buyers toward auditable interoperability that Chinese vendors struggle to satisfy.

V. Canada

Amidst the clash of these three giants, Canada has emerged as a highly influential middle power in the quantum domain, exercising outsized influence—marked by niche excellence but persistent challenges in IP value-capture. Its ecosystem is built on decades of sustained public investment in foundational research at world-renowned institutions like the University of Waterloo's Institute for Quantum Computing (IQC), the Perimeter Institute for Theoretical Physics, and Université de Montréal's quantum institute, led by research luminaries such as the late Raymond Laflamme, and Michele Mosca.[250] This has translated into a remarkably vibrant startup ecosystem with world-leading companies like D-Wave Systems (pioneer in quantum annealing), Xanadu Quantum Technologies (a leader in photonic quantum computing), 1QBit (quantum software), and ISARA Corp (quantum-safe cryptography).[251]

---

[245] Wen-Zhao Liu et al. *supra* note …; and Bo-Wei Lu et al. *supra* note …
[246] *See* Edd Gent, *Alibaba and Baidu Cash Out on Quantum Computing Stakes*, IEEE Spectrum (Feb. 14, 2024).
[247] *See* Regulation (EU) 2024/1252, 2024 O.J. (L) 1 (Critical Raw Materials Act) (noting 2023 gallium/germanium restrictions).
[248] *See* DigiChina, *14th Five-Year Plan for National Informatization* (2021).
[249] *See* Reuters, *China's QuantumCTek sees boost from U.S. Entity List inclusion* (2023).

The Canadian government's National Quantum Strategy, launched in 2023 with a $360 million investment, focuses on computing, communication, and sensing, with a cross-cutting goal of talent and IP commercialization.[252] However, Canada faces a critical strategic challenge, one articulated forcefully by Jim Balsillie, co-founder of Research in Motion (BlackBerry). The challenge is moving beyond an "R&D-only" mindset to master IP value-capture.[253] This is not an isolated issue but a symptom of a deeper, well-documented structural problem.[254] Canada has a history of generating world-class research only to see the resulting high-value IP—foundational patents and trade secrets—commercialized abroad or sold for a fraction of its worth.[255] This paradox is rooted in institutional inertia and a failure to differentiate between IP asset classes. A common institutional reflex is to claim low-value "reputational IP" (like academic copyrights), whose value lies in dissemination, while failing to secure high-value "strategic IP" (like patents), whose value lies in exclusion and proprietary control. This misallocation of strategic focus represents a significant leakage of economic value and sovereign capability. A report by the Council of Canadian Academies found that a large percentage of patents from Canadian public research end up owned by foreign entities, a significant leakage of economic value.[256]

In response, Canada has launched initiatives like the Innovation Asset Collective (IAC), a patent collective to help SMEs pool IP, and the Canada Innovation Corporation (CIC) to help commercialize research and retain benefits in-country.[257] To counter the challenge of a smaller domestic venture capital market, government funds like BDC Capital co-invest in promising firms to keep them anchored in Canada. Strategically, Canada's approach is that of a "smart collaborator." As a member of the Five Eyes intelligence alliance, it aligns closely with allies on security matters. The government has become more cautious of foreign investment in sensitive tech, blocking a Chinese acquisition of a photonics company on national security grounds in 2019 and ordering divestment from critical mineral companies in 2022. Knowing it cannot outspend larger powers, Canada promotes international partnerships, such as the Canada-US Quantum Collaboration, and its experts often play key intermediary roles in international standards bodies. This allows Canada to carve out leadership in niche areas—such as quantum annealing and post-quantum cryptography—and leverage its reputation as a neutral convener to maximize its influence. The challenge of retaining top talent in the face of brain drain is being countered by an open immigration policy for skilled workers, creating a dynamic of "brain circulation" that continues to strengthen its research hubs. An LSI-consistent Canadian playbook fuses procurement-led PQC demand signals with transparent IP/data-rights plans on grants and reciprocity-based license exceptions that keep collaboration partnered rather than one-way.

---

[250] *See* e.g., Council of Canadian Academies (CCA) 2023, *Quantum Potential*. Ottawa, ON: Expert Panel on the Responsible Adoption of Quantum Technologies, Council of Canadian Academies. https://cca-reports.ca/wp-content/uploads/2024/03/Quantum-Potential_Full-Report_March-1-2024.pdf.
[251] *See* e.g., Lars S. Madsen et al., *Quantum Computational Advantage with a Programmable Photonic Processor*, 606 Nature 75 (2022) (Xanadu).
[252] Gov't of Canada, *Canada's National Quantum Strategy*, https://ised-isde.canada.ca/site/national-quantum-strategy/en
[253] Jim Balsillie, *Prosperity & Security: Canada's IP Imperative*, Centre for Int'l Governance Innovation (Mar. 11, 2021), https://www.cigionline.org/multimedia/prosperity-security-canadas-ip-imperative-featuring-jim-balsillie/.
[254] *See* e.g., Robert Asselin & Sean Speer, *A New North Star II Revisited*, Public Policy F. (Sept. 17, 2020), https://ppforum.ca/publications/new-north-star-2-revisited/.
[255] *See* e.g., James W. Hinton Mardi Witzel, Canadian IP Is Draining Abroad: That Needs to Stop, May 10, 2023, CIGI, https://www.cigionline.org/articles/canadian-ip-is-draining-abroad-that-needs-to-stop/
[256] CCA 2023, supra note …
[257] *See* https://www.ipcollective.ca/ and https://cdev.gc.ca/canada-innovation-corporation/.

VI. Middle Powers and Allied Architectures

For middle powers, niche mastery is the route to coalition scale, and allied architectures are the mechanism that converts specialization into systemic advantage. Middle powers face a harder optimization problem and thus a sharper opportunity: their strategies must accept that scale is a coalition property and seek indispensability at distinct control points. Beyond Canada, other technologically advanced nations are pursuing sophisticated strategies. The United Kingdom has leveraged its world-class academic base (at institutions like Oxford, Cambridge, and Imperial College London) to build a strong national quantum program, launched in 2014 with over £1 billion in investment.[258] The UK strategy emphasizes translating this research into commercial success through a vibrant startup scene (e.g., Orca Computing, Quantum Motion) and by attracting inward investment. Its comparative advantage lies in translation and assurance: middleware and benchmarking that render heterogeneous hardware interoperable, paired with a Five Eyes posture that marries cyber assurance to standards diplomacy. Japan has identified quantum technology as critical to its future economic competitiveness and national security. Its strategy is driven by strong collaboration between government, industry (e.g., Fujitsu, NEC, Toshiba), and academia (e.g., RIKEN, University of Tokyo). Japan's industrial base positions it as the coalition's "foundry" for high-precision components—cryo-CMOS, lasers, photonics—where manufacturing prowess becomes a strategic asset for friend-shoring supply chains. A credible Japanese play is to convert maintenance, spares, and field-service rights into explicit policy levers—contract terms that make allied networks resilient in practice, not just on paper. This comparative sweep also flags a broader ring of "focused scale-up" states—Korea, Australia, Singapore, India, Israel, and the UAE—whose playbooks emphasize PQC/QKD networks, sensing, or sovereign compute under investment-screening overlays; their leverage comes from turning procurement and standards participation into coalition scale rather than national autarky.

---

[258] *See* UK National Quantum Technologies Programme (NQTP), *Home*, UK Rsch. & Innovation, https://uknqt.ukri.org/.

VII. Comparative Insights

The various jurisdictions thus exhibit distinct strategic models. The United States focuses on harnessing private innovation with strong security oversight, aiming to outpace adversaries but also shield its technology—an "open but guarded" model. The European Union prioritizes cooperative frameworks, ethical norms, and building capacity collectively—a "multilateral and value-driven" model, though it grapples with internal fragmentation. China employs a high-control, high-investment model that leverages every instrument of state power—a form of "state-capitalist techno-nationalism" where technology is a primary tool for geopolitical ascendancy.

The middle powers and key allies operate with different constraints and objectives. A critical driver shaping these strategies, particularly for U.S. allies, is a growing uncertainty about the credibility of traditional security commitments in a new tripolar world. Just as threats from China and Russia and shifts in U.S. policy have led some allies to reconsider their nuclear posture, so too are they driving a quest for greater technological sovereignty. The push for indigenous quantum capabilities is, in part, a response to fears of abandonment and a desire for more reciprocal and responsive security arrangements in the 21st century. This underscores the need for democratic nations to forge a new extended deterrence compact that integrates both conventional and technological defense capabilities,[259] with quantum treated as a cross-cutting enabler (computing, sensing, navigation/timing, communication) with asymmetric cryptographic implications.[260] Canada functions as a "strategic collaborator," leveraging its deep talent pool and alliances to carve out excellence in specific niches while navigating the IP retention challenge. The United Kingdom and Japan act as "key allied partners," closely aligning their security postures with the United States while building on their unique national strengths in academic research (UK) and industrial capacity (Japan).

Across all of these models, the comparative constant is LSI discipline: measures are justified against capability, audited through benchmarks and certification, and written with visible off-ramps so coalition trust compounds over time rather than eroding in secrecy.

*A. Standards and Governance Arenas: Protocol Politics, FRAND, and Certification*

Standards are not neutral plumbing; they are industrial policy by other means. In communication, this manifests as "protocol politics": the West's PQC software path (NIST FIPS + CMVP) versus China's QKD hardware deployments (leveraged into ITU-T Y.3800

---

[259] *See,* e.g., M. M. Cuellar, E. J. Moniz and M. O'Sullivan, The *Proliferation Problem is Back* - Washington Must Adapt Its Playbook for a New Era of Nuclear Risk, Foreign Affairs (Sept. 2025), https://www.foreignaffairs.com/united-states/proliferation-problem-back

[260] *See,* e.g., Krelina, Michael, Military and Security Dimensions of Quantum Technologies: A Primer, July 2025, SIPRI, https://doi.org/10.55163/ZVTL1529; and Raffaele Cecere & Gaia Raffaella Greco, Quantum Technologies in Defense and Aerospace: A Scientometric Analysis of Emerging Trends and Opportunities (May 21, 2025), https://papers.ssrn.com/sol3/papers.cfm?abstract_id=5265429

frameworks).²⁶¹ Certification is the bridge from paper to procurement: ETSI's QKD profiles and NIST's CMVP validations are the auditable touchpoints that allow ministries and utilities to buy with confidence.²⁶² Because implementation choices create de facto moats, FRAND-honest licensing and defensible essentiality determinations are not niceties but preconditions for interoperable markets. Courts have already supplied a grammar—treat commitments as enforceable contracts and apportion value to the patented contribution. Furthermore, a complementary role is played by competition law: in nascent markets where a handful of platforms can entrench themselves through standards, antitrust doctrine remains the backstop that keeps FRAND commitments honest and prevents certification from calcifying into gatekeeping monopolies. Rather than the standard as a whole—that can travel into quantum without reinvention. The lesson for diplomats is prosaic: do not try to "win" standards by rhetoric—win them by certification programs that audit what counts and by contracts that make openness credible.

*B. Instruments of Economic Statecraft: Export, Investment, Procurement, and Sanctions*

Export controls (like BIS's 2024 quantum rule) set outer boundaries; procurement and standards set the inner topology. One caution merits explicit statement: the recent trend toward monetizing export controls—brokering revenue-sharing deals for licenses into countries of concern—prices security instead of measuring it and risks collapsing the legitimacy of the regime. A second is technical: as compute bottlenecks shift (e.g., toward high-bandwidth memory and inference/deployment stacks), controls and carve-outs must track architecture, or they will miss the locus of capability entirely. PQC migration tied to FIPS 203/204/205 is the cleanest case of procurement-as-policy: it moves markets without restricting research. Sanctions, entity listings, and outbound screens are blunt unless paired with platform governance (cloud access gates, telemetry) and service-contract controls (spares, firmware, field-service rights); together, they close the loopholes that paper rules leave open. Multilateral regimes like Wassenaar offer slow coordination, while TRIPS Article 73 preserves a security valve bounded by good-faith limits and reviewable necessity.²⁶³ In short, legacy multilateral lists remain necessary but insufficient; the real leverage now lies in governed cloud access, certification-anchored procurement, and contractually enforceable service rights.

*C. Supply Chains, Critical Materials, and Enablers*

Quantum's leverage points are upstream: helium value chains, isotopic enrichment, narrow-line lasers, and cryo-electronics. The EU's CRMA and U.S. CHIPS-adjacent programs are industrial policies for these enablers. The 2023 gallium/germanium restrictions were a live-fire exercise in supply-chain coercion.²⁶⁴ An LSI-informed playbook pairs surgical controls and stockpile agreements with license-exception corridors for trusted coalitions, plus explicit service-capacity targets (repair lead times, on-site spares ratios) written into procurement. Treat maintenance contracts as national capability: who can install, calibrate, and service at speed will decide whose networks survive stress.

---

²⁶¹ *See* Nat'l Inst. of Standards & Tech. (NIST), *Announcing Approval of Three Federal Information Processing Standards (FIPS) for Post-Quantum Cryptography* (Aug. 13, 2024) *supra* note …; and ITU-T Y.3800-series Recommendations, *supra* note ….
²⁶² *See* ETSI GS QKD 014 (API); ETSI GS QKD 016 (Common Criteria Protection Profile).
²⁶³ *See* TRIPS Agreement art. 73, Apr. 15, 1994, 1869 U.N.T.S. 299, *supra* note ….

*D. Metrics, Milestones, and Verification: A Comparative Dashboard*

Across blocs, meaningful metrics are converging. For computing, audited benchmarks (QED-C, QBI) are making capability legible to law and procurement.²⁶⁵ For communication, PQC migration is tracked in FIPS-anchored procurement and CMVP validations; QKD deployments are judged by ETSI/CC conformance. For sensing, maturity is measured in SWaP-C against mission profiles and inter-lab reproducibility. For the enabling stack, metrics are capacity and resilience—lead times for dilution refrigerators, lasers, and field-service turnaround. The comparative virtue of these indicators is that they are de-politicizing: they let governments tighten or relax measures as performance and assurance change, rather than as headlines demand.

## VIII. Policy Recommendations and a Coalition Playbook

First, codify LSI audits: every major measure should state the capability targeted, why the lever is the least-restrictive, how it preserves the scientific commons, and what evidence will trigger its sunset. Second, normalize reciprocity-based corridors among trusted partners. Third, treat cloud access and service contracts as control surfaces, with user provenance and performance-tied gates as standard terms. Fourth, anchor standards in certification and FRAND-honest licensing. Finally, invest in enablers through friend-shoring, stockpiles, and service capacity—because in quantum, "spares and service" is strategy. A coalition that internalizes this grammar will set not only the standard but the cadence and will do so in a way that can be explained and reviewed in public—proof against both hype and panic.

## VI. The Nexus of Dual-Use, IP, and Security Controls

The governance core of quantum technology sits precisely where dual-use science meets the law's fragmented toolset. At this junction, intellectual property rights, research-security practices, export and investment screening, and the slower but ultimately decisive machinery of standards and certification do not operate as separate spheres; they interlock into a single system that determines whether discovery becomes capability, capability becomes standard,

---

²⁶⁴ Int'l Energy Agency (IEA), *Announcement on the Implementation of Export Control of Items Related to Gallium and Germanium*, IEA Policies Database (last updated Apr. 25, 2025), https://www.iea.org/policies/17893-announcement-on-the-implementation-of-export-control-of-items-related-to-gallium-and-germanium.

²⁶⁵ *See* e.g., Quantum Econ. Dev. Consortium (QED-C), *Quantum Algorithm Exploration Using Application-Oriented Performance Benchmarks* (Feb. 16, 2024), *supra* note …

and standard becomes leverage.[266] Treating these instruments as substitutes invites two symmetrical mistakes—over-securitization that smothers diffusion and under-securitization that bleeds advantage—each generating costly path dependencies. The practical task is therefore to discipline the interface among them with a single, administrable decision rule that agencies, universities, and firms can share. Throughout this Part, I adopt a least-trade-restrictive, security-sufficient, innovation-preserving posture that I have elsewhere shortened to LSI. The animating idea is straightforward: define the concrete security objective, identify the narrowest legal and contractual tools that credibly achieve it, and structure disclosure, licensing, and standardization so that verification-ready openness is sequenced behind that narrow screen rather than sacrificed altogether. Properly applied, such an LSI discipline yields security-sufficient openness. It rejects both strategic naiveté and a "Silicon Curtain," instead insisting that hard problems—post-quantum cryptography migration, protection of calibration data and device models, assurance of claimed performance, and reduction of switching costs among allies—are solved not by blanket embargo but by verifiable gates that can be audited, reasoned about, and revised.[267]

The entry point is the recognition that quantum is dual-use by construction. The same magnetometer that maps neural currents or archaeological sites can support sensitive anti-submarine warfare; the same error-suppression technique that stabilizes a biomedical simulation can harden a code-breaking workload; the same entanglement-assisted link that secures financial messaging can improve precision targeting. Once a specific capability is identified as sensitive—because it meaningfully diminishes an adversary's cost or time to a strategically relevant threshold—states reach first for trade controls and secrecy. The decision to classify an invention or to place a technology under export licensing is not merely a perimeter measure; it pushes directly on the IP and publication choices available to researchers and firms. A team that would otherwise have drafted a patent application to stake a claim and signal investability may now reconsider. The existence of export controls signals that the technology is highly sensitive; filing a patent could be seen as advertising a critical capability to adversaries. Furthermore, in some jurisdictions like the U.S., patent applications deemed vital to national security can be subjected to a "secrecy order," which prevents the patent from issuing and prohibits the inventor from disclosing the technology.[268] Faced with these risks, a team may instead reserve know-how as a trade secret or delay publication, while a lab that would have released code with permissive licensing may reroute through a controlled-access repository or escrow arrangement. The patent-secrecy dilemma—disclose and risk diffusion, conceal and forgo early financing and standards influence—thus intensifies. The LSI discipline does not wish this dilemma away; it forces precision. It asks decisionmakers to specify the capability they fear in engineering terms, to articulate why narrower mitigations—contractual controls on "technical data," programmatic verification gates, FRAND-compatible licensing commitments at the interface, or license-exception lanes for trusted partners—are inadequate, and to document how the chosen restraint preserves the innovation channels that make the system resilient.[269]

---

[266] *See also* William A. Reinsch, Emily Benson, Thibault Denamiel & Margot Putnam, *Optimizing Export Controls for Critical and Emerging Technologies*, Ctr. for Strategic & Int'l Stud. (CSIS) (May 31, 2023), https://www.csis.org/analysis/optimizing-export-controls-critical-and-emerging-technologies.
[267] Nat'l Inst. of Standards & Tech. (NIST), *Module-Lattice-Based Key-Encapsulation Mechanism (ML-KEM)*, FIPS PUB 203 (2024); NIST, *Module-Lattice-Based Digital Signature Standard (ML-DSA)*, FIPS PUB 204 (2024); NIST, *Stateless Hash-Based Signatures (SLH-DSA)*, FIPS PUB 205 (2024).
[268] 35 U.S.C. § 101; 35 U.S.C. § 112; Invention Secrecy Act, 35 U.S.C. §§ 181–188.

The interaction between IP and security is nowhere more acute than in quantum's software-mediated layers—control stacks, compilers and mappers, error-suppression and error-mitigation routines, and application kernels. After Alice, purely abstract claims are suspect, but software claims that are tethered to concrete, non-conventional improvements in the operation of identified hardware remain patent-eligible when drafted and supported with care. The discipline here is twofold.[270] On the claim side, one must capture the contribution at the right level of generality, avoiding field-preempting formulations while demonstrating that the method improves the functioning of a quantum device or its classical control system. On the disclosure side, one must enable what one claims and no more: hyper-specific calibration constants, device models, and tacit lab techniques—precisely the artifacts most useful for illicit replication—need not be placed in the public domain if they are not necessary to practice the claims as drafted. That residual layer of know-how then migrates behind audited "reasonable measures" as trade secrets while the patent teaches the interface, the validation method, and the performance envelope on which markets, standards bodies, and procurement officials can rely. Sequencing matters: filings, classification decisions, export triage, and pre-publication review should be integrated so that patents, papers, and code releases trail the security screen by design rather than fall into it by accident.

Public funding and procurement frameworks further braid IP with security. Under Bayh–Dole, ownership of subject inventions often remains with the contractor, but the government's data and software rights under FAR and DFARS can be broader—and, for mission delivery, more decisive—than title to any one patent.[271] The category markings set the field of play: "unlimited rights" permit essentially unrestricted government use; "government-purpose rights" fence off competition with the private market but preserve allied interoperability; "limited" or "restricted" rights tighten access and can stall integration. Careless commingling of foreground and background data can default to broader rights than intended; conversely, over-restrictive markings can produce brittle dependencies and delay delivery. A sensible LSI practice insists on early data-rights scoping and negotiated plan annexes in awards, on time-bounded secrecy orders under the Invention Secrecy Act accompanied by scheduled review and compensation for use, and on procurement-grade verification gates—such as DARPA's Quantum Benchmarking Initiative[272], which conditions participation on granting Government Purpose

---

[269] *See* e.g., Export Admin. Regs., 15 C.F.R. pts. 730–774; Wassenaar Arrangement, *Public Documents* (list-update process and alignment across participating states); and 35 U.S.C. §§ 181–188 (Invention Secrecy Act).
[270] *Alice Corp. Pty. Ltd. v. CLS Bank Int'l*, 573 U.S. 208 (2014).
[271] See Bayh–Dole Act, 35 U.S.C. §§ 200–212; FAR 52.227-14 (Rights in Data—General); DFARS 252.227-7013 (Rights in Technical Data—Noncommercial Items); DFARS 252.227-7014 (Rights in Noncommercial Computer Software and Noncommercial Computer Software Documentation), *supra* notes …, … and …
[272] Def. Advanced Research Projects Agency (DARPA), *Quantum Benchmarking (QBI) Program*, *supra* note …

Page 61 of 111

Rights and thus trades a measure of exclusivity for the validation value of satisfying a defense-grade challenge problem.[273] These are not mere administrative niceties; they are the furnaces in which coalition-scale capability is forged.

Trade controls, although the most visible instrument of geoeconomic statecraft, are often the least discriminating. For quantum, BIS has begun to articulate item categories and performance-tied thresholds that better match the technology. That architectural shift should be accelerated. Controls written against measurable engineering attributes—signal fidelity under specified environmental constraints; timing stability; cryogenic performance at given power budgets; portability captured in SWaP-C—are less likely to overreach than family-level bans that sweep in research-grade apparatus and serviceable spares. This logic is vital for university labs, where "deemed export" rules[274] can trigger licensing requirements merely for allowing international postdocs access to controlled technical data, thereby chilling the collaborative exchange that fuels Western innovation.[275] The same applies to QCaaS environments: if what exposes risk is the transfer of "technology" in the form of device models and calibration files rather than mere invocation of a remote service, then access controls, audit-ready logs, and license terms that fence and monitor those artifacts will usually outperform categorical prohibitions. License-exception lanes for allied partners—reciprocity-based, conditioned on aligned controls, and periodically reviewed—preserve the service and spares ecosystem on which reliability depends, and they do so without granting adversaries a free ride. Crucially, this nexus is reactive: Western export controls can trigger asymmetric retaliation through upstream supply-chain chokepoints, notably China's dominance in processing critical minerals like gallium and germanium, creating a feedback loop in which security measures beget supply insecurities.[276] More broadly, the interplay is thus a feedback loop: dual-use capabilities lead to export controls, which influence IP strategies, while vulnerabilities in the underlying physical supply chain create leverage that can be used to counter those very controls. Framed in LSI terms, the burden is to show that narrower instruments fail before resorting to broader denial presumptions.

Investment screening complements trade controls by modulating the financial channels through which sensitive capability can be acquired, built, or steered. On the inbound side, CFIUS has long used jurisdictional expansion, mitigation agreements, and non-notified transaction pursuit to keep strategically relevant assets from drifting into coercible hands. On the outbound side, Treasury's program implementing E.O. 14105 adds a forward-looking layer for quantum information technologies, prohibiting or requiring notification of certain equity, debt, and greenfield investments tied to countries of concern.[277] This is a consequential turn, and it must be handled with the same LSI discipline. The question is not whether financial flows matter—they do—but how to target them so that they interrupt the translation of external capital into sensitive capability without freezing benign cross-border investment in non-sensitive layers of the stack. The more tightly the triggers are coupled to demonstrable capability thresholds and the more predictable the safe harbors for allied or neutral-market transactions, the more likely the regime will be both administrable and durable.

---

[273] *See* 28 U.S.C. § 1498 (government use of patented inventions; compensation), and DFARS 7013, *supra* note …, https://www.acquisition.gov/dfars/252.227-7013-rights-technical-data%E2%80%94other-commercial-products-and-commercial-services.
[274] 15 C.F.R. § 734.13; 15 C.F.R. § 734.2(b)(9) (deemed-export rule for release of controlled "technology" to foreign nationals in the United States).
[275] *See* e.g. Will Nelson (interview with Emily Weinstein), *U.S. Semiconductor Export Controls on the PRC: Prospects for Success*, Nat'l Bureau of Asian Rsch. (Jan. 30, 2023), https://www.nbr.org/publication/u-s-semiconductor-export-controls-on-the-prc-prospects-for-success/.
[276] *See* e.g. Siyi Liu et al., *China to Restrict Exports of Chipmaking Materials as US Mulls New Curbs*, Reuters (July 4, 2023), https://www.reuters.com/markets/commodities/china-restrict-exports-chipmaking-materials-us-mulls-new-curbs-2023-07-04/.

Standards, FRAND licensing, and certification bridge the world of rights and the world of rules. In competitive technology races, the greatest long-run leverage often resides not in the exclusion right but in the shape of the common interface: what conformance looks like, how performance is measured, and which security claims survive falsification. For quantum networking and cryptography, ETSI profiles, protection profiles for evaluation (including Common Criteria for QKD modules), and the Cryptographic Module Validation Program's use of FIPS 140-3 provide templates for how to translate technical culture into legal effect. Application-oriented benchmarks from QED-C and programmatic verification such as DARPA's QBI reinforce that translation by making claims contestable rather than theatrical. Once interfaces harden and patents read on them, FRAND obligations discipline market power: Microsoft v. Motorola and Ericsson v. D-Link teach that royalties must be apportioned to the value of the patented contribution, not the entire standard, while FTC v. Qualcomm warns against the use of standard-essential leverage to distort adjacent markets.[278] Standards cannot be turned into covert trade barriers; where trade restrictions are needed, they should be housed in trade law, justified publicly, and sized to the risk they address.[279] This approach keeps the standards room open to talent and scrutiny while reserving hard perimeter moves for the genuinely hard problems.

Cloud access and open-source posture add two further interfaces where IP and security converge. QCaaS has already dissolved borders in practice; the relevant questions are not nationality of the end-user but the movement of "technology" in export-control terms. Identity-vetted tenancy, auditable logs, and terms that fence calibration files and device models convert what would otherwise be a blunt nationality proxy into a capability-aware control. On the open-source side, the lesson of Jacobsen v. Katzer is that license conditions are enforceable; permissive release is not synonym for abandonment.[280] If a lab or firm wishes to publish code without enabling the transfer of controlled "technology," it can do so by sequencing releases behind classification and export review, by carving out device models and validation artifacts from the open package, and by using contributor license agreements that harmonize with its patent and trade-secret strategy. In other words, openness should be engineered, not presumed.

---

[277] U.S. Dep't of Treasury, *Outbound Investment Security Program—Final Rule* (Oct. 28, 2024) (implementing Exec. Order No. 14,105; effective Jan. 2, 2025); 50 U.S.C. § 4565 (CFIUS), *supra* note …

[278] *Microsoft Corp. v. Motorola, Inc.*, 795 F.3d 1024 (9th Cir. 2015); *Ericsson, Inc. v. D-Link Sys., Inc.*, 773 F.3d 1201 (Fed. Cir. 2014); *FTC v. Qualcomm Inc.*, 969 F.3d 974 (9th Cir. 2020).

[279] *See also* Nigel Cory, *The Biden Administration Overreacts Responding to China's Role in Setting Standards for Quantum Technologies*, Info. Tech. & Innovation Found. (July 29, 2024), https://itif.org/publications/2024/07/29/the-biden-administration-overreacts-responding-to-chinas-role-in-setting-standards-for-quantum-technologies/.

[280] *Jacobsen v. Katzer*, 535 F.3d 1373 (Fed. Cir. 2008); 15 C.F.R. § 772.1 (definitions of "technology" and "technical data").

The international legal order provides an exception space within which democracies can align restraint with reciprocity. TRIPS Article 73, like the security exceptions that preceded it in GATT and other trade instruments, authorizes measures taken to protect essential security interests.[281] The temptation—on all sides—is to read that clause as license for unreviewable discretion. The healthier reading, and the one most likely to preserve the scientific commons on which even national security depends, is LSI-conforming: measures should be tailored to a demonstrable capability-linked risk, time-bounded, subject to reason-giving and review, and scaled back as verification-ready openness becomes feasible. That approach better guards against the corrosion of alliance trust, lowers the fixed costs of switching to an alternative technology stack, and reduces the incentive to retaliate with facially symmetric—but substantively indiscriminate—controls that fracture the standards arena.

The resulting picture is not a patchwork but an operating model. IP drafting is coordinated with classification and export triage so that the patent discloses the interface and the validation method while reserving tacit know-how; government data-rights planning is done *ex ante*, with annexes that specify what will be delivered and when; export and investment controls are written against measurable thresholds and paired with reciprocity-based license-exception lanes for trusted partners; standards and certification translate claims into enforceable procurement obligations; cloud access is fenced so that controlled "technology" does not move merely because a foreign national invokes a remote service; and international security exceptions are used in a manner that can be defended, audited, and revised. The fulcrum is verification. Auditable benchmarks and conformance testing give agencies and allies confidence that openness is not naiveté; procurement gates make marketing claims legally meaningful; and periodic reassessment prevents emergency measures from ossifying into indefinite architecture. In this way, the system resists both panic and drift.[282] It does not ask science to choose between secrecy and the commons, or markets to choose between exclusion and diffusion. It asks institutions to do the harder work of staging openness behind security-sufficient gates and then to prove, repeatedly, that the gates remain fit to purpose.

## VII. Special Issues: DARPA QBI, Government Rights, and Security Carve-Outs

The hinge between national security and intellectual property is sharpest where the state ceases to be a distant rulemaker and becomes a counterparty with real leverage over disclosure, data rights, and timing.[283] Benchmarking programs that adjudicate performance claims, procurement and R&D instruments that buy down technical risk, and exceptional legal authorities that authorize government use or delay publication[284] alter the legal and commercial terrain on which quantum firms decide what to patent, what to publish, and what to keep offstage.[285] Properly framed, these interventions discipline marketing claims, compress learning cycles, and build verifiable trust without collapsing private investment; poorly framed, they extract value without producing durable capability and create predictable incentives to retreat behind secrecy. The task is not to deny the state's need to act, but to translate blunt authority into precise instruments that achieve security-sufficient openness—calibrated access that meets concrete mission needs while preserving the innovation channels on which resilience depends.

---

[281] Agreement on Trade-Related Aspects of Intellectual Property Rights (TRIPS) art. 73, Apr. 15, 1994, 1869 U.N.T.S. 299; *cf.* GATT art. XXI; UK Parliament, Parliamentary Off. of Sci. & Tech., POSTnote 742, *Quantum Computing, Sensing and Communications* (2025).

[282] *See* e.g., Akel Hashim et al., *Practical Introduction to Benchmarking and Characterization of Quantum Computers*, 6 PRX Quantum 030202 (2025), https://link.aps.org/doi/10.1103/PRXQuantum.6.030202.

QBI crystallizes the problem and the promise.[286] The program's animating premise is that quantum advantage cannot be adjudicated by press release; claims must be tethered to application-oriented benchmarks that travel across hardware and across labs. DARPA's solicitation and program materials accordingly force participants to run defense-relevant challenge problems under comparable conditions and to provide the artifacts necessary for independent assessment. The enforcement is epistemic rather than punitive: those who satisfy a defense-grade test earn validation that markets and procurement officials can rely on, while those who cannot are pushed to improve or to recalibrate their claims. As an institutional design choice, QBI functions as a pre-competitive public good that reduces information asymmetry and—when it is tied to auditable logs, reproducible methods, and clear disclosure boundaries—raises the quality of the entire discourse. But QBI also changes the IP gradient. Participation typically entails granting the Government Purpose Rights (GPR) in certain technical data and noncommercial computer software produced with mixed funding, not "inventions," which are governed by Bayh–Dole patent-rights clauses with a separate paid-up government license[287], a concession that expands the class of actors who may lawfully use or build upon those deliverables for government ends (without public release beyond government and authorized contractors).[288] For incumbents with diversified revenue and platform plays, this trade may be straightforward: credibility and access in exchange for bounded rights. For startups whose edge is concentrated in a narrow toolchain or dataset, the risk is starker: overbroad rights markings or careless commingling can allow government customers—and their integrators—to re-use crown-jewel assets without further license for years. Both strategies—participation with bounded rights versus deferral to protect exclusivity—can be rational depending on capitalization, time-to-revenue, and exposure to defense procurement cycles. The LSI resolution is to insist on precision up front: enumerate deliverables, segregate background from foreground IP, limit GPR to the minimum scope necessary for test and evaluation and cabin any sharing to need-to-know contexts with audit trails. That design preserves QBI's verification dividend while avoiding a de facto compulsory license broader than mission needs require. In practice, risk is mitigated by clean separation of background and foreground repositories, precise data-rights markings, and enumerated deliverables that avoid inadvertent rights creep.

---

[283] FAR 52.227-14 (Rights in Data—General) (allocating rights in data and software delivered under federal contracts and grants), *supra* note …

[284] 28 U.S.C. § 1498 (authorizing government use of patented inventions; reasonable and entire compensation; no injunctive relief), *supra* note …

[285] 35 U.S.C. §§ 181–188 (Invention Secrecy Act) (authorizing secrecy orders and associated procedures), *supra* note …

[286] Def. Advanced Research Projects Agency (DARPA), *Quantum Benchmarking (QBI) Program* (program description and materials, updated 2025), *supra* note …

[287] Bayh–Dole Act, 35 U.S.C. §§ 200–212 (contractor election of title; government license; march-in), *supra* note …

[288] DFARS 252.227-7013 (Rights in Technical Data—Other Than Commercial Products and Commercial Services); DFARS 252.227-7014 (Rights in Noncommercial Computer Software & Noncommercial Computer Software Documentation) (defining Government Purpose Rights and related categories), *supra* note …

Government rights in data and software are often outcome-determinative in this space, and they are frequently misunderstood. Bayh–Dole allocates title in subject inventions but does not itself decide who may use source code, calibration corpora, or documentation; those questions turn on FAR and DFARS clauses that pivot on funding source, deliverable status, and markings.[289] Bayh–Dole also includes "march-in rights," a powerful, albeit unused, lever for the government to grant licenses if an invention is not being made available to the public, an authority recently clarified by HHS/NIH guidance. "Unlimited rights" enable essentially unrestricted government re-use, including disclosure outside the government; "Government Purpose Rights" permit government and authorized contractors to use and disclose for government ends but fence off commercial exploitation; "limited" or "restricted" rights constrain both access and field of use. In practice, the distinction is won or lost in program planning: a statement of work that requires delivery of models or code can, absent careful legends and segregated repositories, default those artifacts into broader rights than the contractor intended; conversely, over-asserted restrictions can stall interoperability and invite challenge. An LSI playbook requires that agencies specify the least set of rights necessary to accomplish verification, test, and coalition interoperability, that contractors segregate background from foreground artifacts and avoid commingling that triggers broader rights, and that both sides align publication and release calendars to filing and export-review gates so that openness is sequenced rather than suppressed.[290] Where patents are concerned, the sovereign's parallel authority under 28 U.S.C. § 1498 to authorize government use with compensation remains a backstop.[291] Used casually, it can chill participation, but used transparently, paired with fair remuneration and narrow authorization, it can resolve blockade risk without diluting private exclusion beyond what public missions require. The remedy for such government use lies in "reasonable and entire compensation" in the Court of Federal Claims, as injunctive relief is unavailable against the United States.

Security carve-outs complete the architecture and must be kept exceptional. The Invention Secrecy Act authorizes secrecy orders where disclosure "might be detrimental to national security," permitting time-bounded suppression of issuance and public disclosure.[292] In a quantum context, where tacit technique and parameterizations carry disproportionate value, such orders can protect details that would collapse an adversary's learning curve; but overuse drives research into indefinite opacity, starves standards processes of high-value contributions, and degrades the very verification QBI and similar efforts aim to strengthen. The LSI discipline is to treat secrecy orders as rare and surgical, coupled to periodic review and, where government use ensues, compensation principles apply; to rely in the first instance on modular secrecy and controlled-access data rooms for the most sensitive artifacts; and to draft patent disclosures to enable at the level claimed while reserving hyper-specific calibration assets as trade secrets protected by rigorous reasonable-measures programs. That approach pairs security sufficiency with an innovation-preserving disclosure floor.[293]

---

[289] *Ibid.*
[290] Export Admin. Regs., 15 C.F.R. pts. 730–774 (definitions of "technology," "software," deemed exports; license exceptions).
[291] 28 U.S.C. § 1498, *supra* note …
[292] 35 U.S.C. §§ 181–188, *supra* note …

The international system offers an exception space, but it is not an invitation to carve out everything. TRIPS Article 73 recognizes essential-security measures, with recent WTO panel practice emphasizing that the clause is bounded by good-faith constraints, necessity, and proportionality and must be tailored to demonstrable capability-linked risks.[294] The better reading, consonant with LSI, is to anchor any quantum-related security measures in demonstrable capability thresholds, to time-limit them, and to house explicit trade restrictions in trade law rather than smuggling them into standards. In practical terms, this means that export controls and outbound-investment screens aimed at quantum information technologies should hinge on measurable deployability and performance,[295] preserve reciprocity-based license-exception lanes for trusted partners that have implemented aligned controls, and be revisable as verification improves.[296] Standards rooms remain open to talent and scrutiny; hard perimeter moves are saved for problems that cannot be mitigated by contract, conformance, or controlled access.

These three layers—programmatic verification, calibrated government rights, and disciplined exceptions—mutually reinforce. QBI-like programs convert technical culture into administrative law by tying claims to tests and tests to procurement. FAR/DFARS planning, wielded with care, ensures the state receives the minimum re-use authority necessary to make those tests meaningful without extracting gratuitous private value. And the security exception space, read through TRIPS and WTO practice, remains a safety valve for genuine capability risks rather than an all-purpose rationale for closure.[297] The result is an operating model that raises confidence without raising walls: auditable benchmarks and test-and-evaluation regimes keep advantage theatre in check; data-rights maps and markings translate mission needs into bounded licenses; secrecy orders remain rare, precise, and revisable; and international security exceptions are justified in engineering terms, not asserted as slogans. In such a system, firms can forecast the consequences of cooperation, universities can structure publication and access around clear gates, and agencies can defend their choices as the least trade-restrictive measures that are nonetheless security-sufficient.

---

[293] Agreement on Trade-Related Aspects of Intellectual Property Rights (TRIPS) art. 73, Apr. 15, 1994, 1869 U.N.T.S. 299; Panel Report, *Russia—Measures Concerning Traffic in Transit*, WTO Doc. WT/DS512/R (adopted Apr. 26, 2019).
[294] *Ibid.*
[295] U.S. Dep't of the Treasury, *Outbound Investment Security Program—Final Rule* (Oct. 28, 2024) (implementing E.O. 14105; effective Jan. 2, 2025); 31 C.F.R. § 800.215 (defining "critical technologies" by reference to the Commerce Control List for CFIUS purposes), *supra* note …
[296] NIST, *Cryptographic Module Validation Program (CMVP)* (FIPS 140-3 validations); NIST, *Post-Quantum Cryptography: Selection and Standardization of Public-Key Algorithms*; FIPS PUBS 203–205 (2024), *supra* note …
[297] TRIPS article 73, *supra* note …

Finally, the cloud and the lab are where these abstractions meet practice. QCaaS access models that fence "technology" in the export-control sense—device models, calibration files, characterization corpora—while allowing invocation of remote services transform a blunt nationality proxy into a capability-aware regime. When those fences are combined with QBI-style auditing and with contract clauses that define the government's re-use rights narrowly in time and scope, the system disciplines flows without starving the commons. In these operational settings, capability-aware controls (e.g., role-based access, audit logs, and Technology Control Plans) should fence "technology" in the export-control sense while permitting remote service invocation.[298]

However, hardware controls alone face diminishing returns. Federici's analysis underscores that Beijing is concentrating quantum talent and research infrastructure in state-led hubs designed to provide a shared workforce and platform base, thereby treating expertise, facilities, and system-integration capacity as strategic assets rather than market byproducts.[299] In a domain where tacit know-how and operational integration often determine time-to-capability, "intangibles" become a primary vector of diffusion alongside physical components. Consequently, an LSI-compliant regime cannot stop at the border: it must pair perimeter measures with laboratory- and cloud-level governance, including deemed-export compliance, Technology Control Plans, and access/audit controls that make "technology" (in the export-control sense) legible, bounded, and reviewable without collapsing the research commons into blanket closure.

Conversely, where controls reach beyond what verification and mission sufficiency require, they prompt predictable counter-moves—mineral export restrictions, supply-chain coercion, standards bifurcation—that make everyone less secure.[300] The special issues addressed in this Part thus resolve to a single administrative question: can institutions write down, in advance, the smallest disclosure, rights, and exception package that achieves the targeted security objective, and can they then prove, periodically, that those packages remain fit to purpose? Where the answer is yes, security-sufficient openness is possible; where the answer is no, policy drifts toward slogans and away from science.

## VIII. Grand Strategy and the Weaponization of Interdependence

The geopolitical competition in quantum technology is no longer a simple race for isolated breakthroughs in qubit coherence or sensor precision; it is a struggle over the architectures and infrastructures through which those breakthroughs travel, are governed, and can be withheld.

---

[298] 15 C.F.R. § 734.13 (defining "export" and "release" of technology or software subject to the EAR); 15 C.F.R. § 734.2(b)(9) (deemed exports) (implications for QCaaS and lab-level technical-data controls).
[299] Federici, *supra* note …
[300] Regulation (EU) 2024/1252, 2024 O.J. (L) 1 (*Critical Raw Materials Act*), *supra* note …

In a global economy organized around dense, hub-and-spoke networks of finance, data, and technical standards, states that occupy central nodes can translate connectivity into coercive leverage, turning access and exclusion into instruments of power rather than neutral features of globalization.[301] If mishandled, this contest risks hardening into a bifurcated quantum order—a "Silicon Curtain" of incompatible stacks—that would erode global security and stifle the collaborative science on which the field depends.

In the quantum domain, this "weaponization of interdependence" is structural. The technical "full stack"—spanning critical raw materials, fabrication hardware, control electronics, firmware, cloud platforms, and application-layer services—functions as a web of potential chokepoints. Grand strategy here is not merely about a "Sputnik moment" of domestic capacity; it is about orchestrating these layers so that they generate resilience for one's own ecosystem and friction for rivals. As Henry Farrell and Abraham Newman have demonstrated, the ability to observe and selectively disrupt flows through these hubs creates a new grammar of economic statecraft.[302] A nation's quantum strategy is thus deeply conditioned by its position within this global stack. China's dominance in the processing of critical minerals and its early investment in terrestrial and satellite QKD infrastructure naturally pull it toward a vertically integrated, state-directed model that seeks to secure the foundational layers and key networks. The United States, with unparalleled advantages in the upper layers of the stack—software, governed cloud platforms, and a deep venture-capital and start-up ecosystem—logically emphasizes control over application, platform, and IP-intensive segments of the value chain. The European Union, combining a world-class research base with structural weaknesses in industrial scale-up, has rationally embraced an industrial-policy and regulatory strategy that uses public investment, standards positioning, and coordinated procurement to build sovereignty across the stack rather than at any single layer.

This structural reality intersects with a rapidly shifting legal and policy landscape. In the United States, National Security Memorandum 10 (NSM-10) has crystallized this shift by tying quantum leadership explicitly to the protection of vulnerable cryptographic systems.[303] Subsequent Office of Management and Budget guidance operationalizes that mandate by treating the migration to Post-Quantum Cryptography (PQC) not just as an IT upgrade, but as a coordinated instrument of national resilience against "harvest-now, decrypt-later" strategies.[304] In Europe, the "Brussels Effect"—where internal market rules induce global compliance—is merging with security policy. When EU-defined PQC baselines and Quantum Key Distribution (QKD) security requirements are written into procurement regimes, such as the European Quantum Communication Infrastructure (EuroQCI) initiative, they set de facto global expectations for cryptographic hygiene and supply-chain transparency.[305]

---

[301] Farrell & Newman, *supra* note …
[302] Daniel W. Drezner, Henry Farrell & Abraham L. Newman eds., The Uses and Abuses of Weaponized Interdependence (Brookings Inst. Press 2021), https://www.brookings.edu/books/the-uses-and-abuses-of-weaponized-interdependence/
[303] National Security Memorandum on Promoting United States Leadership in Quantum Computing While Mitigating Risks to Vulnerable Cryptographic Systems (NSM-10), 2022 Daily Comp. Pres. Doc. (May 4, 2022), *supra* note …
[304] Office of Mgmt. & Budget, Memorandum M-23-02, Migrating to Post-Quantum Cryptography (Nov. 18, 2022), *supra* note …

A durable democratic grand strategy, however, requires more than coordinated export controls or investment screening. It must rest on democratic sovereignty: legitimacy derived from public consent and transparent, reviewable criteria for when and how interdependence is constrained. Recent U.S. survey data indicates broad support for quantum development via both public and private investment, coupled with comparatively greater skepticism about "state control," and a large neutral share—an evidentiary basis for treating public education and contestability as core design requirements, not messaging afterthoughts.[306] An alliance of democracies cannot be sustained purely by shared chokepoints; it requires a shared account of why particular flows are restricted. Democratic sovereignty in this sense is not merely a quest for technological independence, but a commitment to ensuring that the deployment and restraint of quantum capabilities remain aligned with public values and contestable through law, rather than delegated entirely to security bureaucracies or corporate gatekeepers. This demands an adherence to the LSI principle established earlier in this Article: measures must be Least trade-restrictive, Security-sufficient, and Innovation-preserving.

States are already leveraging their control over critical nodes in the quantum stack to exert influence. This manifests in three primary domains.

## I. Standards as a Geopolitical Battleground

International standards bodies such as ISO, ITU-T, and ETSI are no longer merely technocratic coordination venues; they are arenas in which rival models of governance and security are negotiated and entrenched. The historical record shows persistent divergence in U.S., EU, and Chinese standards strategies—rooted in institutional design, industrial policy, and conceptions of sovereignty.[307] In quantum technology, standards increasingly function like "eigenstates" of the international order: once a protocol, architecture, or test suite becomes the stable reference configuration, it resists perturbation and shapes both market structure and security assumptions for years. The state whose architectures, protocols, and test suites are written into global standards gains durable economic and strategic advantages: its firms enjoy first-mover economies of scale, its compliance tooling becomes the default, and its values are embedded in code. China's "China Standards 2035" program makes this ambition explicit by seeking to extend state-centric governance preferences into the rulebooks of technical committees, while the United States and its allies have begun to coordinate more systematically to promote standards that embed openness, interoperability, and verifiable security as core design principles rather than afterthoughts.[308]

---

[305] European Commission, European Quantum Communication Infrastructure (EuroQCI) Initiative (2024), *supra* note …
[306] *See* Ruane et al, *supra* note …
[307] *See* Nicholas Zúñiga, Saheli Datta Burton, Filippo Blancato, Madeline Carr, The geopolitics of technology standards: historical context for US, EU and Chinese approaches, *International Affairs*, Volume 100, Issue 4, July 2024, Pages 1635–1652, https://doi.org/10.1093/ia/iiae124

Quantum-specific standardization will increasingly determine not only which protocols survive but also which governance models travel. If PQC migration is tied to NIST FIPS 203, 204, and 205, then those documents will anchor the cryptographic layer of the emerging quantum stack.[309] If quantum networking implementations are validated against Common Criteria protection profiles, the operational meaning of "quantum-safe" will likewise be set by a transatlantic coalition.[310]

Crucially, this domain interacts with intellectual property regimes. Courts have historically treated FRAND (Fair, Reasonable, and Non-Discriminatory) commitments in standards-essential patent (SEP) licensing as enforceable contracts, insisting on damages methodologies that apportion value to the patented contribution rather than the entire standard.[311] Extending these principles into the quantum stack—e.g., for PQC algorithms or compiler intermediate representations—can ensure standards remain open and interoperable, preventing the emergence of patent thickets that stifle innovation. Put differently, the most important IP problem in standard setting is hold-up.[312] Patent pools and cross-licenses can mitigate thickets, but require careful governance to avoid collusion and exclusion.[313] In a weaponized-interdependence environment, disciplined standards governance is a central instrument of security policy: it determines whether chokepoints are built around closed proprietary interfaces or open, audited platforms that trusted vendors can populate.

## II. Supply Chain Chokepoints and the Politics of Materiality

Quantum leverage is often upstream and infrastructural: helium and neon value chains, isotopic enrichment, high-purity substrates, narrow-line lasers, and cryogenics—alongside the component and materials stack that governs deployability, including dilution refrigerators, cryogenic amplifiers, photonic components, vacuum systems, superconducting materials, nanofabrication tools, and rare earth and specialty metals. National industrial policies, such as the U.S. CHIPS and Science Act[314] and the EU Critical Raw Materials Act[315], explicitly target these enablers, and initiatives such as Project Vault signal a complementary turn to strategic

---

[308] *See* e.g., Matt Sheehan, Marjory S Blumenthal, Michael R. Nelson, Three Takeaways From China's New Standards Strategy, Oct 28, 2021, https://carnegieendowment.org/research/2021/10/three-takeaways-from-chinas-new-standards-strategy?lang=en
[309] Nat'l Inst. of Standards & Tech., Announcing Approval of Three FIPS for Post-Quantum Cryptography (FIPS 203/204/205) (Aug. 13, 2024), *supra* note …
[310] ETSI GS QKD 016, Common Criteria Protection Profile for a QKD Module, *supra* note …
[311] *See generally* Microsoft Corp. v. Motorola, Inc., 795 F.3d 1024 (9th Cir. 2015) (contractual enforceability of FRAND), *supra* note …
[312] Farrell et al, *supra* note …
[313] Pitofsky, *supra* note …
[314] CHIPS and Science Act of 2022, Pub. L. No. 117-167, 136 Stat. 1366, *supra* note …
[315] Regulation (EU) 2024/1252, Critical Raw Materials Act, 2024 O.J. (L) 1, *supra* note …

reserves as a buffer against supply shocks and coercive leverage.[316] As a decision aid—without implying governmental adoption—this Article's proposed Quantum Criticality Index (QCI) can help identify which inputs are decisive enough to justify stockpiling rather than broader, more trade-distortive controls.[317] However, the "full stack" reality complicates simple containment strategies. China's dominance in the processing of critical minerals provides it with a credible threat to restrict exports in response to Western controls; conversely, U.S. and allied dominance in high-end semiconductor design allows them to constrain access to the control systems that underwrite quantum capabilities.[318] The result is a dynamic of mutual coercion in which each bloc simultaneously seeks to secure and diversify its own supply chains while probing for vulnerabilities in those of its rivals, even as the underlying networks of trade, logistics, and knowledge remain deeply intertwined.

An LSI-consistent approach to these chokepoints eschews blanket embargoes in favor of targeted, metrics-anchored measures. In practice, this means that governments write explicit capacity targets into procurement contracts to ensure sufficient domestic or allied production of key enablers, negotiate friend-shoring arrangements that treat maintenance and field-service capacity as part of national capability rather than an afterthought, and craft reciprocity-based license exceptions that keep trade open within trusted coalitions while preserving the option to tighten controls if conditions deteriorate. Specifying performance thresholds in operational terms—such as SWaP-C (Size, Weight, Power, and Cost) and reliability metrics for cryogenic systems—allows controls to focus on genuinely strategic capabilities while preserving room for benign commercial scale-up.[319]

### III. Quantum Computing as a Service (QCaaS) and Cloud Governance

In the near term, most users will access quantum computing power via the cloud. QCaaS platforms—typically operated by major U.S. and allied firms—are becoming pivotal hubs in the story of weaponized interdependence. For adversaries, unfettered access to cutting-edge QCaaS compresses development timelines; for the providers' home states, these platforms are natural control surfaces. Because access to the most advanced machines will be mediated through a small number of U.S.-based hyperscalers—firms such as IBM, Google, Amazon, and Microsoft—control over QCaaS begins to resemble control over a strategic maritime chokepoint: in extreme scenarios, home governments could pressure providers to restrict or terminate access for users in adversary jurisdictions, amounting to a "quantum cloud blockade" that conditions access to transformative computational capability on geopolitical alignment. This transforms QCaaS from a purely commercial service into a locus of strategic dependency

---

[316] *See* The White house, Project Vault, *supra* note …. *See* also Project Vault (New York Times report): Alan Rappeport & Tony Romm, Trump Unveils $12 Billion Critical Minerals Stockpile, N.Y. Times (Feb. 2, 2026), https://www.nytimes.com/2026/02/02/business/trump-critical-minerals-stockpile.html

[317] *See* Cho, Kop & Lee, *supra* note …

[318] Bureau of Indus. & Sec., Commerce Control List Additions and Revisions; Implementation of Controls on Advanced Technologies Consistent with Controls Implemented by International Partners, 89 Fed. Reg. 72,926 (Sept. 6, 2024), *supra* note …

[319] U.S. Dep't of Def., Defense Acquisition Guidebook (defining SWaP-C considerations), *supra* note …

for states that lack domestic high-end infrastructure but rely on quantum resources for finance, health, or security applications.

Yet cloud delivery also enables policy bypass. Adversaries can, in effect, "tunnel" through national hardware controls by renting time on remote quantum processing units (QPUs), exporting only classical bits—measurement outcomes, gradients, and model updates—while retaining the learning benefits of access to scarce hardware. Rand and Rand identify this "cloud loophole" as a primary failure mode of traditional export controls: restrictions on physical hardware are rendered moot if knowledge transfer and functional access occur virtually.[320] This dynamic makes the decisive variable not the nationality of the machine, but the provider's identity management, workload classification, logging, and auditability. Accordingly, "secure enclave" approaches to quantum access—coalition-only clouds, purpose-limited APIs, and verified research partnerships, aligned with resilience programs such as Genesis Mission/ASSP—can preserve legitimate scientific use while denying adversaries frictionless acceleration. This logic underpins the Executive Order launching the Genesis Mission, which establishes the American Science and Security Platform (ASSP) as a closed-loop system.[321] By hosting vetted allies within a monitored compute perimeter rather than exporting hardware, the ASSP effectively collapses the "tunneling" vector, allowing innovation to proceed in a decoherence-free subspace.

From a governance perspective, QCaaS offers a more granular alternative to blunt technology denial. However, relying on a few hyperscalers for governance creates a secondary risk: incumbents may "coopt disruption" by using security compliance as a moat to stifle startup dynamism.[322] The Nexus therefore requires antitrust vigilance to ensure the "Quantum Stack" remains open at the application layer. Governance must shift from controlling *items* to controlling *identity*, ensuring that while access is vetted, the platform itself does not become a gatekeeper that precludes the emergence of new entrants. Instead of banning hardware exports, states can require providers to implement user provenance checks and purpose-based access controls, maintain crypto-agility requirements aligned with PQC baselines, and geofence sensitive workloads into intra-coalition frameworks where legal remedies are credible. This shifts strategic leverage to governed cloud access and service contracts—a logic consistent with NATO's emerging quantum strategy, which emphasizes "quantum-readiness" and resilient infrastructure over mere hardware accumulation.[323] While controls on cloud access raise questions under TRIPS Article 73 regarding cross-border services, the security exception, read through an LSI lens, supports calibrated restrictions tied to demonstrable capability thresholds rather than open-ended bans.[324]

---

[320] *See* Rand & Rand *supra* note …
[321] The White House, Launching the Genesis Mission (Nov. 24, 2025), *supra* note …
[322] *See* Lemley, Mark A. and Wansley, Matthew and McGowan, David, Coopting Disruption (February 1, 2024). 105 Boston University Law Review 457 (2025), Cardozo Legal Studies Research Paper No. 2024-24, Stanford Law and Economics Olin Working Paper No. 589, https://www.bu.edu/bulawreview/files/2025/05/LEMLEY-WANSLEY.pdf

[323] NATO, NATO Releases First Ever Quantum Strategy (Jan. 17, 2024), *supra* note …
[324] TRIPS article 73, *supra* note …

IV. Orchestrating Interdependence

Taken together, these domains—standards, supply chains, and cloud governance—show how quantum technologies are embedded in the broader logic of regulatory gravity. The same structural features that enable a "Silicon Curtain" of rival ecosystems can, if disciplined through law and institutional design, be turned toward a more stable form of shared resilience, in which chokepoints serve as levers for deterrence by denial rather than instruments of indiscriminate punishment. The challenge for a democratic coalition is to design controls that are strict enough to avoid materially accelerating adversarial capabilities, yet transparent enough to avoid fracturing the global innovation ecosystem into a "quantum splinternet." This balance is the essence of *values-based deterrence*. Unlike authoritarian coercion, which relies on opaque leverage, democratic deterrence relies on the resilience of open, rule-bound systems that are harder to corrupt and faster to adapt.

Grand strategy in this era is not simply about subsidizing domestic capacity. It is about choreographing standards, supply chains, and cloud governance so that interdependence is disciplined rather than abandoned—leveraged for security-sufficient openness instead of indiscriminate coercion. This governance architecture is the invisible backbone of grand strategic concepts like the "Golden Dome" (integrated missile defense) and "Hellscape" (autonomous systems deterrence, such as the Pentagon's Replicator initiative). Without quantum-assured sensing and communication, "Hellscape" is hackable and "Golden Dome" is blind. This informational advantage is existential, not marginal. Federici's key findings frame quantum supremacy as a mission-essential national asset capable of delivering disproportionate and likely enduring advantages in intelligence collection and precision targeting.[325] On the sensing front in particular, the report notes that quantum sensors could dramatically improve remote sensing—including the ability to detect stealth aircraft and submarines—thereby shifting the credibility of deterrence and the survivability of high-end force postures. Deterrence by denial in this setting therefore plausibly requires a "Policy of Denial" for enabling inputs and chokepoints—not only advanced qubit technology, but also the upstream dependencies (e.g., cryogenics, specialized lasers, and sensor-grade components) that determine whether sensing and cryptanalytic capabilities can be scaled, deployed, and operationalized faster than allied PQC and resilience measures come fully online.

In plain terms: to keep deterrence credible, the U.S. and allies may need to slow China's ability to field quantum sensing/cryptanalysis by restricting not just qubits but the scarce supply-chain "must-haves" (cryogenics, lasers, sensor-grade components, and related know-how), so PQC migration and other defenses can be completed before any breakthrough becomes militarily usable. The Nexus teaches that in a "World Seven" diffusion race (Sullivan-Feldman Eight Worlds uncertainty framework), shielding technology is insufficient.[326] The dominant strategy must shift from control to acceleration: ensuring that the democratic "industrial commons" innovate faster than the authoritarian alternative can appropriate.

---

[325] Federici, *supra* note …
[326] *See* Sullivan & Feldman, *supra* note …

## V. Quantum's Impact on Strategic Stability

Strategic stability is a term of art with contested meanings, but its core concern is concrete: to reduce incentives for rapid escalation or first use of the most destructive capabilities, especially in crisis.[327] Across leading treatments, the concept is commonly decomposed into crisis stability (whether either side has incentives to strike first during a confrontation), arms-race stability (whether either side has incentives to expand or diversify forces in predictably destabilizing ways), and escalation stability (whether actions in one domain create strong pressures to climb the escalation ladder in another).[328]

In the twenty-first century, that diagnosis is harder to sustain. Strategic competition is increasingly multipolar; nuclear risk is entangled with conventional operations and space-enabled systems; and rapid technological change compresses decision time while multiplying potential failure modes.[329] The environment resembles what some describe as a "third nuclear age," in which additional nuclear actors, regional theaters, and alliance commitments create more pathways to miscalculation and inadvertent escalation.[330] A parallel concern, emphasized in both classic and contemporary deterrence scholarship, is that instability can emerge from organizational failure, misperception, and accidents—not only from deliberate aggression.[331] The practical lesson is that stability is not a natural equilibrium; it is an engineered condition whose inputs can be degraded by new technologies, new doctrines, and new information environments.[332]

---

[327] James M. Acton, Reclaiming Strategic Stability, in STRATEGIC STABILITY: CONTENDING INTERPRETATIONS 117 (Elbridge A. Colby & Michael S. Gerson eds., 2014), https://apps.dtic.mil/sti/tr/pdf/ADA572928.pdf

[328] Aaron R. Miles, The Meaning and Implications of Strategic Stability (Lawrence Livermore Nat'l Lab., Ctr. for Global Sec. Rsch., June 2025), https://cgsr.llnl.gov/sites/cgsr/files/2025-06/Strategic-stability-meaning-implications-CGSR.pdf ; Keith B. Payne et al., The Meaning of Strategic Stability, 1 J. POL'Y & STRATEGY (2021), https://nipp.org/wp-content/uploads/2021/10/Stability-Proceedings-1.1.pdf; Ulrich Kühn, Strategic Stability in the 21st Century: An Introduction, 6 J. PEACE & NUCLEAR DISARMAMENT 1 (2023), https://www.tandfonline.com/doi/full/10.1080/25751654.2023.2223804

[329] Mariano-Florentino Cuéllar, Ernest J. Moniz & Meghan L. O'Sullivan, PREVENTING AN ERA OF NUCLEAR ANARCHY: NUCLEAR PROLIFERATION AND AMERICAN SECURITY (Carnegie Endowment for Int'l Peace et al. 2025), *supra* note …

[330] Matthew Kroenig, Strategic stability in the third nuclear age, (Atlantic Council Oct 7, 2024), https://www.atlanticcouncil.org/in-depth-research-reports/issue-brief/strategic-stability-in-the-third-nuclear-age/

[331] Scott D. Sagan, Why Do States Build Nuclear Weapons?: Three Models in Search of a Bomb, International Security, Vol. 21, No. 3 (Winter, 1996-1997), pp. 54-86, The MIT Press, https://www.jstor.org/stable/2539273. *See* also Scott D. Sagan, The Perils of Proliferation: Organization Theory, Deterrence Theory, and the Spread of Nuclear Weapons, 18 INT'L SEC. 66 (1994), *supra* note …

[332] Matthew Bunn, "Preventing Nuclear War in a Multipolar Era of Great-Power Hostility and Evolving Technology," presentation, Norwegian Intelligence School, October 26, 2023, https://matthewbunn.scholars.harvard.edu/nuclear-forces-arms-control-and-disarmament; see also Matthew Bunn, "Deterring without Provoking - and De-Escalating Crises and Conflicts," presentation, Sandia National Laboratory, December 7, 2023, https://matthewbunn.scholars.harvard.edu/file_url/343

Quantum technologies sharpen these pressures because they are, in large part, technologies of information advantage: sensing, timing, communication, optimization, and cryptography.[333] They may stabilize some interactions (by improving secure communication and resilient command and control) while destabilizing others (by enabling detection, targeting, and cyber compromise that compress decision time or threaten survivable forces).[334] Put differently, quantum changes the strategic "information conditions" under which states infer intent, validate warning, and decide whether to de-escalate, absorb, or preempt.[335]

*A. Quantum Pathways that can Destabilize*

First, quantum sensing and quantum-enabled intelligence can erode concealment—the substrate of survivability for both conventional and strategic assets.[336] Quantum magnetometry, gravimetry, and related field-sensing modalities are frequently discussed for anti-submarine warfare and underground mapping.[337] Even incremental gains become strategically salient when fused with autonomous platforms, distributed arrays, and machine learning: the sanctuary value of oceans, hardened facilities, and camouflage can degrade, thereby tightening the perceived "window" for action in crisis.[338] Strategic stability risks emerge when leaders come to believe (rightly or wrongly) that survivable capabilities are becoming trackable, or that hidden preparations are becoming reliably detectable.

Second, quantum timing and navigation can make denial systems more lethal in contested electromagnetic environments. Quantum clocks and quantum inertial sensors can sustain precision in Global Navigation Satellite System (GNSS)-denied or GNSS-degraded settings, enabling swarms, loitering munitions, and undersea vehicles to operate with less dependence on vulnerable external signals.[339] When paired with quantum radio-frequency sensing, quantum transduction, and quantum-enabled electronic warfare concepts, these capabilities can improve

---

[333] Michal Krelina, MILITARY AND SECURITY DIMENSIONS OF QUANTUM TECHNOLOGIES: A PRIMER (SIPRI, July 2025), https://www.sipri.org/publications/2025/other-publications/military-and-security-dimensions-quantum-technologies-primer

[334] Michal Krelina, Quantum Technology for Military Applications, 8 EPJ QUANTUM TECH. 24 (2021), https://link.springer.com/article/10.1140/epjqt/s40507-021-00113-y. See also Weinstein & Rodenburg, *supra* note …

[335] Jake Sullivan & Tal Feldman, Eight Worlds: The Future of AI and American Power, FOREIGN AFF., *supra* note …

[336] Michal Krelina & Denis Dubravcik, Quantum Technology for Defence (1/3): What to Expect for the Air and Space Domains, J. JOINT AIR POWER COMPETENCE CTR., Ed. 35 (2023), https://www.japcc.org/articles/quantum-technology-for-defence/

[337] Krelina, Quantum Technology for Defence (3/3): Quantum for ISR and PNT, J. JOINT AIR POWER COMPETENCE CTR., Ed. 38 (2024), https://www.japcc.org/wp-content/uploads/JAPCC_J38_screen_Art-05.pdf

[338] Michal Krelina, Quantum Technology for Defence (2/3): Quantum-Enhanced Radars and Electronic Warfare, J. JOINT AIR POWER COMPETENCE CTR., Ed. 37 (2024), https://www.japcc.org/wp-content/uploads/JAPCC_J37_screen.pdf

[339] Krelina, MILITARY AND SECURITY DIMENSIONS OF QUANTUM TECHNOLOGIES: A PRIMER, *supra* note …

radar discrimination, increase resistance to jamming, and enhance low-probability-of-intercept and low-probability-of-detection communication. This can strengthen deterrence by denial. But it can also compress decision time and increase the pace of "operate-inside-the-adversary-cycle" strategies, raising escalation risks in a fast-moving crisis.

Third, quantum computing raises a cyber-strategic stability concern that motivates much of the current policy urgency: the prospect of cryptanalytic breakthroughs against widely deployed public-key systems, combined with "harvest now, decrypt later" collection.[340] The destabilizer here is not necessarily an overnight cryptographic collapse; it is erosion of trust in communication integrity, authentication, and critical infrastructure security during crises—particularly if adversaries suspect that sensitive plans or signaling channels have been compromised. Importantly, credible estimates often still place large-scale quantum cryptanalysis beyond the near term, which makes migration governance—not panic—central: a long transition window can itself be exploited if it leaves asymmetric exposure across allies and critical sectors.[341]

Fourth, quantum communication can help—but are not a panacea. Quantum key distribution (QKD) and related techniques may strengthen certain high-value links, yet their security depends on end-to-end system integrity, trusted nodes, supply chains, and realistic threat models. Overconfidence in "unhackable" narratives can itself be destabilizing if it induces risky operational assumptions or diverts attention from classical weaknesses.[342] A related cross-cutting insight is institutional: scientometric work on quantum communication points to rapid growth and diffusion in the field, which tends to expand the competitive surface area (standards, supply chains, talent, trusted components) rather than "solve" security through a single technique.[343]

*B. Quantum Pathways that can Stabilize*

Quantum also offers stabilizing affordances when deployed as resilience rather than as a counterforce enabler. Quantum-enhanced clocks, sensors, and communications can harden command and control (C2) against jamming, spoofing, and deception, preserving the ability to communicate restraint, clarify intent, and terminate hostilities. Improved situational awareness can reduce worst-case inference and inadvertent escalation—provided the underlying systems are trusted and intelligence is shared through credible alliance mechanisms.[344] In this sense,

---

[340] Kop, Mauritz, Presentation on Quantum Supremacy: Technology, Strategy, and International Order at the Oxford Emerging Threats & Technology Working Group, Oxford University, 10 December 2025, https://talks.ox.ac.uk/talks/id/8c717aea-150f-4ed0-bd94-c137d37f3ded/ and https://airecht.nl/blog/2025/mauritz-kop-speaks-at-oxford-university-on-quantum-threats

[341] Weinstein & Rodenburg, *supra* note …

[342] Michal Krelina & Manoj Harjani, "Unhackable" Quantum Communication Is a Myth (RSIS, IDSS Paper No. IP23089, Dec. 15, 2023), https://rsis.edu.sg/wp-content/uploads/2023/12/IP23089.pdf

[343] Raffaele Cecere & Gaia Raffaella Greco, The Technological Emergence of Quantum Communication: A Bibliometric Analysis (2024), https://papers.ssrn.com/sol3/papers.cfm?abstract_id=5265429

[344] Forrest, Tracey, Paul Samson, S. Yash Kalash & Michael P.A. Murphy, (2025), supra note …

quantum's stabilizing contribution is less about "dominance" than about maintaining continuity and interpretability in the fog of a technologically saturated crisis.

The arms-control analogy is instructive. Strategic Arms Limitation Talks (SALT) did not eliminate competition; they structured it through constraints, verification, and a shared interest in avoiding inadvertent catastrophe.[345] New Strategic Arms Reduction Treaty (New START) practice likewise underscores how verification design, transparency measures, and negotiated predictability can preserve crisis communications and reduce worst-case inference even amid adversarial relations.[346] Historical analysis of superpower arms control also cautions that "co-operative competition" has limits: regimes can fail when domestic politics, verification disputes, or asymmetric objectives overwhelm incentives to bound competition.[347] Quantum competition is similarly likely to be managed—not ended—through bounded norms and confidence-building measures, including guardrails that protect crisis communications, authentication, and critical infrastructure from opportunistic exploitation.

*C. Values-Driven Quantum Deterrence by Denial*

Against this background, the Article's broader claim is that responsible quantum technology can be operationalized as values-based deterrence by denial: an integrated legal-technical control plane that reduces adversary payoffs from coercion, aggression, and appropriation while sustaining democratic resilience and the industrial commons. The objective is not to threaten catastrophic punishment; it is to make wrongdoing less likely to succeed and less attractive to attempt—while minimizing escalation pressures and protecting civilian life.[348]

In practical terms, values-driven denial emphasizes (1) resilient, fail-gracefully defenses that preserve civilian life, economic continuity, and alliance cohesion under attack; (2) measurement-driven governance—standards, benchmarks, and auditability—that distinguishes protective deployment from destabilizing counterforce ambition; (3) institutional safeguards (procurement, export controls, investment screening, and supply-chain security) that protect the ecosystem of critical quantum enablers without collapsing into indiscriminate techno-nationalism; and (4) escalation-aware doctrine that deters without provoking—pairing credible denial with reassurance, off-ramps, and crisis communications to reduce misperception and inadvertent escalation.

This is also a realist response to strategic competition. U.S.–China assessments emphasize that quantum advantage will be pursued as a national capability with military relevance, including through civil-military fusion dynamics and upstream supply-chain positioning.[349] A credible denial strategy therefore requires not only operational concepts, but governance of chokepoints in the materials, components, and manufacturing stack that enable quantum sensing, timing, and computing. "Quantum sovereignty" rhetoric can obscure this reality; the empirically relevant question is which dependencies are decisive and how they can be secured without self-defeating restrictions that undermine innovation capacity.[350]

---

[345] James Cameron, Soviet–American Strategic Arms Limitation and the Limits of Co-operative Competition, 33 DIPLOMACY & STATECRAFT 116 (2022), https://www.tandfonline.com/doi/pdf/10.1080/09592296.2022.2041812

[346] Rose Gottemoeller, NEGOTIATING THE NEW START TREATY (Cambria Press 2021), https://cisac.fsi.stanford.edu/publication/negotiating-new-start-treaty. New START expired on February 5, 2026. *See also* CISAC, The Future of Nuclear Proliferation, Panel Discussion with Mariano-Florentino Cuéllar, Rose Gottemoeller and Scott Sagan, moderated by Toby Dalton, FSI, Feb 5, 2026, https://cisac.fsi.stanford.edu/events/future-nuclear-proliferation

[347] Acton, *supra* note …

[348] Bunn, Deterring Without Provoking, *supra* note …

### D. Quantum Deterrence Compared to Classical Nuclear Deterrence

Classical nuclear deterrence is often modeled as deterrence by punishment: the threat of unacceptable damage. Its stabilizing claim rests on the expectation that mutual vulnerability and survivable second-strike forces remove incentives for nuclear first use.[351] Yet the nuclear record also shows persistent instability risks: accidents, misperception, organizational failure, and "stability–instability" dynamics in which high-end deterrence can coexist with frequent lower-level coercion.[352] Contemporary work on nuclear order further emphasizes that as the system becomes more multipolar and less rule-bounded, the "stability margin" can narrow even when major war remains unlikely.[353]

Quantum deterrence is not a substitute for nuclear deterrence; it is an overlay that can either reinforce or undermine it. Where quantum dual-use capabilities threaten survivable platforms (for example, by improving detection of Virginia-class submarines)[354] or compress decision time (for example, by enabling more precise counterforce-adjacent conventional strikes), they can erode crisis stability. Where quantum warfighting capabilities strengthen C2 resilience, authentication, and early-warning integrity, they can reinforce stability by reducing incentives for worst-case preemption and preserving credible crisis signaling. The destabilizing temptation is "weaponizing quantum" in ways that seek unilateral transparency and unilateral precision against an adversary's survivable forces—especially in defense-space architectures where satellites, undersea infrastructure, and electromagnetic dominance interact.[355] The stabilizing opportunity is the opposite: to deploy quantum for resilience, continuity, and defensive denial, while bounding (through doctrine, governance, and confidence-building measures) the applications that most directly threaten crisis stability.

---

[349] Joseph Federici, Vying for Quantum Supremacy: U.S.-China Competition in Quantum Technologies (U.S.-China Econ. & Sec. Rev. Comm'n, Staff Research Report, Nov. 18, 2025), *supra* note …
[350] Michal Krelina, Opinion: Quantum Sovereignty—Reality Check (Oct. 30, 2025) (LinkedIn), *supra* note …
[351] Acton, *supra* note …; Payne et al., *supra* note …
[352] Sagan (1994), *supra* note …
[353] Cuéllar, Moniz & O'Sullivan, *supra* note …; Kroenig, *supra* note …
[354] *See* in this context e.g., Ben Doherty, Possibility of US ever selling Australia nuclear submarines is increasingly remote, Aukus critics say, The Guardian, Feb 5, 2026, https://www.theguardian.com/world/2026/feb/05/aukus-nuclear-submarine-deal-us-australia
[355] Alyssa LaFleur, Navigating the New High Ground: Why Global Defence and Security Depend on the Evolving Space Tech Industry (Space Insider May 9, 2025); NorthStar Earth & Space, An Approach to Modernise Space, Defence, and Security (LinkedIn Sept 7, 2024), https://www.linkedin.com/pulse/approach-modernise-space-defence-innovation-tufdc

*E. Taiwan Use Case: A Quantum-Enabled Deterrence-by-Denial Architecture*

Consider a Taiwan Strait crisis in which the People's Republic of China (PRC) evaluates whether a fait accompli seizure is feasible before allied reinforcements arrive. Deterrence by denial in this setting is fundamentally a contest over time, transparency, and attrition: the attacker seeks to move fast under cover of deception, electronic warfare, and mass; the defender seeks to slow, expose, and fragment the assault, buying time for mobilization and coalition response.[356] A quantum-enabled denial architecture aims to widen the attacker's error bars while narrowing the defender's—without relying on escalatory threats of punishment.

A single integrated narrative illustrates the logic. As the PRC concentrates forces, quantum-enhanced sensing—spanning space-enabled detection, distributed undersea arrays, and quantum radio-frequency and gravity-field modalities—can raise the probability that invasion-relevant signatures are detected early enough to trigger political, economic, and operational measures that blunt the prospects of a Pearl Harbor–style surprise attack in a quantum-enabled battlespace.[357] DARPA's emphasis on taking quantum sensors out of the lab and into defense platforms reinforces that the near-term strategic effect is likely to come from ruggedized, platform-integrated sensing that survives vibration, temperature variation, and contested electromagnetic environments—i.e., operationalized detection and timing resilience rather than headline laboratory sensitivity.[358] As electronic warfare intensifies, quantum clocks and quantum inertial sensors help distributed forces maintain coherent operations in GNSS-degraded conditions, sustaining swarm coordination and precision navigation despite jamming and spoofing.[359] This is not speculative as a strategic problem: Ukraine's operational environment demonstrates how pervasive spectrum contestation can become, and how navigation and timing resilience translates into battlefield viability. In parallel, quantum radio-frequency sensing and quantum-enabled electronic warfare concepts can improve the defender's ability to locate emitters, resist jamming, and sustain low-probability-of-intercept communications—reducing the attacker's confidence that a blackout or deception campaign will hold.[360]

At the operational edge, denial turns on scalable, attritable autonomy. The Department of Defense's Replicator initiative—explicitly oriented toward fielding large numbers of autonomous systems across domains—maps onto denial concepts that impose cost and uncertainty through dispersion, redundancy, and rapid replacement.[361] Public reporting suggests that autonomy firms, including Anduril, are positioning software and uncrewed systems for such architectures.[362] The "unmanned hellscape" concept, associated with contemporary U.S. Indo-Pacific thinking, similarly frames deterrence as saturating the attacker's decision calculus with distributed systems that make a rapid seizure unreliable.[363] Quantum's contribution, in this framing, is not a magic weapon; it is the enabling substrate for resilient timing, sensing, and communication that keeps denial architectures functional under heavy electronic and cyber attack.[364]

---

[356] Larry Diamond & James O. Ellis Jr., Deterring a Chinese Attack on Taiwan (Hoover Inst., Working Paper, 2022), https://thebulletin.org/premium/2023-03/deterring-a-chinese-military-attack-on-taiwan/
[357] *Ibid.*
[358] Def. Advanced Rsch. Projects Agency, *Taking Quantum Sensors Out of the Lab and Into Defense Platforms* (Feb. 7, 2025), https://www.darpa.mil/news/2025/quantum-sensors-defense-platforms
[359] Michal Krelina, AN INTRODUCTION TO MILITARY QUANTUM TECHNOLOGY FOR POLICYMAKERS (SIPRI, Background Paper, Mar. 2025), https://www.sipri.org/publications/2025/sipri-background-papers/introduction-military-quantum-technology-policymakers
[360] Krelina, Quantum Technology for Defence (2/3), *supra* note …

The vignette also clarifies why values matter for stability. The same quantum tools that aid denial can tempt destabilizing counterforce ambitions (for example, persistent tracking of strategic assets) and can invite asymmetric responses grounded in maskirovka (systematic military deception), cyber sabotage, or escalation in secondary theaters.[365] Russia's operational Maskirovka doctrine and recent experience in electronic warfare underscore the centrality of deception and spectrum contestation, while Iranian navigation interference provides a reminder that manipulation of signals and attribution ambiguity are enduring features of modern crises.[366] Values-driven denial therefore requires governance discipline: deploy quantum to preserve life, continuity, and alliance credibility; maintain escalation-aware guardrails; and resist the drift from defensive denial toward destabilizing transparency-and-precision strategies that undermine strategic stability.

## IX. Governance Blueprints from Past Technological Revolutions

The challenge of governing quantum technology is not without precedent. Across the last century, states, firms, and scientific communities have repeatedly confronted technologies whose benefits and risks were tightly intertwined, whose value depended on cross-border collaboration, and whose misuse could generate systemic or even existential harm. Those episodes do not supply an off-the-shelf template for quantum, but they do furnish a set of governance blueprints—nuclear non-proliferation, nanotechnology and patent thickets, the open-standards architecture of the classical internet, and the risk-based and precautionary regimes that emerged around artificial intelligence and biotechnology—that help to discipline analogy and calibrate ambition. Read through the lens of security-sufficient openness and the LSI test, they suggest how far treaty-based constraint can reach, where standards and IP must shoulder more of the load, and when adaptive, community-driven guardrails outperform rigid *ex ante* rules. They also underscore that "technology governance" is now itself a domain of grand strategy, as coalitions compete not only to build frontier capabilities but also to shape the legal and institutional architectures within which those capabilities will be deployed.[367]

---

[361] U.S. Department of Defense / Defense Innovation Unit, Replicator initiative, https://www.diu.mil/replicator
[362] Anduril Industries, DIU Selects Anduril to Enable Collaborative Autonomy for Replicator Systems, Anduril News (Oct. 2024), https://www.anduril.com/news/diu-selects-anduril-to-enable-collaborative-autonomy-for-replicator-systems.
[363] Carter Johnston, *Breaking Down the U.S. Navy's "Hellscape" in Detail*, Naval News (June 16, 2024), https://www.navalnews.com/naval-news/2024/06/breaking-down-the-u-s-navys-hellscape-in-detail/.
[364] *See* Kop, Mauritz, A Bletchley Park for the Quantum Age (2025), *supra* note …
[365] Matthew J. Fecteau, Q-Day Approaches: Preparing the NATO Alliance for the Quantum Era, (Mad Scientist Laboratory Dec 3, 2025), https://madsciblog.tradoc.army.mil/554-q-day-approaches-preparing-the-nato-alliance-for-the-quantum-era/
[366] *Ibid.*

I. Nuclear Technology

The most obvious historical parallel for quantum's dual-use dilemma lies in the governance of nuclear technology, and the relevant governance blueprint is the non-proliferation model: controlled sharing with safeguards and verification. The initial impulse, embodied in the secrecy of the Manhattan Project, was to ring-fence fissile expertise and materials within a small coalition of states. That policy soon proved unsustainable as the basic science diffused and additional states approached the nuclear threshold. The turning point, both conceptually and institutionally, came with President Eisenhower's 1953 "Atoms for Peace" address to the United Nations, which reframed nuclear governance around a shift from absolute secrecy and zero-sum rivalry to managed international cooperation for peaceful uses.[368] That vision underwrote the creation of the International Atomic Energy Agency (IAEA) and, later, the 1968 Treaty on the Non-Proliferation of Nuclear Weapons (NPT), whose "grand bargain" traded non-nuclear-weapon states' renunciation of nuclear arms for access to civilian technology under international safeguards, inspections, and materials accounting.[369] The non-proliferation regime thus married a discriminatory prohibition on certain end-uses (weapons programs) with an affirmative commitment to share certain capabilities, mediated by a specialized verification institution that possessed both technical sophistication and political legitimacy, including a mandate not only to police diversion but also to provide technical assistance and confidence-building services that made adherence more attractive than defection for many states.

Yet the nuclear template is itself under strain. A regime architected for a bipolar Cold War is now tasked with managing an emerging tripolar nuclear order, in which China is rapidly expanding and modernizing its arsenal and arms-control cooperation with Russia has deteriorated, even as Russia engages in open nuclear signaling in the context of its war in Ukraine.[370] The NPT review process has repeatedly failed to deliver consensus outcome documents, and analysts warn that the normative "grand bargain" at the treaty's core—non-proliferation in exchange for disarmament progress and peaceful-use access—is fraying.[371]

---

[367] *See* e.g., *Strategy and Grand Strategy in an Age of Artificial Intelligence* (Inst. of Glob. Pol., Columbia Univ. event highlight, July 28, 2025), https://igp.sipa.columbia.edu/news/strategy-and-grand-strategy-age-artificial-intelligence
[368] Dwight D. Eisenhower, Address Before the General Assembly of the United Nations on Peaceful Uses of Atomic Energy ("Atoms for Peace") (Dec. 8, 1953), *in* 1 Pub. Papers of the Presidents of the United States: Dwight D. Eisenhower 813 (1958), https://archive.org/stream/publicpapersofth015184mbp/publicpapersofth015184mbp_djvu.txt
[369] Treaty on the Non-Proliferation of Nuclear Weapons, July 1, 1968, 21 U.S.T. 483, 729 U.N.T.S. 161.
[370] *See, e.g.*, Oliver Meier et al., , *Protecting the Nuclear Non-Proliferation Treaty in Turbulent Times*, Eur. Leadership Network (31 July 2024), https://europeanleadershipnetwork.org/report/protecting-the-nuclear-non-proliferation-treaty-in-turbulent-times-commentary-collection/

For quantum governance, the lesson is double-edged. On the one hand, the nuclear experience shows that treaty-based regimes with intrusive verification can be built around technologies whose misuse threatens catastrophic harm, and that such regimes can durably shape state behavior even amid profound strategic competition. On the other hand, it highlights the fragility of architectures that hinge on great-power consensus and on control of scarce physical commodities. Qubits, error-correcting codes, and quantum-enhanced algorithms do not lend themselves to the same kind of ledger-based, materials-accounting safeguards that the IAEA applies to enriched uranium. A quantum "non-proliferation" regime, if it emerges at all, is more likely to focus on the most explosive applications—cryptanalytic capabilities at scale, quantum-enhanced command-and-control, or highly sensitive sensing of nuclear forces—while leaning heavily on domestic export controls, standards, and procurement to manage the broader technology stack. Any future quantum accord will have to be designed from the outset for a more fragmented and contested geopolitical landscape in which stable great-power consensus cannot be taken for granted and minilateral arrangements may carry much of the practical load. For technologies whose failure modes implicate the stability of the international system, the nuclear episode thus suggests that some combination of top-down, treaty-based commitments and specialized verification bodies will remain indispensable, even if their remit is narrower and more modular than the NPT and IAEA.

Crucially, Scott Sagan's analysis of nuclear proliferation warns that governance must account not only for state intentions but for *organizational* behaviors—bureaucratic politics, standard operating procedures, and the risk of accidents.[372] This internal fragility is now compounded by a macro-strategic collapse. As the 2025 Carnegie-NTI Task Force led by Cuéllar, Moniz, and O'Sullivan warns, the post-Cold War order is dissolving into an "era of nuclear anarchy" defined by cascading proliferation and the erosion of legal guardrails.[373] In this volatile environment, emerging technologies like AI, quantum sensing and cryptanalysis act as accelerants that could destabilize deterrence; therefore, governance must evolve from static treaty compliance to dynamic, multi-domain risk reduction. A robust quantum governance regime must therefore police internal organizational routines within labs and firms, not just the high politics of interstate treaties.

II. Nanotechnology

---

[371] *Ibid.*
[372] Scott D. Sagan, *The Perils of Proliferation: Organization Theory, Deterrence Theory, and the Spread of Nuclear Weapons*, 18 Int'l Sec. 66 (1994), https://doi.org/10.2307/2539178
[373] Mariano-Florentino Cuéllar, Ernest J. Moniz & Meghan L. O'Sullivan, *Preventing an Era of Nuclear Anarchy: Nuclear Proliferation and American Security* (Carnegie Endowment for Int'l Peace et al., Report of the Task Force on Nuclear Proliferation and U.S. National Security, Sept. 2025), https://carnegieendowment.org/research/2025/09/preventing-nuclear-anarchy-nuclear-proliferation-and-american-security

Where nuclear governance teaches how to contain tail risks for a technology built on scarce physical inputs, the nanotechnology episode is best read as a cautionary tale about patent thickets: governance must prevent the IP system itself from becoming a barrier to diffusion. As nanoscale science took off in the 1990s and early 2000s, commentators warned that aggressive patenting of basic materials, fabrication methods, and platform technologies risked creating a "patent thicket"—dense, overlapping portfolios that would render freedom to operate prohibitively complex and expensive for later entrants.[374] In the nanotechnology context, this concern was sharpened by evidence that multiple firms and institutions were acquiring rights over related nanoscale building blocks and characterization techniques, raising the specter of "nanothickets" in which innovators would face not a single blocking patent, but a fragmented landscape of rights that each commanded hold-up power.[375] The quantum stack is structurally vulnerable to a similar dynamic: if foundational compiler intermediate representations, widely used error-mitigation routines, or generic architectures for trapped-ion or superconducting platforms are each fenced off by broad, overlapping patents, startups and public laboratories could find themselves negotiating a gauntlet of licenses before even testing new ideas. However, recent empirical data suggests this fear may be overstated in the current quantum landscape, and that the patent system as is provides optimal results. Patent filings have grown at a compound annual growth rate of 42% between 2001 and 2020 without yet evidencing systematic hold-up, suggesting that the sector is currently driven by robust entry rather than gridlock.[376] This growth is also highly concentrated: roughly 88% of quantum computing patents have been filed via China's office, the USPTO, or the WIPO route, and China accounts for about 60% of quantum-computing patent filings—suggesting that "thicket" risk will be shaped by a small number of jurisdictions and their institutional choices.[377] The governance challenge is therefore not to dismantle the patent system, but to manage the interface between patents and trade secrets—the "IP in superposition"—to prevent the enclosure of foundational scientific tools.[378]

The nanotechnology experience, however, also reveals that dire forecasts of gridlock are not destiny. In practice, firms and universities experimented with portfolio design, cross-licensing, defensive publication, and, in some cases, patent pools to manage overlapping rights.[379] The more successful arrangements treated patents as instruments for structuring collaboration—signaling where contributions would be rewarded and how downstream uses would be licensed—rather than as weapons for maximizing short-term extraction. As commentators have emphasized, a legal regime that relies on private intellectual property rights to stimulate innovation must be continuously calibrated so that those rights reward genuine technological advance without hardening into exclusionary market power that deters rivals from experimenting at the frontier.[380] For quantum governance, the analogue is an early, deliberate move toward disciplined claim drafting, standards-aligned interface patents, and interoperable licensing models for genuinely standard-essential innovations. If the most basic "keystone" elements of the stack—such as calibration interfaces, error-syndrome formats, or quantum-safe network handshakes—are encumbered by incompatible or aggressively priced rights, a "quantum thicket" will be hard to unwind. By contrast, if those elements are placed in pools with predictable fair, reasonable, and non-discriminatory (FRAND) terms or in carefully curated open-source regimes, while more differentiated, application-specific advances remain subject to conventional proprietary protection, the patent system can reinforce rather than undermine security-sufficient openness.

---

[374] Carl Shapiro, *Navigating the Patent Thicket: Cross Licenses, Patent Pools, and Standard Setting*, in 1 Innovation Policy and the Economy 119 (Adam B. Jaffe et al. eds., 2001), supra note …
[375] *See* Mark A. Lemley, *Patenting Nanotechnology*, 58 Stan. L. Rev. 601 (2005), https://law.stanford.edu/publications/patenting-nanotechnology/; and Graeme B. Clarkson, *The Problem of Patent Thickets in Convergent Technologies*, 1093 Ann. N.Y. Acad. Sci. 180 (2006), https://pubmed.ncbi.nlm.nih.gov/17312259/
[376] *See e.g.,* Mateo Aboy, Timo Minssen & Mauritz Kop, *Mapping the Patent Landscape of Quantum Technologies: Patenting Trends, Innovation & Policy Implications*, 53 INT'L REV. INTELL. PROP. & COMP. L. 853 (2022), https://link.springer.com/article/10.1007/s40319-022-01209-3
[377] *See* Ruane et al, *supra* note …
[378] *See* Gabriela Lenarczyk, Timo Minssen & Mateo Aboy, *IP in Superposition: Patents, Trade Secrets and Open Innovation in Quantum Information Technology*, in Quantum Technology Governance: Law, Policy and Ethics in the Quantum Era, (Mateo Aboy et al. eds., Springer forthcoming 2026)
[379] Shapiro, *supra* note 257, at 122–24.

## III. The Internet

The governance of the classical internet offers a blueprint grounded in open standards: interoperability through transparent, multi-stakeholder processes that reward implementability and diffusion. Rather than being designed by a single sovereign or corporate hierarchy, the internet's core protocols emerged from a distributed process coordinated by the Internet Engineering Task Force (IETF)[381] and related bodies, formalized through "Requests for Comments" (RFCs) that were debated in public, tested in the wild, and revised iteratively under a norm of "rough consensus and running code."[382] RFC 2026's description of the Internet Standards Process captures a distinctive institutional choice: standards were to be built on open participation, technical merit, and widespread implementation, with intellectual-property encumbrances managed so that protocols could be implemented by multiple competing vendors.[383] That architecture drove what economists would later describe as "permissionless innovation" that shaped the Web's growth trajectory: any actor who could speak the protocol could deploy new applications without seeking prior authorization from a central gatekeeper.[384]

For quantum networking and communication there is a powerful temptation to depart from this open-standards model. States and incumbents, conscious of quantum's security implications, may prefer vertically integrated, nationally controlled stacks—each with its own proprietary interfaces and cryptographic assumptions. The result would be a "quantum splinternet" in which cross-border interoperability is sacrificed to short-term control, amplifying the weaponization of interdependence and raising switching costs for users. The internet's open-standards story argues for a different path: a core of non-proprietary, publicly specified protocols for quantum key establishment, entanglement management, and quantum-safe routing, developed through transparent, multi-stakeholder processes and designed from the outset for crypto-agility and post-quantum interoperability. That path does not preclude national security overlays—classified profiles for defense networks, for example—but it keeps the default architecture aligned with security-sufficient openness, ensuring that firms and research institutions across allied and partner states can plug into a common quantum internet without navigating a maze of incompatible standards or discriminatory licensing. It also illustrates how standards bodies, when properly insulated from overt securitization, can serve as engines of long-run competitiveness for entire innovation systems rather than as instruments of narrow industrial policy.

---

[380] Herbert J. Hovenkamp, *Intellectual Property and Competition*, in Research Handbook on the Economics of Intellectual Property (Ben Depoorter, Peter Menell & David Schwartz eds., 2019), https://papers.ssrn.com/sol3/papers.cfm?abstract_id=2569129
[381] The Internet Engineering Task Force (IETF) (last visited Feb. 3, 2026), https://www.ietf.org/
[382] Scott Bradner, *The Internet Standards Process—Revision 3*, RFC 2026 (Oct. 1996), https://www.rfc-editor.org/rfc/rfc2026.html
[383] *Id.* § 10.
[384] *See* e.g., Tim Berners-Lee, Information Management: A Proposal (CERN, Mar. 1989), https://www.w3.org/History/1989/proposal-msw.html .

## IV. Artificial Intelligence

If nuclear and internet governance illustrate, respectively, the outer and inner edges of the control spectrum, the emerging regulatory regimes for artificial intelligence provide a blueprint for risk-based regulation: tiered duties keyed to context and use case rather than to an entire technology class. The European Commission's proposed Artificial Intelligence Act famously organizes obligations around tiers of risk—from "unacceptable" systems that are prohibited, through "high-risk" systems subject to *ex ante* conformity assessment and ongoing oversight, to lower-risk systems that face mainly transparency and post-market monitoring duties.[385] Rather than attempting to regulate "AI" in the abstract, the Act targets specific use cases—credit-scoring engines, biometric identification in public spaces, safety-critical systems—where systemic or rights-based harms are most acute. That approach has obvious attractions for quantum. A risk-based quantum governance regime would focus its heaviest scrutiny on applications that compress strategic timelines (for example, large-scale cryptanalysis against critical infrastructure, quantum-enhanced targeting and sensing in nuclear command-and-control, or highly invasive quantum neuro-sensing), while subjecting more benign applications—such as quantum-assisted materials discovery for clean energy or drug design—to lighter-touch documentation and audit. It also highlights that the same classes of computational capability that can accelerate weapons development, disinformation, or coercive surveillance can be redeployed for governance purposes—improving anomaly detection, system monitoring, and verification—so that AI and, in time, quantum–AI hybrids will sit simultaneously on the threat surface and in the enforcement toolkit.

Empirical work further suggests that AI/ICT diffusion can hinder democratization by strengthening incumbents' coercive and informational advantages—an effect that may be amplified if quantum-enabled sensing, secure communications, and quantum-AI integration

---

[385] Proposal for a Regulation of the European Parliament and of the Council Laying Down Harmonised Rules on Artificial Intelligence (Artificial Intelligence Act), COM(2021) 206 final (Apr. 21, 2021), *supra* note …

Page 86 of 111

expand state capacity faster than legal and institutional constraints.[386] Amodei's "technological adolescence" warning sharpens the timing problem: rapid AI capability gains can compress quantum TRL timelines and diffusion cycles, pulling forward national security threat horizons (including PQC migration and crypto-agility) while also expanding the toolset available for monitoring, verification, and defensive resilience.[387]

The AI experience also underscores the importance of building governance into the development lifecycle rather than bolting it on *ex post*. High-risk AI systems under the EU proposal must be designed with data-quality management, human oversight mechanisms, and post-deployment monitoring from the outset.[388] Translating that logic to quantum suggests that, for certain classes of systems, LSI-guided controls should be embedded directly into research and engineering workflows: capability thresholds that trigger internal red-team reviews, pre-deployment testing against agreed performance and safety benchmarks, and, where appropriate, external certification aligned with PQC and other security standards. Rather than categorically constraining qubit counts or algorithmic families, regulators and institutional review processes would focus on the contexts in which those tools are deployed and the combinations—quantum-AI hybrids, for example—that create qualitatively new risk profiles, thereby reducing the incentive to hide sensitive projects behind an impenetrable wall of secrecy while still addressing the most concerning modes of misuse.

V. Biotechnology

Biotechnology, finally, illustrates both the promise and the limits of precautionary, adaptive, and IP-based governance: staged containment and iterative oversight that can evolve with scientific practice and observed failure modes. The recombinant DNA revolution of the 1970s prompted a remarkable episode of voluntary self-regulation: the Asilomar Conference brought together molecular biologists, physicians, and lawyers to debate potential biohazards, leading to guidelines that temporarily constrained certain high-risk experiments and established graded containment standards for others.[389] That process was not a treaty and lacked formal enforcement, yet it had real effect: laboratories adjusted their practices, funders incorporated Asilomar-inspired conditions into grants, and national regulators built on the framework when crafting more formal biosafety regimes. In parallel, the rise of gene-sequencing and other upstream tools fueled debate over the "tragedy of the anticommons," with Heller and Eisenberg arguing that extensive patenting of basic research tools could fragment control rights and deter downstream innovation.[390] The biomedical 'anticommons' literature warns that fragmented upstream rights can deter downstream experimentation—an analogue risk if quantum enabling patents become excessively layered across measurement, control, and error-mitigation toolchains.[391]

---

[386] *See* e.g., C.Y.C. Chu et al., Why Does AI Hinder Democratization?, Proc. Nat'l Acad. Sci. U.S.A. (2025), https://pmc.ncbi.nlm.nih.gov/articles/PMC12088401/
[387] *See* Dario Amodei, *supra* note 2.
[388] *See* EU AI Act, art. 9.
[389] *See* Kathleen E. Hanna, *Asilomar and Recombinant DNA: The End of the Beginning*, in Biomedical Politics 261 (Kathi E. Hanna ed., 1991), https://www.ncbi.nlm.nih.gov/books/NBK234217/
[390] Michael A. Heller & Rebecca S. Eisenberg, Can Patents Deter Innovation? The Anticommons in Biomedical Research, 280 Science 698 (1998), https://www.science.org/doi/10.1126/science.280.5364.698

Subsequent empirical work has painted a more nuanced picture. Surveys of academic biomedical researchers suggest that, while patents and material transfer agreements do impose frictions, they rarely block projects outright; most investigators report that they are able to "invent around," negotiate licenses, or rely on informal norms of tolerance, with only modest delays.[392] Large-scale studies of biotechnology patenting similarly find limited evidence that dense webs of upstream rights systematically choke off follow-on research, even as they caution that vigilance remains warranted in subfields where access to particular materials or datasets is tightly held.[393] The lesson for quantum governance is not that patent-induced gridlock is impossible, but that innovation ecosystems possess adaptive capacities that law can either reinforce or erode. If clear licensing pathways, research exemptions, and shared repositories of non-rival inputs—benchmark circuits, open-access datasets, reference implementations—are deliberately cultivated, quantum researchers may be able to navigate dense IP landscapes without paralyzing deadlock. If, by contrast, foundational elements of the stack are locked behind opaque, rigid licensing or broad secrecy, adaptive workarounds will be correspondingly harder, and the system will lose exactly the kind of distributed problem-solving capacity that allowed biotechnology to avoid the worst anticommons scenarios.

Taken together, these historical cases suggest that governance for transformative technologies is better conceived as a dynamic lifecycle than as a single legislative event. Nuclear governance evolved from tight secrecy to controlled sharing and, eventually, to a formal treaty regime with specialized verification institutions. Internet governance moved from research-network experimentation to a mature, procedurally rich standards ecosystem that nonetheless must now retrofit stronger security properties onto protocols initially optimized for openness. AI regulation is currently mid-stream, as lawmakers and regulators test risk-based models in real time. Biotechnology has oscillated between episodes of precaution, waves of IP expansion, and a more recent appreciation of the system's own adaptive resilience. Across these domains, early legal and institutional choices created path dependencies that conditioned who could participate in innovation, how easily knowledge traveled, and where rents accrued. Quantum governance will be no different.

Comparative work on national innovation systems shows that states embed technology policy in broader institutional architectures—spanning education, finance, competition law, and industrial strategy—so the same formal governance blueprint can yield sharply divergent

---

[391] *Ibid.*

[392] John P. Walsh, Charlene Cho & Wesley M. Cohen, *View from the Bench: Patents and Material Transfers*, 309 Science 2002 (2005), https://pubmed.ncbi.nlm.nih.gov/16179461/

[393] David E. Adelman & Kathryn L. DeAngelis, *Patent Metrics: The Mismeasure of Innovation in the Biotech Patent Debate*, 85 Tex. L. Rev. 1677 (2007), https://www.law.berkeley.edu/files/Adelman_patent_metrics.pdf

Page **88** of 111

outcomes in the United States, Europe, China, and other jurisdictions.[394] Analyses of innovation systems in both advanced and developing economies likewise emphasize that policy instruments do not operate on a blank slate, but interact with legacy institutions, political coalitions, and existing sectoral strengths, which will shape how export controls, standards, IP rules, and public procurement actually bite in the quantum domain.[395] Recent comparative studies of the innovation systems of the United States, Korea, China, Japan, and Taiwan further underline that foundational and emerging technologies now sit at the core of national development models and security strategies, raising the stakes of aligning quantum governance with long-term objectives rather than with short-term political pressures.[396] Decisions now being made about export-control thresholds, IP strategies, standards participation, and public–private research contracts are not merely tactical moves in a transient race; they are constitutive choices that will set the trajectory of the quantum ecosystem for decades to come, determining whether security-sufficient openness and the LSI principle remain live options or are crowded out by a more brittle equilibrium of secrecy, fragmentation, and zero-sum control. Empirically, cross-national differences now show up in measurable indicators—patents, QPU availability, corporate signaling, workforce demand, and testbed deployment—making "innovation systems" analysis operational rather than merely descriptive.[397]

## X. Strategic Navigation for Quantum Innovators

The complex nexus of security controls and intellectual property strategies creates a perilous and uncertain landscape for the primary drivers of quantum innovation: startups, scaleups, and their investors. For these actors, the governance paradox is not an abstract policy debate but a matter of survival. The capital-intensive nature of quantum hardware and sensing platforms, the path dependence of software and control stacks, and the unusually long timelines to technical and commercial maturity force founders to raise large amounts of relatively patient capital while operating under deep scientific and regulatory uncertainty. Investors, in turn, typically condition such capital on the existence of a defensible portfolio of rights—patents, trade secrets, data, and know-how—capable of supporting future product lines, collaborations, and exit. Recent data on the quantum sector confirm that this pressure is intensifying: global funding for quantum-technology startups roughly doubled year-on-year to around two billion dollars in 2024, with a growing share coming from governments and strategic corporate investors who themselves must answer to national-security and economic-security mandates.[398] A young quantum firm thus confronts not only the standard start-up challenge of convincing backers that its technology is real and its business model credible, but also the additional burden of doing so in an environment where each disclosure, each partnership, and each licensing decision can trigger scrutiny from export-control authorities, security services, or foreign-investment review boards.

---

[394] Vyacheslav V. Volchik, Sergey S. Tsygankov & Artem I. Maskaev, *Evolution of the National Innovation Systems of the United States, the United Kingdom, China and Iran*, 16 Econ. & Soc. Changes: Facts, Trends, Forecast 284 (2023), https://www.semanticscholar.org/paper/Evolution-of-the-National-Innovation-Systems-of-the-Volchik-Tsygankov/8239494f89cfec6bbecde4dab672c2d074bfb795

[395] Stephen Feinson, *National Innovation Systems Overview and Country Cases* (Ctr. for Sci., Pol'y & Outcomes, 2003), https://cspo.org/library/national-innovation-systems-overview-and-country-cases/

[396] Robert D. Atkinson et al., *Understanding and Comparing National Innovation Systems: The U.S., Korea, China, Japan, and Taiwan* (Info. Tech. & Innovation Found. & Chey Inst. for Advanced Stud., Feb. 20, 2025), https://www2.itif.org/2025-itif-chey-national-innovation-systems.pdf

[397] *See* Ruane, J., Kiesow, E., Galatsanos, J., Dukatz, C., Blomquist, E., Shukla, P., "The Quantum Index Report 2025", MIT Initiative on the Digital Economy, Massachusetts Institute of Technology, Cambridge, MA, May 2025, https://qir.mit.edu/wp-content/uploads/2025/06/MIT-QIR-2025.pdf

I. The Dilemma of Patent Disclosure: Signaling versus Security

Within the United States and European Union, this dynamic crystallizes into a familiar but sharpened dilemma. Patent filings remain the canonical way to signal technical capability, stake out a slice of the emerging value chain, and build an asset base that can be priced, diligenced, and collateralized. Yet for dual-use technologies near the frontier, patenting is no longer a purely private act. In the United States, every application is screened under the Invention Secrecy Act, and those that, in the judgment of designated agencies, risk revealing capabilities detrimental to national security can be placed under secrecy orders that delay publication, restrict foreign filing, or effectively halt commercialization until the order is lifted.[399] In the European Union, patent disclosures increasingly intersect with a tightening dual-use regime under Regulation 2021/821 and its delegated updates, which now encompass a widening array of quantum-related items and associated "technology."[400] Where quantum computers, cryogenic systems, control electronics, or specialized materials fall within these lists, even the act of elaborating an enabling invention can, in practice, invite questions about whether the know-how described should itself be treated as controlled technology. The result is that a start-up that patents too aggressively may find that the very disclosures designed to attract investment also provide a roadmap for security agencies to fence, delay, or redirect its technology. In practice, this can mean that the same patent that secures a critical funding round simultaneously functions as a beacon for security agencies, triggering controls that, in the worst case, cripple the firm's ability to engage in international collaboration or access global markets and talent.

---

[398] OECD, *A Quantum Technologies Policy Primer*, OECD Digital Economy Papers No. 371 (2025), https://doi.org/10.1787/fd1153c3-en (pdf: https://www.oecd.org/content/dam/oecd/en/publications/reports/2025/01/a-quantum-technologies-policy-primer_bdac5544/fd1153c3-en.pdf), *supra* note ...
[399] Invention Secrecy Act of 1951, 35 U.S.C. §§ 181–188, https://www.law.cornell.edu/uscode/text/35/181 *supra* note ...
[400] Regulation (EU) 2021/821 of the European Parliament and of the Council of 20 May 2021 setting up a Union regime for the control of exports, brokering, technical assistance, transit and transfer of dual-use items, 2021 O.J. (L 206) 1, https://eur-lex.europa.eu/legal-content/EN/TXT/?uri=CELEX:32021R0821 (and subsequent Commission delegated updates to Annex I).

Abandoning the patent system is rarely an attractive alternative. A strategy premised entirely on trade secrecy may avoid some of the *ex ante* visibility that triggers security review, but it makes valuation more speculative, due diligence more difficult, and future enforcement against imitators or independent inventors more uncertain. Investors cannot search trade-secret registries; they must instead rely on representations, technical audits, and fragile covenants. The risk that a competitor will independently patent an overlapping invention or that a key employee will depart to a better-funded rival adds additional fragility. It is therefore unsurprising that sophisticated quantum firms increasingly adopt hybrid approaches: ring-fencing the most sensitive "crown-jewel" algorithms, calibration routines, or materials recipes as trade secrets while patenting non-core, enabling technologies and interfaces that showcase their technical depth without revealing the full path to performance and build a visible portfolio that reassures investors that there is something durable to underwrite.[401] In some cases, they supplement this with defensive publications or narrowly drafted claims designed to secure freedom to operate without giving away the last mile of engineering detail. The aim is not to retreat into opacity, but to manage a controlled gradient of disclosure—enough to unlock capital, talent, and partners, yet not so much that either competitors or regulators can trivially appropriate, weaponize, or immobilize the firm's capabilities.

## II. State Entanglement and Geopolitical Divergence

The same tension recurs in the firm's engagement with public funding, benchmarking programs, and government R&D contracts. Bayh–Dole–style frameworks, which assign ownership of federally funded inventions to private actors but reserve significant residual rights for the state, have become central to quantum, as governments deploy grants and cooperative agreements to accelerate platform development and anchor domestic industrial capacity.[402] For founders, the allure of such non-dilutive funding must constantly be weighed against the cost of granting broad government-purpose rights over what may later prove to be their most valuable assets. March-in rights, government-purpose licenses, export-control riders, and data-rights clauses all shape the conditions under which a startup can later license or sell its technology, especially if foreign partners or buyers are involved. At the same time, national quantum initiatives and flagship programs in the United States, Europe, and key allies offer non-dilutive capital, access to testbeds, and visibility through participation in benchmark and standards efforts that few early-stage firms can afford to decline.[403] A founder deciding whether to place a core subsystem inside a publicly funded consortium, or to keep it in a privately financed subsidiary, is therefore making a governance choice as consequential as any technical or commercial one. Those choices determine which elements of the stack will be encumbered by government rights, which will be free to travel under cross-border licensing arrangements, and which will be exposed to future political pressure should security concerns harden or alliances shift.

---

[401] Mark A. Lemley, *The Surprising Virtue of Treating Trade Secrets as IP Rights*, 61 STAN. L. REV. 311 (2008), https://www.stanfordlawreview.org/print/article/the-surprising-virtues-of-treating-trade-secrets-as-ip-rights/

[402] Bayh–Dole Act, 35 U.S.C. §§ 200–212, https://www.law.cornell.edu/uscode/text/35/part-II/chapter-18, *supra* note ...

[403] OECD, *Quantum Technologies as a New Paradigm for Digital Economies and Societies* (Policy Brief, 2025), https://doi.org/10.1787/e6664d58-en (pdf: https://www.oecd.org/content/dam/oecd/en/publications/reports/2025/02/quantum-technologies-as-a-new-paradigm-for-digital-economies-and-societies_0f18bb86/e6664d58-en.pdf).

For European firms, these questions are compounded by the EU's evolving economic-security toolbox. The same institutions that provide substantial research funding, infrastructure investment, and standardization platforms now also coordinate export controls, outbound investment screening, and foreign-subsidy disciplines, often explicitly naming quantum computing and communication as sensitive domains.[404] A quantum-hardware startup in, say, the Netherlands or Germany that joins a flagship pilot project may find that, as its systems mature, they are swept into new EU-level control lists or risk-based screening procedures designed to prevent leakages of critical capabilities to adversarial jurisdictions. For founders, the long-run implications are double-edged: EU-level controls can offer a more predictable and harmonized regulatory environment than a patchwork of national rules, but they can also narrow the field of permissible counterparties and limit the scope for bespoke solutions with individual member states. Strategic navigation thus requires both a granular understanding of the evolving legal instruments and a realistic assessment of the political trajectory behind them.

For startups in China and in jurisdictions closely intertwined with Chinese supply chains and capital, the calculus is different but no less fraught. On the one hand, Chinese firms benefit from extensive state support, long-term strategic planning, and a large protected domestic market, all embedded within a Military–Civil Fusion framework that explicitly targets quantum computing, quantum communication, and related enabling technologies as priority areas for integrated civilian–military development.[405] On the other hand, they face the mirror-image vulnerability of being cut off from high-end Western components, design tools, and cloud-based services through export controls, entity listings, and informal pressure on third-country suppliers. A Chinese quantum-hardware startup that relies on imported cryogenic refrigerators, specialized lasers, control electronics, or EDA software must assume that these inputs are contingent, subject to unilateral or coalition-based restrictions that may tighten with little warning. Its survival depends on the success of national drives for indigenous alternatives, its ability to design around Western-controlled chokepoints, and the willingness of intermediary states to absorb secondary sanctions risk by continuing to sell or re-export controlled items. In this environment, scientific excellence is necessary but not sufficient; survival depends equally on an ability to read and anticipate the shifting legal and geopolitical constraints that define which partnerships, markets, and exits will remain open over time.

This divergence in starting conditions—American and European firms operating in comparatively open innovation ecosystems but under increasingly dense security overlays,

---

[404] *EU Consolidates Export Control Powers to Bypass Russia*, FIN. TIMES (Sept. 16, 2025), https://www.ft.com/content/6b0648b4-3e7c-41d1-b96a-3da2b41e4fd3.

[405] Elsa B. Kania & John K. Costello, *Quantum Hegemony? China's Ambitions and the Challenge to U.S. Innovation Leadership*, Center for a New American Security (Sept. 12, 2018), https://www.cnas.org/publications/reports/quantum-hegemony, *supra* note …

Chinese firms operating in a more statist and securitized environment but with strong domestic industrial policy support—feeds back into the global geography of the quantum "full stack." The layers from materials, fabrication, and cryogenics through control systems, cloud access, middleware, and domain-specific applications are not evenly distributed; they cluster in a small number of hubs, under the jurisdiction of states that are also expanding their economic-security reach. Export-control coalitions, EU-level measures to bypass multilateral vetoes, and national scrutiny of outbound investment create de facto chokepoints in access to high-end hardware and manufacturing know-how, particularly for actors perceived as aligned with adversarial strategic projects.[406] Startups that anchor themselves in these hubs gain privileged access to talent, customers, and public funding, but they also become more tightly coupled to the security and industrial policies of their host states. Those that seek to operate from peripheral jurisdictions may enjoy more formal neutrality but can find themselves effectively excluded from the highest-performance infrastructure and the densest standards-setting networks.

III. Compute Scarcity as a Structural Determinant

Against this backdrop, the constraint that matters most for many quantum innovators is not only capital but compute, understood broadly to include access to high-quality qubits or qubit-equivalents, low-noise measurement chains, and reliable, scalable control and calibration cycles. For quantum-AI hybrids in particular, limited access to advanced quantum hardware functions as the analogue of GPU scarcity in classical machine learning: it restricts the size and frequency of final experimental runs, constrains the ability to explore alternative architectures and error-mitigation schemes, and, most importantly, throttles the feedback loops through which deployed systems generate empirical insight and structured problem instances for the next generation of designs and, in AI-adjacent settings, high-quality synthetic data for training and stress-testing subsequent generations of more capable systems.[407] The AI literature on compute scaling, from early analyses of exponential growth in training-run compute to more recent work on scaling laws and algorithmic efficiency, underscores how access to large, well-instrumented experiments can accelerate learning curves and widen the gap between those who can afford to run them and those who cannot.[408] In quantum, the same dynamic plays out under tighter physical and geopolitical constraints: high-end systems are few, expensive, and often embedded in national testbeds that privilege domestic or allied participants. Being compute-constrained therefore becomes not just an engineering inconvenience but a structural drag on innovation, a systemic brake on the pace at which new architectures, materials stacks, and application domains can be explored and deployed, one that interacts with export controls, cloud-access policies, and transnational standards efforts to shape the universe of feasible business models. Recent U.S. policy signals—accelerated federal permitting for data-center infrastructure and an executive push to promote exports of the U.S. AI technology stack—illustrate how compute and infrastructure are being treated as strategic enablers rather than neutral market inputs,[409] including state-directed efforts to close the AI-chip gap.[410]

---

[406] Henry Farrell & Abraham L. Newman, *Weaponized Interdependence: How Global Economic Networks Shape State Coercion*, 44 INT'L SEC. 42 (2019), https://doi.org/10.1162/isec_a_00351, *supra* note ...

[407] Dario Amodei & Danny Hernandez, *AI and Compute*, OpenAI (May 16, 2018), https://openai.com/blog/ai-and-compute/.

[408] Jared Kaplan et al., *Scaling Laws for Neural Language Models*, arXiv:2001.08361 (2020), https://arxiv.org/abs/2001.08361.

## IV. Governance as Design Space

Strategic navigation for quantum innovators must therefore treat the governance stack—IP, contracting, supply-chain architecture, location and corporate structuring, and participation in consortia and standards bodies—as a design space co-equal with the technical stack. The firm that reflexively maximizes short-term valuation by patenting everything, taking every grant on offer, and entering every available joint venture may discover that it has fragmented its rights among multiple sovereigns, encumbered its most valuable technology with far-reaching government-purpose licenses, and narrowed its exit options to a handful of politically acceptable acquirers. By contrast, a firm that consciously sequences disclosures, tailors its IP filings to avoid unnecessary technical detail, separates sensitive subsystems into entities with different ownership and jurisdictional profiles, and engages selectively with public programs can preserve strategic degrees of freedom even as it taps into public support. OECD work on anticipatory governance and international cooperation in quantum reinforces this point: the state is simultaneously a funder, regulator, and customer, and firms that understand how these roles interact can help co-produce guardrails that are both security-sufficient and innovation-preserving.[411]

For boards and founders, the practical implication is that governance cannot be relegated to a compliance function consulted late in the product cycle. It must be integrated into strategic planning from the outset, informed by scenario analysis that asks how the firm's options change under different evolutions of export-control coalitions, data-protection rules, security classifications, and standards trajectories. Corporate governance reforms can operationalize this by assigning board-level accountability for quantum-safe migration, export-control compliance, and supply-chain due diligence as enterprise-risk functions rather than ad hoc engineering tasks.[412] The heuristics developed earlier in this Article—security-sufficient openness and the LSI test, which counsels choices that are least trade-restrictive, security-sufficient, and innovation-preserving—are as applicable at the firm level as at the level of national or allied policy. A start-up deciding whether to host its primary IP in one jurisdiction rather than another, whether to license a subsystem broadly or keep it as an in-house differentiator, whether to join an international testbed or focus on a domestic niche, is in effect making LSI choices in miniature. Some of those choices will be forced by the surrounding legal environment; many, however, remain within the firm's control.

---

[409] *See* The White House, Exec. Order on Accelerating Federal Permitting of Data Center Infrastructure (July 2025), https://www.whitehouse.gov/presidential-actions/2025/07/accelerating-federal-permitting-of-data-center-infrastructure/; The White House, Exec. Order on Promoting the Export of the American AI Technology Stack (July 2025), https://www.whitehouse.gov/presidential-actions/2025/07/promoting-the-export-of-the-american-ai-technology-stack/; and America's AI Action Plan, AI.gov (last visited Feb. 3, 2026), https://www.ai.gov/action-plan

[410] Potkin, *supra* note …

[411] OECD, *Framework for the Anticipatory Governance of Emerging Technologies*, OECD Science, Technology and Industry Policy Papers No. 165 (2024), https://doi.org/10.1787/0248ead5-en.

[412] *See* e.g., Michael Siebecker, Revitalizing Corporate Governance for the Quantum Age, JUST SECURITY (May 21, 2025), https://www.justsecurity.org/113334/revitalizing-corporate-governance-quantum-age/

In this sense, the "strategic navigation" challenge for quantum innovators is not to find a magic jurisdictional arbitrage that escapes politics altogether, but to recognize that governance is now part of the competitive landscape. Firms that internalize this early, building cross-functional teams that bring together legal, technical, and geopolitical expertise, can turn what might otherwise appear as a thicket of constraints into a source of resilience and advantage, helping to shape the emerging norms and standards on which their own long-term viability depends. Those that continue to treat export controls, secrecy orders, and public-funding conditions as exogenous shocks to be managed case-by-case risk discovering, too late, that the most consequential design decisions in their journey were not only about qubits and algorithms, but about where, how, and under whose rules those qubits and algorithms were allowed to operate.

## XI. Pathways for Governance: From Incremental Reform to Systemic Change

The preceding chapters have argued that quantum governance cannot be reduced to a single regulatory instrument, because the most consequential risks and benefits of quantum systems arise from a moving intersection of uncertainty, dual-use dynamics, and strategic competition that is itself shaped by intellectual property and supply-chain structure. That is why a credible governance agenda must be built as a portfolio: it must deliver near-term risk reduction within existing legal architectures while preserving the political and institutional option value to adopt deeper reforms as technical capability clarifies and strategic behavior hardens. The point is not to draft a constitution for an unknowable technology, but to assemble a framework that is robust under surprise, resilient to capture, and disciplined by the consistent design principle developed throughout this Article: the LSI test. This rubric requires measures to be least trade-restrictive, security-sufficient to mitigate demonstrable threats, and innovation-preserving for the scientific commons, so that the governance response remains interoperable across borders to avoid a fragmented quantum order that is simultaneously less safe and less productive.

The near term already supplies an anchor for that portfolio. Post-quantum cryptography (PQC) is no longer a speculative concern but a concrete, time-bound migration problem with direct implications for financial stability, critical infrastructure, and the confidentiality of long-lived sensitive data.[413] The importance of that migration is underscored by the reality that the relevant standards are now fixed in implementable form, including NIST's lattice-based KEM standard

---

[413] Off. of Mgmt. & Budget, Memorandum M-23-02, Migrating to Post-Quantum Cryptography (Nov. 22, 2022), https://www.whitehouse.gov/wp-content/uploads/2022/11/M-23-02-Migrating-to-Post-Quantum-Cryptography.pdf, *supra* note …

and its lattice- and hash-based digital signature standards, which are designed for broad deployment across federal systems and, by extension, the vendors and supply chains that serve them.[414] That fact alone forces a governance choice: states can treat PQC as a narrow compliance exercise, or they can treat it as a template for a broader "standards-first" mode of quantum governance in which technical baselines, assurance mechanisms, and procurement levers do a large share of the work before any new treaty or comprehensive statute is even on the table.

What follows develops this portfolio through two pathways. The first focuses on incremental reforms that can be pursued immediately through patent administration, standards strategy, multilateral interpretive practice, and targeted security policy. The second examines systemic proposals that would reconfigure the institutional landscape, including new approaches to the public domain, treaty modernization, global research infrastructure, and durable verification capacity. In both pathways, the same question recurs: whether a given measure reduces strategic risk without triggering a race to the bottom in openness or a race to the top in restriction, and whether it can be implemented in a way that preserves scientific legitimacy and democratic accountability.

I. Incremental Reforms

Incremental reform is often dismissed as inadequate for technologies that carry the aura of existential disruption, yet the history of dual-use governance shows that durable regimes frequently begin as modest institutional upgrades that later become the scaffolding for more ambitious coordination. In the quantum context, the first such upgrade is administrative rather than legislative: patent offices remain the practical gatekeepers of the innovation system, and the quality of quantum patents will materially shape whether the field evolves through competitive entry and diffusion or through litigation-driven congestion. The U.S. Patent Public Advisory Committee has repeatedly emphasized that examination quality is inseparable from examiner training, search capacity, and the systematic identification of emerging-technology pressure points.[415] The European Patent Office has likewise framed quality as an institutional objective that depends on consistent search and examination practices and on feedback loops that reduce error and variance in grant decisions.[416] In practice, this points to a governance move that is both modest and high-leverage: patent offices should deepen specialized capacity in quantum-related art, not merely by hiring technically trained examiners, but by building examination protocols that treat quantum claims—especially those framed at a high level of abstraction—as unusually susceptible to overbreadth and strategic drafting.

The goal is not to weaken patent protection for quantum innovation, but to preserve the patent system's legitimacy by reducing the probability that low-quality grants become strategic assets in geopolitical competition or barriers to entry for smaller firms and academic spinouts. The United States has already institutionalized "patent quality" as a policy objective through sustained USPTO initiatives that emphasize clarity, search robustness, and stakeholder feedback, and those efforts provide a ready administrative channel for quantum-specific quality measures without requiring statutory change.[417] Patent quality is less about maximizing grant counts and more about reducing the fragility and noise of enforceable claims in downstream disputes. Recent shifts in PTAB inter partes review (IPR) denial patterns are reshaping challenger and patent-owner strategies and should be treated as a practical constraint on how 'patent leverage' plays out in quantum-adjacent disputes.[418] A complementary step is procedural rather than substantive: work-sharing arrangements such as the Global Patent Prosecution Highway can be leveraged more aggressively for quantum technologies so that credible prior art and search results in one jurisdiction are less easily arbitraged by filing strategies in another.[419] This is an incremental reform with systemic implications, because it strengthens the international "information layer" that helps the patent system function as a global filter rather than a set of siloed backlogs.

---

[414] Nat'l Inst. of Standards & Tech., Post-Quantum Cryptography (PQC) Project, https://csrc.nist.gov/projects/post-quantum-cryptography, *supra* note …
[415] U.S. Patent Pub. Advisory Comm., Annual Report 2024 (2024), https://www.uspto.gov/about-us/organizational-offices/patent-public-advisory-committee/ppac-annual-reports.
[416] Eur. Patent Off., Quality Report 2024 (2024), https://www.epo.org/en/about-us/statistics/quality/quality-report.

Standards governance provides a second domain in which incremental steps can produce disproportionate gains. The creation of ISO/IEC JTC 3 on quantum technologies is a signal that quantum standardization is moving from exploratory coordination to formal global infrastructure, including terminology, performance, interoperability, and assurance practices that will be difficult to unwind once deployed.[420] Yet standard-setting is no longer a neutral technical exercise; it is increasingly a theater of strategic influence, particularly where standards can embed architectural choices that favor certain supply chains, patent portfolios, or security models. NATO's articulation of quantum technologies as a priority area for alliance resilience and defense innovation reflects an allied recognition that interoperability and trusted ecosystems will be determined as much by standards participation as by R&D spending.[421] For democratic states, the incremental reform here is not to "politicize" technical committees, but to fund participation, coordinate technical positions *ex ante*, and treat standards as a governance instrument that can embed security and openness simultaneously—especially where assurance and certification frameworks can reduce the temptation to substitute sweeping restrictions for verifiable technical safeguards.

---

[417] U.S. Patent & Trademark Off., USPTO Launches Data-Driven Quality Initiative to Address Areas of Highest Deviation (Nov. 21, 2025), https://www.uspto.gov/subscription-center/2025/uspto-launches-data-driven-quality-initiative-address-areas-highest.

[418] *See* e.g., William Meunier et al., The PTAB Pendulum Swings: How IPR Denials are Reshaping Patent Owner and Challenger Strategies, Mintz (Aug. 28, 2025), https://www.mintz.com/insights-center/viewpoints/2231/2025-08-28-ptab-pendulum-swings-how-ipr-denials-are-reshaping

[419] Global Patent Prosecution Highway, About the GPPH (n.d.), https://www.uspto.gov/patents/basics/international-protection/patent-prosecution-highway-pph-fast-track.

[420] ISO/IEC JTC 3, Quantum Technologies (n.d.), https://www.iso.org/committee/10046463.html.

[421] N. Atl. Treaty Org., Summary of NATO's Quantum Technologies Strategy (Jan. 16, 2024), https://www.nato.int/cps/en/natohq/official_texts_222089.htm, *supra* note …

This standards strategy also interacts with intellectual property in ways that are both familiar and newly salient. As quantum technologies mature, the likelihood grows that key interoperability layers will depend on patents that are, in effect, unavoidable for implementers, which in turn raises the governance stakes of licensing commitments and dispute resolution mechanisms. WIPO's current strategy on standard essential patents is best understood as an effort to strengthen the institutional tools available to manage licensing frictions before they harden into systemic barriers to diffusion.[422] In parallel, WIPO's work on patent pools illustrates a pragmatic path for reducing transaction costs in fields where overlapping rights can otherwise translate into litigation risk and innovation delay.[423] The incremental governance opportunity is therefore twofold: states can encourage FRAND-consistent licensing practices through procurement and public-funding conditions, and international institutions can provide neutral venues to reduce the probability that IP congestion becomes an instrument of strategic pressure.

WIPO also has a distinct role that is not reducible to standards: it can serve as a multilateral convening point for patent-quality norms and for shared infrastructure that lowers the cost of high-quality examination. Its ongoing work on frontier technologies—including quantum—signals an institutional willingness to treat emerging technology governance as part of the IP ecosystem rather than as an external security add-on.[424] A particularly concrete incremental reform would be to expand multilingual, curated prior-art repositories that are examiner-friendly and that treat preprints and technical disclosures as first-class inputs into novelty and obviousness analysis, thereby reducing the gap between fast-moving science and slow-moving administrative search. The point of such repositories is not to suppress patenting, but to reduce the strategic advantage of filing broad claims in a domain where the underlying knowledge often appears first in non-patent literature.

Trade law provides a fourth incremental lever, though one that requires careful design to avoid normalizing broad security rationales. The TRIPS Agreement already contains a security exception that, on its face, can be invoked for measures a state considers necessary for the protection of its essential security interests.[425] The question is not whether that clause exists, but how it should be interpreted and operationalized for quantum measures in a way that preserves the credibility of the trading system and avoids carte blanche invocation. WTO dispute settlement has begun to give structure to security exceptions, including by emphasizing the relevance of good-faith application and the need to connect measures to articulated security interests rather than to disguised protectionism.[426] The WTO's TRIPS-specific jurisprudence has likewise clarified that security exceptions are not an evidentiary vacuum, even where deference remains substantial.[427] An incremental and realistic governance step, therefore, would be interpretive rather than textual: states could pursue an authoritative clarification—through statement or practice—that certain narrowly tailored quantum-related measures, such as restrictions on patentability or enforcement for demonstrably cryptography-breaking capabilities, can be evaluated under a structured good-faith rubric rather than treated as inherently illegitimate or inherently unreviewable.

---

[422] World Intell. Prop. Org., WIPO Strategy on Standard Essential Patents 2024–2026 (2024), https://www.wipo.int/seps/en/.

[423] World Intell. Prop. Org., Patent Pools and Antitrust—A Comparative Analysis (Mar. 2014), https://www.wipo.int/export/sites/www/ip-competition/en/documents/patent_pools_report.pdf.

[424] World Intell. Prop. Org., WIPO Conversation on Intellectual Property and Frontier Technologies (n.d.), https://www.wipo.int/en/web/frontier-technologies/frontier_conversation.

[425] Agreement on Trade-Related Aspects of Intellectual Property Rights art. 73, Apr. 15, 1994, Marrakesh Agreement Establishing the World Trade Organization, Annex 1C, 1869 U.N.T.S. 299, https://www.wto.org/english/docs_e/legal_e/27-trips_05_e.htm, *supra* note …

[426] Panel Report, Russia—Measures Concerning Traffic in Transit, WTO Doc. WT/DS512/R (Apr. 5, 2019), https://docs.wto.org/dol2fe/Pages/SS/directdoc.aspx?filename=q:/WT/DS/512R.pdf.

The most immediate security pathway, however, is not doctrinal interpretation but the operational migration of cryptographic infrastructure. U.S. executive policy has already treated the quantum threat to cryptography as a national-security planning problem rather than a distant research question, and it has done so in terms that explicitly connect quantum capability to the vulnerability of widely deployed cryptographic systems.[428] The Office of Management and Budget has reinforced that frame by directing agencies to plan and inventory cryptographic dependencies as part of a transition to quantum-resistant standards.[429] The Department of Defense has also issued guidance that treats PQC migration as a near-term readiness obligation with procurement and programmatic implications rather than as a discretionary modernization choice.[430] Incremental reform here is governance in its most practical form: it is the creation of inventory discipline, implementation timelines, vendor pressure, and testing regimes that convert a technical standard into a real-world baseline across critical systems.

Export controls remain unavoidable in a domain where certain quantum components, architectures, and know-how will be treated as enabling strategic capability, but incremental improvement is possible even in a high-friction field. The Wassenaar Arrangement continues to provide a multilateral template for dual-use controls that can be updated iteratively to match technical reality, thereby reducing unilateral overreach while limiting obvious gaps.[431] The United States has shown that quantum computing can be treated as part of broader advanced-technology control updates, though the governance challenge remains to define thresholds and scope in ways that do not inadvertently control basic research tools, widely available software, or generic scientific components.[432] The European Union's dual-use regime, including its periodic control list updates, illustrates a parallel effort to keep control definitions synchronized with technical change while maintaining the legal basis for targeted restrictions.[433] Incremental reform in this area is therefore less about creating new lists than about making controls more technically legible, more regularly updated, and more paired with license exceptions or cooperative research channels for allied collaboration, so that controls do not become a blunt substitute for strategy.

---

[427] Panel Report, Saudi Arabia—Measures Concerning the Protection of Intellectual Property Rights, WTO Doc. WT/DS567/R (June 16, 2020), https://docs.wto.org/dol2fe/Pages/SS/directdoc.aspx?filename=q:/WT/DS/567R.pdf.

[428] Nat'l Sec. Memorandum-10, Promoting United States Leadership in Quantum Computing While Mitigating Risks to Vulnerable Cryptographic Systems (May 4, 2022), https://www.whitehouse.gov/briefing-room/presidential-actions/2022/05/04/national-security-memorandum-on-promoting-united-states-leadership-in-quantum-computing-while-mitigating-risks-to-vulnerable-cryptographic-systems/, *supra* note …

[429] Off. of Mgmt. & Budget, Memorandum M-23-02, Migrating to Post-Quantum Cryptography (Nov. 22, 2022), https://www.whitehouse.gov/wp-content/uploads/2022/11/M-23-02-Migrating-to-Post-Quantum-Cryptography.pdf, *supra* note …

[430] Dep't of Def., Chief Info. Officer, Preparing for Migration to Post-Quantum Cryptography (Nov. 20, 2025), https://dodcio.defense.gov/Portals/0/Documents/Library/PreparingForMigrationPQC.pdf.

[431] Wassenaar Arrangement, List of Dual-Use Goods and Technologies and Munitions List (2024), https://www.wassenaar.org/control-lists/.

Outbound investment screening is emerging as a further incremental instrument, with direct relevance to quantum given its dependence on specialized talent and its sensitivity to capital-enabled ecosystem building. The United States has begun to formalize restrictions and notification mechanisms for certain national-security technologies, including quantum information technologies, in a way that treats outbound investment as a vector of capability transfer rather than merely as finance.[434] Treasury's implementing materials make clear that this is intended as a targeted, sector-specific tool rather than as a generalized capital control, which is crucial if the instrument is to retain legitimacy and predictability.[435] The European Commission has likewise moved toward an outbound investment review framework that explicitly includes quantum among technology areas deemed critical for economic security, signaling that the policy convergence is no longer hypothetical.[436] The incremental governance challenge is to design these regimes so that they focus on high-leverage transfers—large investments, control rights, and capability-building partnerships—while preserving lawful academic exchange and avoiding a chilling effect that would undermine allied innovation ecosystems.

Counterintelligence and law enforcement are often treated as separate from "governance," yet in a dual-use field they are part of the baseline conditions under which any governance framework operates. U.S. criminal law already treats economic espionage directed by foreign governments as a distinct and serious offense, and that legal infrastructure is technology-neutral

---

[432] Bureau of Indus. & Sec., Dep't of Com., Department of Commerce Implements Controls on Quantum Computing and Other Advanced Technologies Alongside International Partners (Sept. 5, 2024), https://www.bis.gov/press-release/department-commerce-implements-controls-quantum-computing-other-advanced-technologies-alongside, *supra* note …

[433] Eur. Comm'n, 2025 Update of the EU Control List of Dual-Use Items (2025), https://policy.trade.ec.europa.eu/enforcement-and-protection/trade-control-and-compliance/dual-use-controls_en, *supra* note …

[434] Exec. Order No. 14,105, Addressing United States Investments in Certain National Security Technologies and Products in Countries of Concern (Aug. 9, 2023), https://www.federalregister.gov/documents/2023/08/14/2023-17354/addressing-united-states-investments-in-certain-national-security-technologies-and-products, *supra* note …

[435] U.S. Dep't of the Treasury, Outbound Investment Security Program—Final Rule Materials (2025), https://home.treasury.gov/policy-issues/international/outbound-investment-program, *supra* note …

[436] Eur. Comm'n, Commission Recommendation on Reviewing Outbound Investments in Technology Areas Critical for the Economic Security of the Union (Jan. 2025), https://eur-lex.europa.eu/eli/reco/2025/63/oj, *supra* note …

enough to apply to quantum IP theft without new statutes.[437] What is frequently missing is not legal authority but prioritization, expertise, and trusted engagement channels between government and quantum labs. U.S. counterintelligence strategy has specifically identified critical and emerging technologies as focal points for hostile collection, and the inclusion of quantum information technologies in that category provides an institutional rationale for dedicated capacity-building and allied information sharing.[438] The FBI has also explicitly framed the protection of quantum science as part of a broader effort to prevent strategic technology theft, suggesting that operational attention is already shifting toward the sector.[439] Incremental reform, accordingly, is the normalization of security-by-design practices in quantum research environments—without collapsing academic openness into blanket suspicion—and the creation of alert and response channels that treat attempted infiltration as a shared ecosystem threat.

Supply-chain strategy is another incremental lever that is frequently overlooked because it sits between industrial policy and security policy, yet quantum hardware is unusually sensitive to fragile, specialized inputs. The EU's Critical Raw Materials Act reflects an explicit legal and policy attempt to address strategic dependencies and to structure diversification and resilience planning through a combination of monitoring, permitting, and strategic project designation.[440] The United States has pursued parallel supply-chain mapping and resilience strategies through executive-led frameworks that treat dependency reduction as a national economic security objective, even where quantum-specific tools remain underdeveloped.[441] The incremental governance move here is not to claim autarky, but to build transparent, periodically updated maps of single-point vulnerabilities—potentially formalized as a "Quantum Criticality Index" that guides stockpiling (e.g. through Project Vault) and substitution strategies for scarce isotopes like Helium-3—and then to align procurement, grants, and allied coordination to reduce those vulnerabilities over time.[442]

Public procurement itself can function as a governance instrument precisely because it translates abstract norms into enforceable contract requirements. PQC migration illustrates this mechanism in practice: once federal agencies are required to inventory, plan, and deploy quantum-resistant cryptography, vendors and integrators face strong incentives to deliver interoperable implementations and to internalize the relevant assurance practices.[443] Procurement can similarly require conformance to emerging quantum standards, documented security testing, and transparent claims about performance and error rates, which would reduce the space for "quantum-washing" and would strengthen trust in early deployments. This is increasingly salient as corporate communications referencing quantum computing have moved sharply into the mainstream, heightening incentives for premature capability signaling absent independently verifiable benchmarks.[444] Quantum-washing is not novel in form; it is analogous to how cryptographic module validation and standards-based procurement have historically shaped cybersecurity markets, and it can be adapted to quantum components as certification ecosystems mature.

---

[437] Economic Espionage Act of 1996, 18 U.S.C. § 1831, https://www.law.cornell.edu/uscode/text/18/1831.
[438] Nat'l Counterintelligence & Sec. Ctr., National Counterintelligence Strategy 2024–2028 (2024), https://www.dni.gov/index.php/ncsc-featured-content/ncsc-national-counterintelligence-strategy.
[439] Fed. Bureau of Investigation, Protecting Quantum Science and Technology (Apr. 12, 2024), https://www.fbi.gov/news/stories/protecting-quantum-science-and-technology.
[440] Regulation (EU) 2024/1252 of the European Parliament and of the Council establishing a framework for ensuring a secure and sustainable supply of critical raw materials (Apr. 11, 2024), https://eur-lex.europa.eu/eli/reg/2024/1252/oj, *supra* note …
[441] Exec. Order No. 14,017, America's Supply Chains, 86 Fed. Reg. 11,849 (Feb. 24, 2021), https://www.federalregister.gov/documents/2021/02/26/2021-04059/americas-supply-chains.
[442] *See* Cho, Kop & Lee, *supra* note …

Finally, a credible incremental agenda depends on human capital and interdisciplinary literacy. Labor-market indicators suggest the constraint is already binding: U.S. demand for quantum skills has almost tripled since 2018 (though the growth curve has recently moderated), reinforcing that governance capacity depends on scaling "quantum-literate" professionals across technical, legal, and security roles.[445] The National Quantum Initiative's annual reporting requirements and program descriptions underscore that quantum leadership is being pursued not only through R&D funding, but through coordination across agencies and through workforce development that spans academia, government, and industry.[446] The NSF's Quantum Leap Challenge Institutes program embeds education and training as a core institutional objective, reflecting an understanding that governance capacity is inseparable from the cultivation of practitioners who can navigate technical systems and their legal, ethical, and security constraints.[447] Work at Stanford RQT (Responsible Quantum Technology) and UNESCO similarly signals that the legitimacy of quantum innovation will depend on proactive engagement with societal implications rather than post hoc reassurance after deployment.[448] The incremental reform implication is straightforward: quantum curricula should integrate IP strategy, export control literacy, research security practices, and ethics as routine professional competencies, while legal and policy training should integrate enough technical grounding to reduce the probability that governance choices are driven by metaphors rather than mechanisms.

## II. Transformative Ideas and Big-Picture Proposals

---

[443] Off. of Mgmt. & Budget, Memorandum M-23-02, Migrating to Post-Quantum Cryptography (Nov. 22, 2022), https://www.whitehouse.gov/wp-content/uploads/2022/11/M-23-02-Migrating-to-Post-Quantum-Cryptography.pdf.
[444] *See* Ruane et al, *supra* note …
[445] *See* Ruane et al., *supra* note …
[446] Nat'l Quantum Initiative, NQI Annual Report FY 2024 (2023), https://www.quantum.gov/national-quantum-initiative/annual-reports/.
[447] Nat'l Sci. Found., Quantum Leap Challenge Institutes (QLCI) (Aug. 26, 2024), https://www.nsf.gov/funding/pgm_summ.jsp?pims_id=505770.
[448] Stanford Center for Responsible Quantum Technology Scholarship Repository, Stanford University 2025, https://purl.stanford.edu/hp536nb5631; UNESCO, Concept Note of the World Commission on the Ethics of Scientific Knowledge and Technology (COMEST) (2024), https://www.unesco.org/en/ethics-science-technology/comest.

Incremental reforms can reduce friction and risk, but they do not answer the deeper question of what quantum governance should become if the technology reaches a mature phase in which certain capabilities are strategically decisive and widely deployable. Systemic proposals therefore matter less as immediate blueprints than as directional commitments that can shape institutional investment now, before path dependence hardens. The first and perhaps most fundamental systemic question concerns the boundary between the public domain and proprietary control. Patent law already contains doctrinal limits that exclude abstract ideas and mathematical relationships from patentability,[449] and those limits take on heightened importance in quantum computing because the difference between an "algorithm" and an "implementation" can become a drafting exercise rather than a technical one.[450] The governance objective should be to keep foundational quantum algorithms, core mathematical insights, and basic scientific discoveries meaningfully available as building blocks, both to preserve innovation dynamics and to reduce the probability that critical interoperability layers become fragmented by exclusionary rights.

This public-domain orientation does not imply hostility to commercial return. Rather, it implies a structural distinction between layers of the stack: the closer a contribution is to core scientific knowledge or to cross-platform interoperability, the stronger the case that it should be governed as a public good, while more application-specific engineering—especially in hardware—can remain a domain in which patents, trade secrets, and investment incentives play a larger role. In high-risk domains, moreover, there is a plausible case that the patent system may be the wrong venue entirely. The United States already retains statutory mechanisms for imposing secrecy orders on inventions that would be detrimental to national security, and that authority provides an institutional basis for keeping certain capabilities out of public disclosure channels when the security case is compelling and narrowly bounded.[451] The systemic challenge is to ensure that such tools are used as exceptions under disciplined criteria rather than as normal governance, because overuse would erode scientific openness and would likely accelerate parallel capability development elsewhere.

A second systemic proposal would move beyond interpretive practice toward formal trade-law modernization. If quantum technologies become central to strategic stability and to the integrity of global digital infrastructure, the current ambiguity surrounding TRIPS security exceptions may become a persistent source of dispute, forum shopping, and retaliatory trade measures. A formal amendment that clarifies the scope of security exceptions—specifically adding a sub-paragraph "relating to quantum technology" to Article 73(b)—would provide legal transparency, but it would also risk normalizing a broader security carve-out that could be abused as industrial policy.[452] For that reason, any treaty-based modernization would need to be paired with procedural guardrails—such as notification expectations, periodic review, and a disciplined conception of good faith—so that the exceptional status of quantum does not become a precedent for generalized techno-nationalist fragmentation.

---

[449] *See* e.g., Burk, Dan L. & Mark A. Lemley, Policy Levers in Patent Law, 89 VA. L. REV. 1575 (2003) 29, https://www.jstor.org/stable/3202360
[450] 35 U.S.C. § 101, https://www.law.cornell.edu/uscode/text/35/101, *supra* note …
[451] Invention Secrecy Act of 1951, 35 U.S.C. § 181, https://www.law.cornell.edu/uscode/text/35/181, *supra* note …
[452] Kim Moloney & Saif Al-Kuwari, Public Policy Considerations of Quantum Computing, Sci. & Pub. Pol'y, scaf065 (advance article Nov. 30, 2025), https://academic.oup.com/spp/advance-article/doi/10.1093/scipol/scaf065/8361814.

A third systemic pathway is the creation of a normative "Quantum Accord" or "Qubits for Peace" initiative that functions less as enforceable law than as coordinated expectation-setting.[453] In the near term, the most plausible version of such an accord would be built around commitments that are politically and technically defensible: mutual restraint in using quantum capabilities to compromise the confidentiality of civilian critical infrastructure; cooperation on PQC migration support for less-resourced states; and shared baseline norms on the integrity of global financial and communications systems. The logic would resemble other domains where nonbinding commitments and iterative review have shaped behavior even without traditional enforcement, particularly when reputational costs and the desire to preserve collaborative access create real incentives to comply. The promise of such an accord is not that it eliminates rivalry, but that it clarifies red lines and lowers the probability that states treat quantum exploitation of civilian systems as "business as usual."

A fourth systemic proposal concerns research infrastructure and access. If large-scale quantum systems remain scarce and concentrated in a few corporate or national laboratories, the risk of capability concentration will interact with competition policy, scientific equity, and geopolitical leverage. One response is a "global quantum research zone" modeled on the institutional logic of shared big-science infrastructure, in which states pool resources to build and operate facilities that no single state would finance at optimal scale on its own.[454] The CERN model demonstrates that international laboratories can produce scientific excellence, technical diffusion, and durable collaboration even in a world of strategic competition, although it also illustrates that such institutions require carefully designed governance, access rules, and political insulation.[455] The key governance claim is not that quantum should mimic particle physics, but that shared infrastructure can be an instrument of stability by reducing the perceived need for exclusivity and by embedding collaboration in institutional form.

The most ambitious institutional pathway is creation of an international agency for quantum technologies, modeled loosely on the IAEA's safeguards and verification functions.[456] A safeguards-inspired International Quantum Agency[457] can be grounded in public international law accounts of strategic stability, transparency obligations, and verification—while remaining compatible with differentiated national security exceptions.[458] Building on existing institutional experimentation[459], and consistent with broader "control plane" governance framings in the Just Security series, such an agency could be given a mandate that extends beyond research collaboration to include safeguards, assurance, and (where feasible) verification.[460] The International Atomic Energy Agency provides the closest historical analogue for how an international body can combine peaceful-use promotion with inspection, standards, and compliance mechanisms under treaty authority.[461] A quantum agency would face distinctive obstacles, including the intangibility of software and know-how, the difficulty of verifying capabilities without revealing sensitive information, and the reality that private-sector actors may control a large share of relevant infrastructure. Yet those obstacles do not eliminate the need; they shift the focus toward what verification can realistically mean in the quantum domain. An early-phase agency could begin with technical baselines, incident reporting coordination, and assurance frameworks for quantum networks and cryptographic migration, and only later evolve toward more intrusive verification functions if a treaty-based nonproliferation logic emerges.

---

[453] Kop, A Bletchley Park for the Quantum Age (2025), *supra* note … *See also* Dwight D. Eisenhower, "Atoms for Peace," Address Before the Gen. Assembly of the United Nations (Dec. 8, 1953), https://www.eisenhowerlibrary.gov/sites/default/files/file/atoms_for_peace.pdf

[454] Convention for the Establishment of a European Organization for Nuclear Research (CERN) (July 1, 1953), https://home.cern/about/who-we-are/our-governing-bodies/cern-convention.

[455] CERN, About CERN and Its Mission (n.d.), https://home.cern/about/who-we-are/our-mission.

[456] *See* Treaty on the Non-Proliferation of Nuclear Weapons, July 1, 1968, 21 U.S.T. 483, 729 U.N.T.S. 161, https://treaties.un.org/doc/Publication/UNTS/Volume%20729/volume-729-I-10485-English.pdf

[457] *See* Mauritz Kop & Tracey Forrest, Global Quantum Governance: From Principles to Practice, CIGI Policy Brief No. 222, Feb 5, 2026, https://www.cigionline.org/publications/global-quantum-governance-from-principles-to-practice/

Intellectual property governance can also be reframed systemically through the creation of a quantum IP commons that lowers transaction costs while preserving incentives. Patent pools and cross-licensing mechanisms already provide precedents for structuring access in technology domains where overlapping rights create collective-action problems, and WIPO's analysis of patent pools highlights their potential to reduce licensing friction when carefully structured.[462] A more ambitious approach would be to adapt the logic of open-source governance to patents through structured licensing commitments, whether through defensive patent licenses or through alliances that pledge nonassertion against compliant participants.[463] The Open Invention Network illustrates that patent nonaggression commitments can support ecosystem growth when they are tethered to clear membership conditions and a defined commons.[464] The systemic question is how to design these mechanisms so that they do not become cartel-like tools that entrench incumbents, but instead function as entry-enabling infrastructure that accelerates diffusion and reduces litigation risk.

---

[458] *See* e.g., Elija Perrier & Mateo Aboy, Quantum Information Technologies and Public International Law: Strategic, Legal, and Geopolitical Dimensions, in QUANTUM TECHNOLOGY GOVERNANCE: LAW, POLICY AND ETHICS IN THE QUANTUM ERA (Mateo Aboy, Marcello Corrales Compagnucci & Timo Minssen eds., 2026), https://papers.ssrn.com/sol3/papers.cfm?abstract_id=5710562 .

[459] *See* e.g., Open Quantum Institute (OQI) Progress Report 2024, Geneva Sci. & Diplomacy Anticipator (GESDA) (last visited Feb. 3, 2026), https://www.gesda.global/report/oopen-quantum-institute-oqi-progress-report-2024

[460] *See* Michael Karanicolas, Governing the Quantum Revolution, JUST SECURITY (May 1, 2025), https://www.justsecurity.org/111044/series-governing-quantum-revolution/

[461] Statute of the International Atomic Energy Agency (Oct. 26, 1956), https://www.iaea.org/about/statute.

[462] World Intell. Prop. Org., Patent Pools and Antitrust—A Comparative Analysis (Mar. 2014), https://www.wipo.int/export/sites/www/ip-competition/en/documents/patent_pools_report.pdf.

[463] Defensive Patent License, Project Overview (n.d.), https://defensivepatentlicense.org/.

[464] Open Invention Network, License and Community Overview (n.d.), https://openinventionnetwork.com/joining-oin/license-agreement/.

Transparency infrastructure offers another systemic lever that can be built without waiting for treaties. A "quantum governance observatory" could consolidate standards, export control thresholds, major patent landscape indicators, and procurement baseline requirements into a single navigable architecture, making the governance environment legible to researchers and firms who currently face a fragmented and often opaque compliance terrain. The point of such an observatory is not surveillance but predictability: the more predictable and navigable the governance environment, the less likely it is that actors respond to uncertainty by over-classifying, over-patenting, or over-complying in ways that reduce openness without improving security. Over time, such transparency infrastructure would also produce data that can support evidence-based governance, including identifying the early formation of patent thickets, the diffusion of standards adoption, and the emergence of supply-chain choke points that warrant targeted diversification.

Ethical governance, finally, must be treated as systemic rather than ornamental if quantum technologies are to retain social legitimacy during rapid deployment. UNESCO's engagement with quantum ethics signals an emerging expectation that quantum innovation will be evaluated not only by technical performance but by its implications for privacy, fairness, and human agency, particularly as quantum sensing and quantum-AI convergence expand the scope of inference and surveillance.[465] The OECD has likewise emphasized that quantum policy must address not only innovation incentives but also digital security, privacy risks, supply-chain constraints, and the distribution of benefits and access.[466] A systemic ethical initiative would therefore link technical assurance, human-rights safeguards, and access considerations into an integrated governance agenda, rather than relegating ethics to voluntary principles that are invoked only after harms materialize.

A particularly compelling systemic frame for integrating these elements is resilience planning for "Q-Day" as a global public-security project. The procedural logic is familiar: treat the cryptographic transition as a shared infrastructure upgrade with target dates, measurable milestones, and assistance channels for states and sectors that would otherwise lag. The novelty is not the idea of coordinated migration, but the scale and consequence of failure: uneven adoption could create weak-link vulnerabilities and could widen security divides, while premature claims of quantum breakthroughs could trigger instability if they are not met with calibrated response mechanisms. That is why PQC migration governance should be paired with simulation exercises, verification discipline for extraordinary capability claims, and international support mechanisms that reduce the temptation for opportunistic exploitation of lagging systems.

---

[465] UNESCO, Concept Note of the World Commission on the Ethics of Scientific Knowledge and Technology (COMEST) (2024), https://www.unesco.org/en/ethics-science-technology/comest.
[466] *See* OECD, A Quantum Technologies Policy Primer, OECD Digital Economy Papers No. 371 (Jan. 28, 2025), https://www.oecd-ilibrary.org/science-and-technology/a-quantum-technologies-policy-primer_7f33e2b6-en, *supra* note …

III. Open Research Questions

The portfolio described above is implementable, yet it is also incomplete in the way any serious governance framework for an emerging strategic technology must be incomplete. Quantum capability remains difficult to measure in ways that map cleanly onto strategic effect, and the field's rapid evolution makes static legal categories brittle. A central research frontier is therefore the development of governance-relevant metrics: measures that translate laboratory performance into application-level capability and that can serve as a basis for procurement decisions, export-control thresholds, and, in a more ambitious future, confidence-building measures between rivals. The challenge is not simply technical benchmarking; it is the creation of metrics that are resistant to gaming, meaningful across architectures, and interpretable by institutions that must make real decisions under uncertainty. One practical starting point is to build these governance metrics on publicly indexed QPU and benchmarking datasets (covering >200 QPUs across 17 countries) and to translate "core" device metrics into application-style performance measures—analogous to lap-times rather than marketing proxies—so procurement and controls are tied to verifiable capability, not slogans.[467]

Strategic stability raises parallel research demands. The most acute concern remains the relationship between quantum-enabled code-breaking, communications security, and nuclear command and control, yet the stability problem is broader than deterrence alone. It includes the integrity of global financial messaging, the resilience of critical infrastructure, and the governance of quantum sensing in domains where strategic advantage could be gained without crossing traditional thresholds of armed attack. Research is needed on how quantum capabilities interact with existing doctrines of espionage, sovereignty, and the law of state responsibility, and on whether the emerging quantum threat environment pushes toward new norms of restraint or instead accelerates a shift toward covert exploitation as a default strategic posture.

Economic governance is another underdeveloped frontier. If quantum computing becomes widely accessible primarily through a small number of cloud providers, traditional competition-policy tools may prove inadequate to address the risks of ecosystem lock-in, discriminatory access, and monopoly pricing in markets where switching costs are high and where interoperability is constrained by standards and IP. The governance challenge will not be solved by generic antitrust rhetoric; it will require careful analysis of how quantum markets differ from classical cloud markets, how standard essential patents and licensing practices shape entry, and how procurement and public funding can be used to preserve contestability.

Legal institutions will also confront novel evidentiary and procedural questions as quantum capability increases. Courts and regulators will be forced to decide what weight to give to evidence derived from quantum-enhanced inference or from the breaking of previously secure encryption, and whether such evidence should be treated as inherently unreliable, presumptively admissible, or governed by new procedural safeguards. These questions sit at the boundary of criminal procedure, privacy law, and cybersecurity policy, and they will require doctrinal development that does not yet exist in mature form.

---

[467] *See* Ruane et al., *supra* note …

Finally, convergence governance demands sustained attention. Quantum technologies will not mature in isolation; they will converge with advanced AI systems, with biotechnology, and with networked autonomy in ways that blur sector-specific governance lines. That convergence complicates institutional design, because it raises the likelihood that governance failure in one domain propagates into another, and it increases the probability that the most significant harms emerge from system interactions rather than from any single technology. The implication is that quantum governance should be designed to interoperate with AI governance, cybersecurity governance, and export-control governance, and that institutional silos should be treated as governance risks rather than as mere bureaucratic inconvenience.

Quantum governance should proceed through layered instruments that combine high-quality patent administration, strategic standard-setting, disciplined security policy, and operational cryptographic migration in the near term, while investing in the institutional capacity—ethical, legal, and technical—to adopt deeper reforms if and when quantum capability reaches a threshold where systemic risk cannot be managed through incremental tools alone. The governance choices made now will shape whether quantum technology becomes primarily a domain of fragmented rivalry and covert exploitation, or whether it can evolve—despite rivalry—within a framework that preserves innovation, protects security, and distributes benefits in ways that sustain legitimacy. The pathway to that outcome is narrow but real, and it begins with governance that is technically grounded, legally disciplined, and strategically self-aware.

## XII. Conclusion: Implementing the LSI Test to Secure the Quantum Industrial Commons

This Article has argued that the nexus of quantum technology, intellectual property, and national security is no longer best understood as a clash of abstract principles, but as a problem of institutional engineering under uncertainty. If powerful AI is entering a capability-acceleration phase consistent with Amodei's "adolescence" warning—and if AI can shorten quantum TRL timelines and diffusion cycles—then LSI is best read as a time-critical control plane: it keeps restrictions narrow, auditable, and revisable before deployment dynamics lock in irreversible harms and strategic instability. The central contribution—the least-trade-restrictive, security-sufficient, innovation-preserving (LSI) test—operationalizes values-based deterrence by denial in the quantum era by translating competing imperatives into an administrable decision rule.

Applied across the quantum stack and its enabling inputs, LSI keeps the system coherent across IP doctrine and disclosure strategy, research security and publication controls, export and services controls, procurement baselines (including PQC and crypto-agility), standards and certification, and resilience measures such as Project Vault for quantum-decisive critical-minerals reserves. The objective is not maximal restriction. It is to secure the industrial commons by preserving coalition-scale innovation in an "allowed subspace," while imposing default denial and heightened assurance only for the narrow set of flows that would predictably accelerate adversarial capability in "red-zone" domains. As a stability measure, that same LSI logic operates as an intervention: by narrowing and hardening only the specific quantum-enabled pathways that would most predictably compress warning time, erode second-strike survivability, or enable coercive gray-zone faits accomplis—including a Taiwan scenario where denial hinges on resilient swarm-scale autonomy under GNSS and communications attack—it aims to reduce incentives for preemption while sustaining coalition-scale innovation.

The strategic stakes now justify coalition architecture rather than piecemeal national measures. Recent U.S. government assessments underscore that quantum sensing and cryptanalysis could yield disproportionate operational advantages—including by eroding stealth and accelerating intelligence exploitation—making clear that coalition deterrence by denial must target enabling inputs and chokepoints, not only headline qubit breakthroughs, during the window in which PQC migration and allied resilience measures are still being brought fully online. More broadly, current U.S. strategic reporting and planning—including the U.S.-China Economic and Security Review Commission's "Quantum First" by 2030 framing and the 2025 U.S. National Security Strategy's emphasis on U.S. dual-use technology and standards leadership—treat quantum as emerging infrastructure rather than a distant research horizon. The practical risk is not "fragmentation" in the abstract, but the splintering of technical standards, certification regimes, QCaaS access rules, and PQC interoperability baselines into competing blocs—raising transaction costs for allies, lowering the price of adversarial learning, and hardening markets into incompatible ecosystems. To mitigate that trajectory, this Article supports a standards-first approach and "one test, many markets" assurance pathways, and it advances an institutional option: a fit-for-purpose International Quantum Agency that couples access to safeguards—focused on standards profiles, assurance and audit mechanisms for high-risk capabilities, and transparency commitments—mirroring IAEA-style safeguards logic without importing nuclear exceptionalism wholesale.

Finally, quantum governance should be empirically anchored, deliberately reflexive and historically literate. The last three decades of governing AI, biotechnology, nanotechnology, nuclear technology, and the internet offer recurring lessons: the costs of under-enforcement and regulatory capture; the strategic damage of poorly designed IP incentives; and the long-term lock-in effects of early standards and platform choices. In the quantum domain, the imperative is to force timely, calibrated "measurement" rather than drift in superposition: to build the legal, technical, and institutional control plane before capability, diffusion, and systemic rivalry collapse into a strategic reality that is far more expensive to reverse.

The deeper objective at the nexus of quantum technology, intellectual property, and national security is therefore governance by design: embedding democratic values into the architecture of quantum systems before they harden into durable infrastructure and hard-power assets. Done well, responsible quantum technology by design strengthens standards leadership and coalition interoperability, reinforcing democratic resilience in the quantum era. It is also an allied diffusion strategy: by making trust properties verifiable and procurement-relevant, RQT-by-design can shape adoption pathways beyond the coalition, including in the majority world. If privacy, civil liberties, auditability, interoperability, and open inquiry are incorporated into technical architectures and the IP arrangements that shape them, authoritarian approaches—often optimized for surveillance and selective interoperability— face higher assurance burdens and integration costs in high-trust markets, narrowing their pathways to coalition standards and procurement, and therefore to global adoption.

Deterrence by denial in the quantum era thus depends on implementing the LSI test as a coalition control plane to secure the industrial commons before today's probability cloud collapses into an irreversible strategic fact.

* * *

Appendix A. LSI Playbook (One-Pager): Deterrence by Denial for Democratic Resilience

**Implementing the LSI Test to Secure the Quantum Industrial Commons**

*Portable workflow for applying the least-trade-restrictive, security-sufficient, innovation-preserving (LSI) test across the quantum stack's physical, digital, and collaborative fronts.*

**Strategic objective:** Operationalize security-sufficient openness: accelerate coalition innovation in trusted pathways, while denying or tightly constraining narrowly defined red-zone flows that predictably accelerate adversarial capacity. Values-based guardrail: embed privacy, civil liberties, auditability, interoperability, and open inquiry into architectures, standards, and IP arrangements.

**LSI decision logic (apply before publication, contracting, procurement, and cross-border access).**

| L - Least trade-restrictive | S - Security-sufficient | I - Innovation-preserving |
|---|---|---|
| Objective & scope. Prefer standards/procurement/contract tools. Design for mutual recognition. | Threat model + control point. Thresholds + verification. QCI snapshot + "catch-up" filter. | Preserve diffusion & contestability. Use secure enclaves/tiered access. Review + sunset triggers. Values-by-design: prefer auditable, interoperable, privacy-preserving architectures. |

**Rapid workflow (first-pass LSI determination).**
1. Map the governed flow (hardware, software, data, services, know-how) and locate the control point.
2. Assign zone defaults across fronts: Green (open/accelerate), Amber (guarded), Red (default denial with bounded exceptions).
3. Quantify exposure: QBoM + QCI snapshot (substitutability, concentration, qualification lead time) for critical inputs.
4. Select the least restrictive lever that is still security-sufficient; record an LSI Statement; implement assurance (identity, access, logging/audit) and set a review date.

**LSI Governance Matrix (checklist).**

| Zone (apply LSI before classify/restrict/license). | Strategic profile (examples; catch-up). | Policy posture | Key instruments (authority check). |
|---|---|---|---|
| Green | Foundational/high diffusion; e.g., PQC Algorithms, Basic Theory, Metrology Standards; Catch-up: easy. | ACCELERATION: open publication; standardization; interoperability. | Standards-first; PQC-by-default procurement; disciplined IP disclosure; FRAND/SEP + interoperability profiles; mutual recognition of PQC assurance with allied/Bletchley Park partners. |
| Amber | Dual-use/high integration; e.g., Quantum Sensing for Navigation, QKD Hardware, Cryo-Coolers; Catch-up: mixed. | MONITORED ACCESS: trust-but-verify; end-use checks; KYC cloud gates. | Tiered access + secure enclaves; controlled technical data; logging/audit; selective licensing; Genesis Mission/ASSP closed-loop R&D; License Exception IEC; QCI supply-chain monitoring; QCaaS IAM/KYC gates. |
| Red | Mission-critical/low substitutability; e.g., Error-Corrected Qubits, Isotopically Enriched Silicon, Cryptanalysis; Catch-up: hard. | POLICY OF DENIAL: strict scrutiny; presumption of denial for non-allies. | Default denial; heightened assurance; pre-publication review; services controls/QCaaS gates; export + services controls; investment screening; time-bounded Invention Secrecy Act orders; supply-chain denial/stockpiles; IP: modular trade secrecy + narrow patents. |

**Minimum artifacts:** LSI Statement; QBoM + QCI snapshot; IP/disclosure posture memo; contracting/data-rights addendum; QCaaS access plan; procurement baseline (PQC/crypto-agility/interoperability); standards participation protocol.

**Authority anchors (as cited in the Article):** USCC "Quantum First" by 2030; 2025 U.S. National Security Strategy; Genesis Mission/ASSP; Section 232 critical-minerals actions; NIST PQC standards; BIS and allied export-services controls; coalition mutual-recognition pathways.

**End state:** a coalition control plane that secures the quantum industrial commons by raising adversary costs where it matters most, while preserving standards-first interoperability and innovation at scale.